\documentclass{jfm}

\pdfoutput=1

\usepackage{graphicx}
\usepackage[usenames]{color}
\usepackage{amssymb}

\newcommand{\go}{\mathcal{O}}
\newcommand{\be}{\begin{equation}}
\newcommand{\ee}{\end{equation}}
\newcommand{\bea}{\begin{eqnarray}}
\newcommand{\eea}{\end{eqnarray}}
\newcommand{\bfig}{\begin{figure}}
\newcommand{\efig}{\end{figure}}
\newcommand{\bc}{\begin{center}}
\newcommand{\ec}{\end{center}}
\newcommand{\btab}{\begin{tabular}}
\newcommand{\etab}{\end{tabular}}
\newcommand{\dr}{\partial}

\let\oldepsilon\epsilon
\let\epsilon\varepsilon
\let\varepsilon\oldepsilon

\newcommand{\Fr}{{\mathcal{F}}}
\newcommand{\fr}{{\mathcal{F}}}
\newcommand{\Rey}{{\mathcal{R}}}

\newcommand{\dz}{\partial_z}

\title[]{Bedforms in a turbulent stream.\\Part 1: Turbulent flow over topography} 
\author[A. Fourri\`ere, P. Claudin and B. Andreotti]
{A\ls N\ls T\ls O\ls I\ls N\ls E \ns F\ls O\ls U\ls R\ls R\ls I\ls \`E\ls R\ls E ,  \ns P\ls H\ls I\ls L\ls I\ls P\ls P\ls E \ns C\ls L\ls A\ls U\ls D\ls I\ls N \and \ns B\ls R\ls U\ls N\ls O\ns A\ls N\ls D\ls R\ls E\ls O\ls T\ls T\ls I \ns} 
\affiliation{
Laboratoire de Physique et M\'ecanique des Milieux H\'et\'erog\`enes\\
PMMH UMR 7636 CNRS-ESPCI-P6-P7,\\
10 rue Vauquelin, 75231 Paris Cedex 05, France.}
\pubyear{???}
\volume{???}
\pagerange{??--??}
\date{\today}
\begin{document}
\maketitle

\begin{abstract}
In the context of subaqueous ripple and dune formation, we present here a Reynolds averaged calculation of the turbulent flow over a topography. Using a Fourier decomposition of the bottom elevation profile, we perform a weakly non-linear expansion of the velocity field, sufficiently accurate to recover the separation of streamlines and the formation of a recirculation bubble above the some aspect ratio. The normal and tangential basal stresses are investigated in details; in particular, we show that the phase shift of the shear stress with respect to the topography, responsible for the formation of bedforms, appears in an inner boundary layer where shear stress and pressure gradients balance. We study the sensitivity of the calculation with respect to (i) the choice of the turbulence closure, (ii) the motion of the bottom (growth or propagation), (iii) the physics at work in the surface layer, responsible for the hydrodynamic roughness of the bottom, (iv) the aspect ratio of the bedform and (v) the effect of the free surface, which can be interpreted in terms of standing gravity waves excited by topography. The most important effects are those of points (iii) to (v), in relation to the intermixing of the different length scales of the problem. We show that the dynamical mechanisms controlling the hydrodynamical roughness (mixing due to roughness elements, viscosity, sediment transport, etc) have an influence on the basal shear stress when the thickness of the surface layer is comparable to that of the inner layer. We evidence that non-linear effects tend to oppose linear ones and are of the same order for bedform aspect ratios of the order of $1/10$. We show that the influence of the free surface on the basal shear stress is dominant in two ranges of wavelength: when the wavelength is large compared to the flow depth, so that the inner layer extends throughout the flow and in the resonant conditions, when the downstream material velocity balances the upstream wave propagation. 
\end{abstract}

\section{Introduction}
The formation of ripples and dunes at the surface of an erodible sand bed results from the interplay between the relief, the flow and the sediment transport. The aim of these two companion papers is to propose a coherent and detailed picture of this phenomenon in the generic and important case of a unidirectional turbulent stream. This first part is devoted to the study of the stationary flow over a wavy rough bottom. In the second part we propose a common theoretical framework for the description of the different modes of sediment transport. Hydrodynamics and transport issues at hand, we then revisit the linear instability of a flat sand bed submitted to a water shear flow and show that, in contrast to ripples, subaqueous dunes cannot form by a primary linear instability.

It has long been recognised that the mechanism responsible for the formation and growth of bedforms is related to the phase-lag between sediment transport and bed elevation (\cite{K63,R65,K69,S70,H70,P75,EF82,McL90}). It has been shown in the context of aeolian dunes that this lag comes from two contributions, which can be considered as independent as the time scale involved in the bed evolution is much slower than the hydrodynamics relaxation (\cite{ACD02,KSH02,V05}). First there is a shift between the bed and the basal shear stress profiles. This shift purely results from the hydrodynamics and its sign is not obvious \emph{a priori}, i.e. the stress maximum can be either upstream or downstream the bed crest depending on the topography or the proximity of the free surface. The second contribution comes from the sediment transport: the sediment flux needs some time/length to adapt to some imposed shearing. This relaxation mechanism induces a downstream lag of the flux with respect to the shear. When the sum of these two contributions results in a maximum flux upstream the bed crest, sediment deposition occurs on the bump, leading to an unstable situation and thus to the amplification of the disturbance. In part 1, we shall focus on the first of these contributions, the second one being treated in part 2.
\begin{figure}
\includegraphics{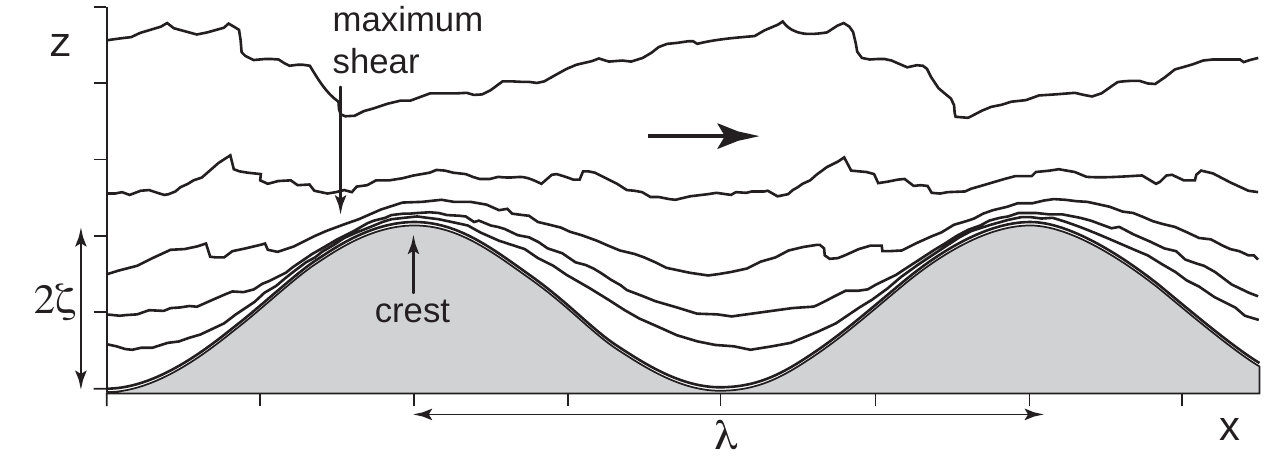}
\caption{Iso-velocity lines over a wavy bottom (data after \cite{PKAR07}). The fluid is flowing from left to right along the $x$-axis. $z$ is perpendicular. The bottom profile is $z=Z(x)=\zeta \cos(kx)$. In that experiment, the wavelength and amplitude of the bumps are $\lambda=2\pi/k=3.2$~m and $\zeta=0.08$~m, for a water depth $H=0.6$~m). The point of maximum shear on the bump (where the lines are squeezed) is located upstream the crest.
\label{WavySchematic}}
\end{figure}

We consider here the generic case of a flow over a fixed sinusoidal bottom of wavelength $\lambda$ (see figure~\ref{WavySchematic} for an illustration of the geometry and some notations). In order to obtain the basal shear stress and in particular its phase shift with respect to the topography, the equations of hydrodynamics must be solved in this geometry. The case of viscous flows has been investigated by \cite{B59,B78,CFKMcLSY82,CH00,L03,VL05}. The first attempts to model the high Reynolds number regime in the context of ripples and dunes in rivers have dealt with potential flows (\cite{K63,R65,CF00}), for which the velocity field does not present any lag with respect to the bottom. The shallow-water approximation (\cite{G70}) implies that the bedforms spread their influence on the whole depth of the flow. However, patterns only have a significant influence within a vertical distance on the order of their wavelength. It is then crucial to compute explicitly the vertical flow structure, taking into account the turbulent fluctuations.

In order to overcome the flaws of the perfect flow, constant eddy viscosity closures have been tried to improve Kennedy's original model (\cite{E70,S70,F74}). Further progress has been made by \cite{R80}, who used a more sophisticated modelling with an additional equation on the turbulent energy and a closure which involves a Prandtl mixing length in the expression of the eddy viscosity. \cite{SB84} made use of the same turbulent modelling, but in the case of an infinite water depth. A mixing length approach was also used by \cite{KM85} to improve Benjamin's laminar description.

In the meteorological context of atmospheric flows over low hills, a deep and fundamental understanding of the physics of turbulent flows over a relief has been developed from the 70's (see the review by \cite{BH98}). Starting with the seminal work of \cite{JH75}, further refined by \cite{S80} and \cite{HLR88}, the gross emerging picture is that the flow can be thought of as composed of two (or more) layers, associated with different physical mechanisms and different length scales. \cite{JH75} have been able to compute analytically the basal shear stress for asymptotically large patterns, under an infinite flow depth assumption. Their ideas have been discussed in a rather vast literature. The predictions of these calculations, and in particular this layered structure of the flow, has been compared with experiments (see e.g. \cite{BHR81,GI89,FRBA90}), or field measurements on large scale hills (see e.g. the review paper by \cite{TMB87}), with a good degree of success, especially on the upstream side of the bumps. Moreover, they have been tested against the results of the numerical integration, in various configurations, of Navier-Stokes equations closed with different turbulent closures (\cite{T77a,T77b,RT81,AXT94}). The relevance of this approach for the description of the flow and the stresses around aeolian sand dunes has also been investigated (see e.g. \cite{WHCWWLC91}), and is amongst the current directions of research in that community (\cite{W01}).

Because the prediction of the stable or unstable character of a flat sand bed submitted to a turbulent shear flow is very sensitive to the way both hydrodynamics and transport issues are described and intermixed, we find it useful to discuss at length, in the two parts of this paper, the different mechanisms and scales involved at the different steps of the modelling. It is indeed particularly revealing that, despite the fact that the approach presented here is very close to those of \cite{R80} and \cite{C04} and has been motivated by these works, we basically disagree with their conclusions, especially that river dunes are initiated by the linear instability of a flat bed. The detail discussion we provide here gives also the opportunity to revisit the still debated question of the subaqueous ripple size selection (see \cite{C06} and references therein), as well as the important issue of the classification of bedforms (\cite{A90}).

This article is structured as follows. In the next section, we briefly recall the equations for the base flow over a uniform bottom. We then study the linear solution in the case of wavelengths much smaller than the flow depth. Importantly, the sensitivity of these linear results with respect to various changes in the modelling is tested in sections \ref{controlzzero} and \ref{robust}. Section \ref{NL} is devoted to the derivation of the first non-linear corrections. In section \ref{FS}, we investigate the effect of the free surface in the case of wavelengths comparable or larger than the flow depth and interpret it in terms of topography induced standing gravity waves. Finally, we provide in section \ref{qualitativesummary} a qualitative summary of the main results of the paper. The most technical considerations are gathered in appendices.

\section{Turbulent flow over a uniform bottom}
\label{unif}

\subsection{The logarithmic law}
We consider a turbulent flow over a relief. Following Reynolds' decomposition between average and fluctuating (denoted with a prime) quantities, the equations governing the mean velocity field $u_i$ can be written as:
\begin{eqnarray}
\partial_i u_i  & = & 0,
\label{NS1}\\
D_t u_i=\partial_t u_i+u_j \partial_j u_i  & = & -\partial_j \tau_{ij}-\partial_i p,
\label{NS2}
\end{eqnarray}
where $\tau_{ij}=\overline{u'_i u'_j}$ is the Reynolds stress tensor (\cite{R74}). For the sake of simplicity, we omit the density factor $\rho$ in front of the pressure $p$ and the stress tensor. The aim of this paper is to describe quantitatively the average flow over a fixed corrugated boundary within this framework. The reference state is the homogeneous and steady flow over a flat bottom, submitted to an imposed constant shear stress $\tau_{xz}=-u_*^2$. The turbulent regime is characterised by the absence of any intrinsic length and time scales. At a sufficiently large distance $z$ from the ground, the only length-scale limiting the size of turbulent eddies --~the so-called mixing length $L$~-- is precisely $z$; the only mixing time-scale is given by the velocity gradient $|\partial_z u_x|$. As originally shown by \cite{P25}, it results from this dimensional analysis that the only way to construct a diffusive flux is a turbulent closure of the form:
\begin{equation}
\tau_{xz} = -\kappa^2 L^2 |\partial_z u_x| \partial_z u_x,
\label{tauxzPrandtl}
\end{equation}
where the mixing length is $L=z$ and $\kappa \simeq 0.4$ is the (phenomenological) von K\'arm\'an constant. After integration, one obtains that the velocity has a single non zero component along the $x$-axis, which increases logarithmically with $z$ (\cite{T88}):
\begin{equation}
u_x=\frac{u_*}{\kappa} \ln \left(\frac{z}{z_0} \right).
\label{uzero}
\end{equation}
where $z_0$ is a constant of integration called the hydrodynamical roughness. This expression does not apply for  $z \rightarrow 0$. There should be layer of thickness $h_0$ close to the bottom, called the surface layer, matching the logarithmic profile to a null velocity on the ground. 

\subsection{Hydrodynamical roughness}
The hydrodynamical roughness $z_0$ should be distinguished from the geometrical (or physical) roughness of the ground, usually defined as the root mean square of the height profile variations. $z_0$ is defined as the height at which the velocity would vanish, when extrapolating the logarithmic profile to small $z$. The physical mechanism controlling $z_0$ can be of different natures. If the ground is smooth enough, a viscous sub-layer of typical size $\go(\nu/u_*)$ must exist, whose matching with the logarithmic profile determines the value of $z_0$. On the contrary, if the geometrical roughness is larger than the viscous sub-layer, turbulent mixing dominates at small $z$ with a mixing length controlled by the ground topography. In the case of a static granular bed composed of grains of size $d$, reported values of the hydrodynamical roughness are reasonably consistent ($z_0 \simeq d/30$ in \cite{B41}, $z_0 \simeq d/24$ in \cite{SG00} and $z_0 \simeq d/10$ (\cite{K74,A04}). In section~\ref{NL}, we will justify the connection between geometrical and hydrodynamical roughness on a rigourous basis and show that they are not simply proportional.

The situation is of course different in the presence of sediment transport, which may (or not) induce some negative feedback on the flow. In this case, the hydrodynamical roughness $z_0$ may directly be controlled by the transport characteristics (e.g. mass flux and grain trajectories). Nature presents many other physical processes controlling the roughness: for instance, the flexible stems of wetland plants in low marshes or, for the wind, the canopy or the waves over the ocean. In all these cases, it can be assumed that the logarithmic law is a good approximation of the velocity profile above the surface layer, with a single known parameter $z_0$.

We will first consider the asymptotic limit in which the typical relief length --~say, the dune wavelength $\lambda$~-- is much larger than the surface layer thickness $h_0$. The relief is locally flat at the scale $h_0$, so that there must be a region close to the ground where the velocity profile shows a logarithmic vertical profile. We will then discuss the case of moderate values of the ratio $\lambda/h_0$, for which the flow becomes sensitive to the details of the mechanisms controlling the roughness.

\subsection{A turbulent closure}
In the logarithmic boundary layer, the normal stresses can be written as:
\begin{equation}
\tau_{xx} = \tau_{yy}=\tau_{zz}=\frac{1}{3} \tau_{ll}
\qquad \mbox{with} \qquad
\tau_{ll} = \kappa^2 \chi^2 L^2 |\partial_z u_x|^2,
\label{taullPrandtl}
\end{equation}
where $\chi$ is a second phenomenological constant estimated in the range $2.5 - 3$. Note that $\chi$ does not have any influence on the results as it describes the isotropic component of the Reynolds stress tensor, which can be absorbed into the pressure terms. Normal stress anisotropy is considered in section~\ref{robust} and appendix~\ref{AvI}. Introducing the strain rate tensor $\dot \gamma_{ij}=\partial_i u_j+\partial_j u_i$ and its squared modulus $|\dot \gamma|^2=\frac{1}{2} \dot \gamma_{ij} \dot \gamma_{ij}$, we can write both expressions (\ref{tauxzPrandtl}) and (\ref{taullPrandtl}) in a general tensorial form:
\begin{equation}
\tau_{ij} =\kappa^2 L^2 |\dot \gamma| \left(\frac{1}{3} \chi^2 |\dot \gamma| \, \delta_{ij} - \dot \gamma_{ij} \right).
\label{tauijPrandtl}
\end{equation}

In this paper, we focus on 2D steady situations, i.e. on geometries invariant along the transverse $y$-direction, see figure~\ref{WavySchematic}. As they are of permanent use for the rest of the paper, we express the components of the velocity and stress equations in the $x$- and $z$-directions. The Navier-Stokes equations read:
\begin{eqnarray}
\partial_x u_x +\partial_z u_z=0,
\label{NScont}\\
u_x \partial_x u_x+ u_z \partial_z u_x&=&-\partial_x p-\partial_z \tau_{xz} -\partial_x \tau_{xx},
\label{NSx}\\
u_x \partial_x u_z+ u_z \partial_z u_z&=&-\partial_z p-\partial_z \tau_{zz} -\partial_x \tau_{zx}.
\label{NSz}
\end{eqnarray}
The stress expressions are the following:
\begin{eqnarray}
\tau_{xz} & = & -\kappa^2 L^2 |\dot \gamma| \dot \gamma_{xz},
\label{relaxxz}\\
\tau_{xx} & = & -\kappa^2 L^2 |\dot \gamma| \dot \gamma_{xx}+ \frac{1}{3} \kappa^2\chi^2 L^2 |\dot \gamma|^2,
\label{relaxxx}\\
\tau_{zz} & = & -\kappa^2 L^2 |\dot \gamma| \dot \gamma_{zz}+ \frac{1}{3} \kappa^2\chi^2 L^2 |\dot \gamma|^2.
\label{relaxzz}
\end{eqnarray}
In these expressions, the strain tensor components are given by
\begin{equation}
\dot \gamma_{xz} =\dot \gamma_{zx}=\partial_z u_x+\partial_x u_z, \quad \dot \gamma_{xx} =2\partial_x u_x  \quad {\rm and } \quad \dot \gamma_{zz} =2\partial_z u_z=-\dot \gamma_{xx},
\label{strain}
\end{equation}
and the strain modulus by:
\begin{equation}
|\dot \gamma|^2 =2(\partial_x u_x)^2+2(\partial_z u_z)^2+(\partial_z u_x+\partial_x u_z)^2=4(\partial_x u_x)^2+(\partial_z u_x+\partial_x u_z)^2.
\label{strainmod}
\end{equation}
%

\section{Unbounded turbulent boundary layer over a wavy bottom}
\label{UnboundedCase}

We now consider the turbulent flow over a wavy bottom constituting the floor of an unbounded boundary layer. In rivers, this corresponds to the limit of a flow depth $H$ much larger than the bed-form wavelength $\lambda$. The solution is computed as a first order linear correction to the flow over a uniform bottom, using the first order turbulent closure previously introduced.

\subsection{Linearised equations}
For small enough amplitudes, we can consider a bottom profile of the form
\begin{equation}
Z(x)=\zeta \cos(kx)
\label{bottomprofile}
\end{equation}
without loss of generality. $\lambda=2\pi/k$ is the wavelength of the bottom and $\zeta$ the amplitude of the corrugation, see figure~\ref{WavySchematic}. The case of an arbitrary relief can be deduced by a simple superposition of Fourier modes. We introduce the dimensionless variable $\eta=k z$, the dimensionless roughness $\eta_0=k z_0$ and the function:
\begin{equation}
\mu(\eta)=\frac{1}{\kappa}  \, \ln \left( \frac{\eta}{\eta_0} \right).
\label{Mu2Eta}
\end{equation}
We also switch to the standard complex number notation: $Z(x)=\zeta e^{ikx}$ (real parts of expressions are understood).

We wish to perform the linear expansion of equations (\ref{NScont})-(\ref{strainmod}) with respect to the small parameter $k \zeta$. The mixing length is still defined as the geometrical distance to the bottom: $L=z-Z$. We introduce the following notations for the two first orders:
\begin{eqnarray}
u_x & = & u_* \left[\mu+k\zeta e^{ikx} U \right],
\label{defU}\\
u_z & = & u_* k\zeta e^{ikx} W,
\label{defW}\\
\tau_{xz} & = & \tau_{zx}= - u_*^2 \left[1+k\zeta e^{ikx} S_t\right],
\label{defSt}\\
p+\tau_{zz} & = & p_0+u_*^2 \left[ \frac{1}{3}\chi^2 + k\zeta e^{ikx} S_n\right],
\label{defSn}\\
\tau_{zz} & = & u_*^2  \left[ \frac{1}{3}\chi^2 + k\zeta e^{ikx} S_{zz}\right],
\label{defSzz}\\
\tau_{xx} & = & u_*^2  \left[ \frac{1}{3}\chi^2 + k\zeta e^{ikx} S_{xx}\right].
\label{defSxx}
\end{eqnarray}
The quantities $U$, $W$, etc, are implicitly considered as functions of $\eta$. An alternative choice is to consider functions of the coordinate $\xi=\eta-kZ$. Such alternative functions are denoted with a tilde to make the distinction. This important --~but somehow technical~-- issue of the choice of a representation is discussed in appendix~\ref{rep}. Although the curvilinear and Cartesian systems of coordinates are equivalent, the distinction between the two is of importance when it comes to the expression of the boundary conditions, and for the range of amplitudes $\zeta$ for which the linear analysis is no more valid (see section \ref{NL}). In particular, vertical profiles in the forthcoming figures will be mostly plotted as a function of the shifted variable $\xi$.

The linearised strain rate tensor reads
\begin{eqnarray}
\dot \gamma_{xz} &=&\dot \gamma_{zx}=k u_* \mu' + u_* k^2 \zeta e^{ikx}  (U'+iW), \\
\dot \gamma_{xx} &=&2 i u_* k^2 \zeta e^{ikx} U, \\
\dot \gamma_{zz} &=&2 u_* k^2 \zeta e^{ikx} W', \\
|\dot \gamma| &=& |\dot \gamma_{xz}|,
\end{eqnarray}
and the stress equations can be simplified into
\begin{eqnarray}
\mu' S_t &=&2  (U'+iW)-2 \kappa^2 \eta \mu'^3, \label{relaxSt}\\
\mu' S_{xx} &=& -2 i U+ \frac{2}{3}\chi^2(U'+iW) - \frac{2}{3}\chi^2 \kappa \mu'^2, \label{relaxSxx}\\
\mu' S_{zz} &=&- 2 W'+\frac{2}{3}\chi^2(U'+iW) - \frac{2}{3}\chi^2\kappa \mu'^2. \label{relaxSzz}
\end{eqnarray}
Finally the Navier-Stokes equations lead to
\begin{eqnarray}
W' &=&-i U, \\
S_t' &=&  \mu i U+  \mu' W +i S_n+i S_{xx}-i S_{zz}, \label{NSxLin}\\
S_n' &=&-  \mu i W +i S_t. \label{NSzLin}
\end{eqnarray}

Taking the difference of equations (\ref{relaxSxx}) and (\ref{relaxSzz}), one can compute
\begin{equation}
S_{xx}-S_{zz}= \frac{-4 iU}{\mu'}
\end{equation}
to obtain four closed equations:
\begin{eqnarray}
U'&=&- i W +\frac{1}{2}\mu' S_t+\kappa \mu'^2, \label{equaU'}\\
W'&=&-i U, \label{equaW'}\\
S_t' &=& \left(i\mu +\frac{4}{\mu'}\right) U + \mu' W + i S_n, \label{equaSt'}\\
S_n' &=&-  i \mu  W +i S_t. \label{equaSn'}
\end{eqnarray}
Introducing the vector $\vec{X}=(U,W,S_t,S_n)$, we finally get at the first order in $k\zeta$ the following compact form of the equation to integrate:
\begin{equation}
\frac{d}{d\eta} \vec{X} = {\mathcal{P}} \vec{X}+\vec{S},\,\, {\rm with}\,\,
{\mathcal{P}} =  \left ( \!\!\!\!
\begin{tabular}{cccc}
$0$ & $-i $ & $\frac{1}{2}\mu' $ & $0$ \\
$-i$ & $0$ & $0$ & $0$ \\
$\left(i\mu +\frac{4}{\mu'}\right)$ & $\mu'$ & $0$ & $i$ \\
$0$ & $-\mu i$ & $i$ & $0$
\end{tabular}
\! \right )
\, \mbox{and} \,\,
\vec{S} =\left ( \!\!
\begin{tabular}{c}
$\kappa \mu'^2$ \\ $0$ \\ $0$ \\ $0$
\end{tabular}
\!\! \right ).
\label{systlinplaque}
\end{equation}
The general solution of this equation is the linear superposition of all solutions of the homogeneous system (i.e. with $\vec{S}=\vec{0}$), and a particular solution $\vec{X}_s$.

\subsection{Boundary conditions}
Four boundary conditions must be specified to solve the above equation (\ref{systlinplaque}). The upper boundary corresponds to the limit $\eta \to \infty$, for which we ask that the vertical fluxes of matter and momentum vanish asymptotically. This means that the first order corrections to the shear stress and to the vertical velocity must tend to zero: $W(\infty)=0$ and $S_t(\infty)=0$. In practice, a boundary at finite height $H$ (at $\eta_H=kH$) is introduced, at which we impose a null vertical velocity $W(\eta_H)=0$ and a constant tangential stress $\rho u_*^2$ so that $S_t(\eta_H)=0$. This corresponds to a physical situation where the fluid is entrained by a moving upper plate, for instance a stress-controlled Couette annular cell. Then, we consider the limit $H \to +\infty$, i.e. when the results become independent of $H$.

The lower boundary condition must be specified on the floor ($\eta \to kZ$). We consider here the limit in which the surface layer thickness $h_0$ is much smaller than the wavelength $\lambda$. This allows to perform an asymptotic matching between the solution and the surface layer, whatever the dynamical mechanisms responsible for the hydrodynamical roughness $z_0$ are. Indeed, focusing on the surface layer, we know that in the limit  $z \gg h_0$, the asymptotic behaviour of the local tangential velocity $u$ should be a logarithmic profile controlled by the local shear stress $\tau$ and the roughness $z0$. The solution of (\ref{systlinplaque}) should thus match this asymptotic behaviour as $\eta \to kZ$. Thus, $z_0$ is the only parameter inherited from the surface layer in the limit $h_0\ll\lambda$. We will investigate the situation where this approximation is not valid anymore in section~\ref{controlzzero}.

In the limit $z-Z \ll \lambda$, the homogeneous solution of (\ref{systlinplaque}) can be expanded in powers of $\eta$ and $\ln \frac{\eta}{\eta_0}$ and expressed as the sum over four modes. Adding the asymptotic behaviour of the particular solution $\vec{X}_s=(-\frac{1}{\kappa \eta}, \frac{i}{\kappa} \ln \frac{\eta}{\eta_0}, 0, 0)$, the full solution writes:
\begin{equation}
{\vec{X}} \mathop{\sim}_{\eta \to 0}  a_1\left (\!
\begin{tabular}{c}
$\mu^2/4$\\
$1$\\
$\mu$\\
$- \eta^2 \mu^3/4$
\end{tabular}
\right ) + a_2\left (\!
\begin{tabular}{c}
$\mu/2$\\
$-i\,\eta\,\mu/2$\\
$1$\\
$i\,\eta$
\end{tabular}
\right ) + a_3\left (\!
\begin{tabular}{c}
$1$\\
$-i\,\eta$\\
$i\,\eta\,\mu(\eta)$\\
$-\,\eta^2\,\mu(\eta)$
\end{tabular}
\right ) + a_4\left (\!
\begin{tabular}{c}
$i\eta/(2\kappa)$\\
$\eta^2/(4\kappa)$\\
$i\,\eta$\\
$1$
\end{tabular}
\right )
+ \vec{X}_s.
\label{AssymptotExpansion}
\end{equation}
The next terms in this expansion are $\go(\eta \ln^2\frac{\eta}{\eta_0})$.
\begin{figure}
\includegraphics{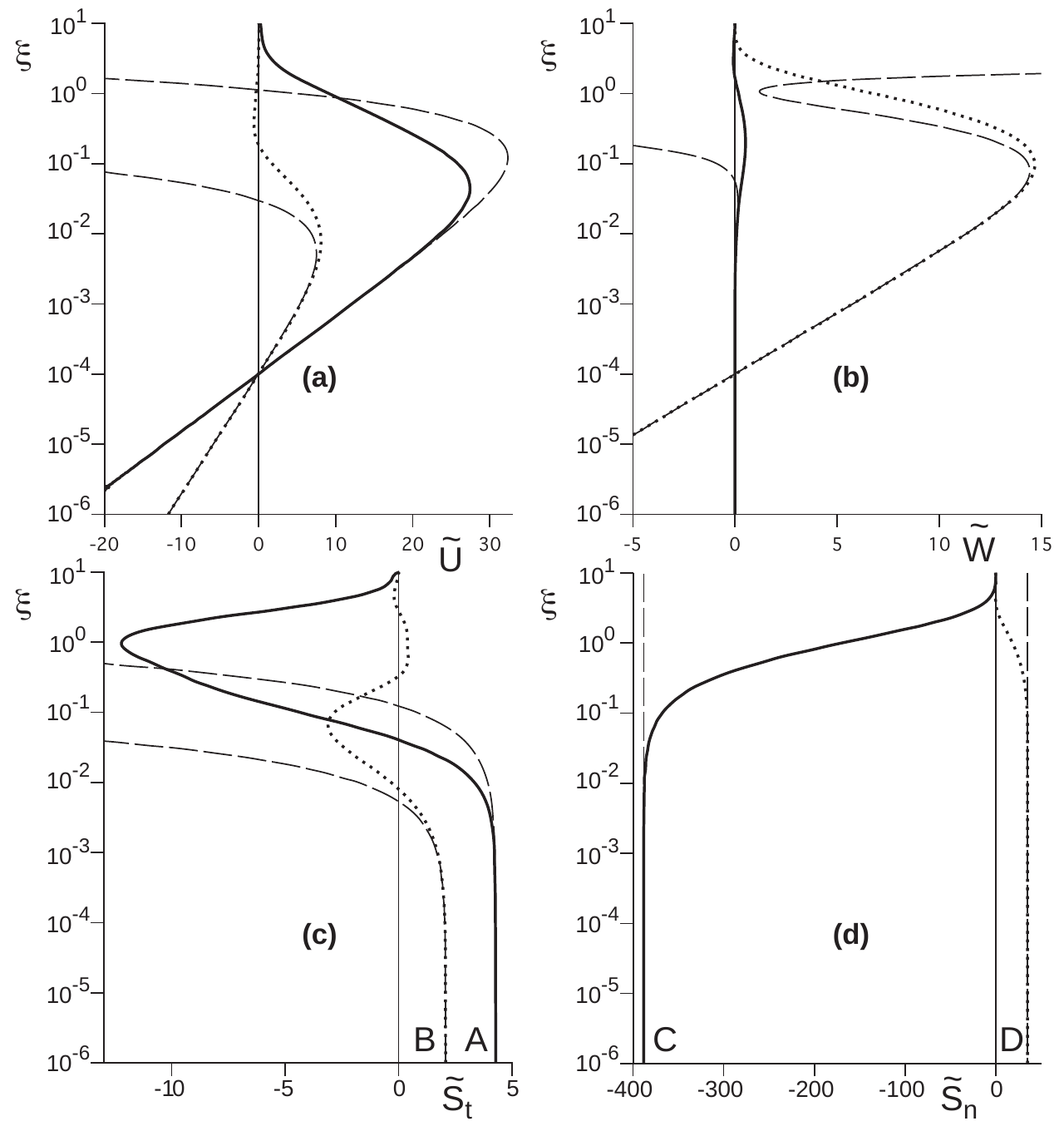}
\caption{Vertical profiles of the first order corrections to velocities and stresses for $\eta_0=10^{-4}$. $\xi=\eta-kZ$ is the distance to the bottom, rescaled by the wavenumber. In all panels, the solid lines represent the real parts of the functions, whereas the dotted lines represent the imaginary ones. Dashed lines show the asymptotic behaviours (\ref{Uasymp}-\ref{Snasymp}) used as boundary conditions. They match the solutions in the inner layer, which extends up to $\eta \simeq k\ell \simeq 10^{-2}$ here. We note $\tilde{S}_t(0)=A+iB$ and $\tilde{S}_n(0)=C+iD$. Close to the boundary, a plateau of constant shear stress can be observed, which corresponds to the logarithmic zone. It is embedded into a slightly larger zone of constant pressure in which the shear stress varies linearly.}
\label{ModesVerticaux}
\end{figure}

The values of the four coefficients $a_1$, ..., $a_4$ are selected by the matching with the surface layer. $a_1$ would correspond to a non vanishing normal velocity through the surface layer and should thus be null. $a_2$ precisely corresponds to the logarithmic profile with a roughness $z_0$ and a basal shear stress modulation $a_2$. This gives $a_2=S_t(0)$. $a_3$ would correspond to a modulation of the local roughness --~more precisely of its logarithm. We do not consider such a modulation so that $a_3=0$. $a_4$ corresponds to a sub-dominant behaviour associated to the basal pressure modulation ($a_4=S_n(0)$). In summary, the functions $U$, $W$, $S_t$ and $S_n$ should follow the following asymptotic behaviour:
\begin{eqnarray}
U(\eta) & = & \frac{S_t(0)}{2 \kappa} \, \ln \frac{\eta}{\eta_0} +\frac{i S_n(0)}{2 \kappa} \eta- \frac{1}{\kappa \eta}, \label{Uasymp} \\
W(\eta) & = & -\frac{i S_t(0)}{2 \kappa} \eta \left (\ln \frac{\eta}{\eta_0} - 1 \right )+\frac{S_n(0)}{4 \kappa} \eta^2 +\frac{i}{\kappa} \, \ln \frac{\eta}{\eta_0} , \label{Wasymp} \\
S_t(\eta) & = & S_t(0)+i S_n(0) \eta, \label{Stasymp} \\
S_n(\eta) & = & S_n(0). \label{Snasymp}
\end{eqnarray}

The region of thickness $\ell$ in which this asymptotic behaviour constitutes a good approximation of the flow field is called the inner layer. Equation~(\ref{Snasymp}) means that the total pressure  $\wp=p+\tau_{ll}/3$ is constant across this boundary layer:
\begin{equation}
\dz \wp = 0
\end{equation}
and equation~(\ref{Stasymp}) that the shear stress decreases linearly with height according to:
\begin{equation}
\partial_x \wp + \dz \tau_{xz}  =  0,
\end{equation}
The tangential pressure gradient is balanced by the normal shear stress, which means that inertial terms are negligible or equivalently that the fluid is in local equilibrium. In terms of energy, the space variation of the internal energy (pressure) is dissipated in turbulent "friction". These two equations correspond to the standard lubrication approximation for quasi-parallel flows. 

\subsection{Equations solving}
In practice, we solve the equations using a fourth order Runge-Kutta scheme with a logarithmic step. The integration is started at an initial value of $\eta$ inside the inner layer i.e. which verifies $\eta \ln^2\frac{\eta}{\eta_0} \ll 1$). We write the solution as a linear superposition of the form $\vec{X} = \vec{X}_s + S_t(0) \vec{X}_t + S_n(0) \vec{X}_n$, where the different terms verify:
\begin{eqnarray}
\frac{d}{d \eta}\vec{X}_s= {\mathcal P} \vec{X}_s+\vec{S}
& \qquad \mbox{starting from} \qquad &
\vec{X}_s (\eta) =\left (\begin{tabular}{c}
$-\frac{1}{\kappa \eta}$ \\ $\frac{i}{\kappa} \ln \frac{\eta}{\eta_0}$ \\ $0$ \\ $0$
\end{tabular}\right ), \label{equaXs}\\
\frac{d}{d \eta}\vec{X}_t= {\mathcal P} \vec{X}_t
& \qquad \mbox{starting from} \qquad &
\vec{X}_t (\eta) =\left (\begin{tabular}{c}
$\frac{1}{2\kappa} \ln \frac{\eta}{\eta_0}$ \\ $-\frac{i\eta}{2\kappa} \left ( \ln \frac{\eta}{\eta_0}-1 \right )$ \\ $1$ \\ $0$
\end{tabular}\right ), \label{equaXt}\\
\frac{d}{d \eta}\vec{X}_n= {\mathcal P} \vec{X}_n
& \qquad \mbox{starting from} \qquad &
\vec{X}_n (\eta) =\left (\begin{tabular}{c}
$\frac{i\eta}{2\kappa}$ \\ $\frac{\eta^2}{4\kappa}$ \\ $\eta$ \\ $1$
\end{tabular}\right ). \label{equaXn}
\end{eqnarray}
The boundary conditions on the bottom are then automatically satisfied, and the top ones give algebraic equations on the real and imaginary parts of $S_t(0)$ and $S_n(0)$, which can be solved easily. We have checked that the result is independent of the initial value of $\eta$, as long as it remains in the announced range.
\begin{figure}
\includegraphics{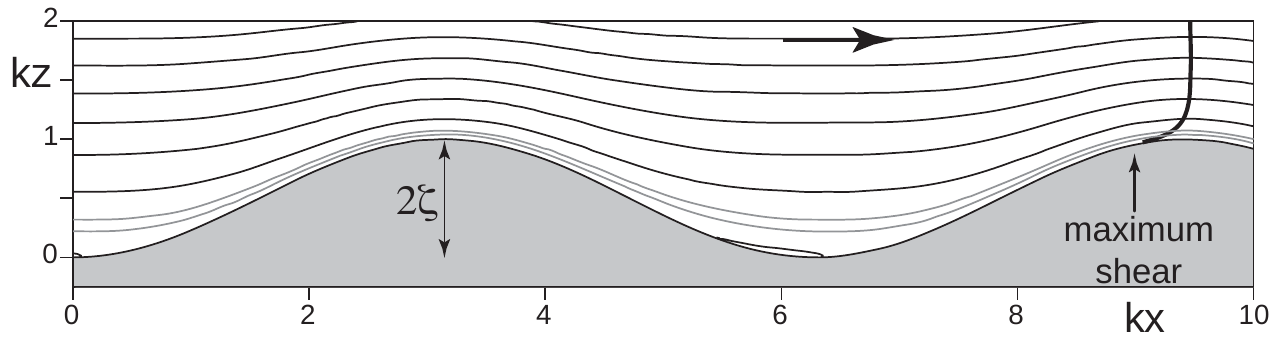}
\caption{Flow streamlines above a wavy bottom of rescaled amplitude $k\zeta=0.5$ (aspect ratio $\sim 1/6$), computed from the linearised equations ($\eta_0=10^{-4}$). The flow direction is from left to right. Note the left-right asymmetry of the streamlines around the bump in the inner layer (grey lines). Note also the onset of emergence of a recirculation bubble in the troughs. The thick line in the top right corner shows the positions that maximises the velocity along a streamline.}
\label{Streamlines}
\end{figure}
%

\subsection{Results}
The velocity and stress profiles resulting from the integration of equation (\ref{systlinplaque}) are displayed in figure~\ref{ModesVerticaux}. Looking at panel (c), one can clearly see the region close to the bottom where the shear stress is constant, while the horizontal velocity component (panel a) exhibits a logarithmic behaviour. This plateau almost coincides with the inner layer, which is the zone where the solution is well approximated by the asymptotic behaviour derived above. The inner layer is embedded in a wider region characterised by a constant pressure (panel d). The estimate of the thickness $\ell$ is of crucial importance for the transport issue (see next section and Part 2). $\ell$ is the scale at which inertial terms are of the same order as stress ones in the Reynolds averaged Navier-Stokes equations. The original estimation of $\ell$ given by \cite{JH75} was further discussed in several later papers (see e.g. \cite{TMB87,C88,BT89,FRBA90}). Our data are in good agreement with the scaling proposed by \cite{TMB87}
\begin{equation}
\frac{\ell}{\lambda}\,\frac{1}{\kappa^2}\,\ln^2 \frac{\ell}{z_0} =\go(1).
\label{ellscaling}
\end{equation}
Consistently, this scaling relationship is precisely that of the first neglected terms in the asymptotic expansion (\ref{AssymptotExpansion}). Away from the bottom, all profiles tend to zero, so that one recovers the undisturbed flow field (\ref{uzero}) at large $\eta$. The shape of these profiles are very consistent with the work of \cite{AXT94}, who have compared the influence of the closure scheme on the linear flow over a relief, which means that the precise choice of the turbulent closure does not affect significantly the results.
\begin{figure}
\includegraphics{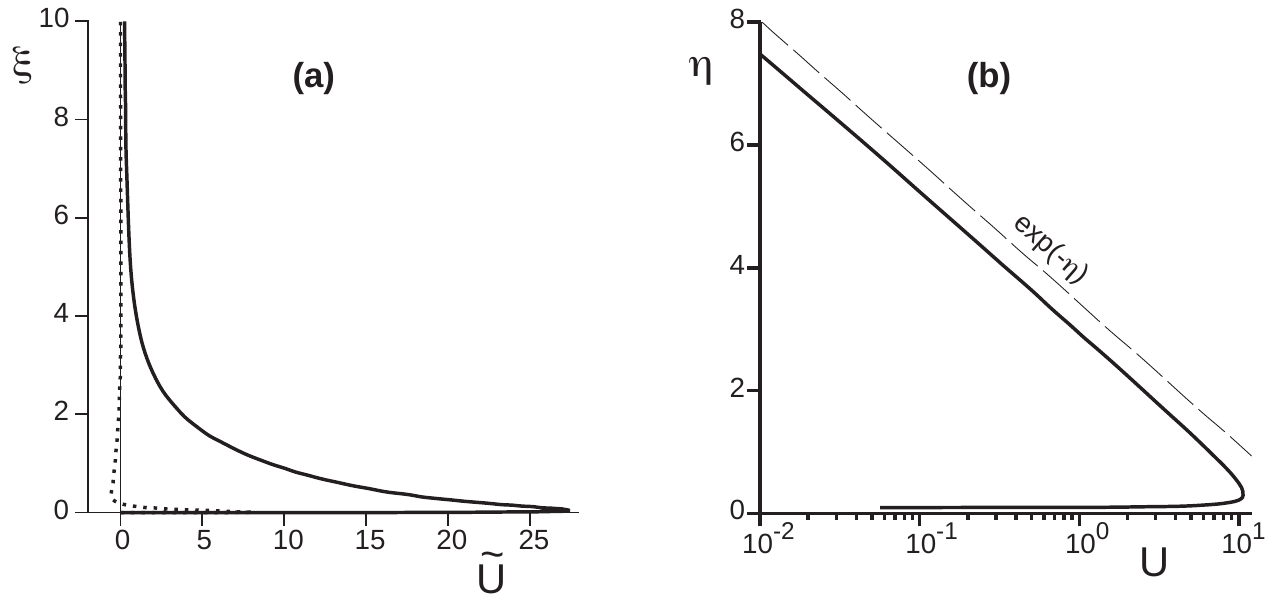}
\caption{Vertical profiles of the first order correction to the horizontal velocity for $\eta_0=10^{-4}$. (a) Lin-lin plot in the shifted representation. (b) Lin-Log plot in the non-shifted representation. The solid lines correspond to the real part and the dotted line to the imaginary one. The velocity disturbance decreases exponentially over one wavelength (dashed line). In this outer region, the Reynolds stress can be neglected.}
\label{PerfectFluidInner}
\end{figure}
\begin{figure}
\includegraphics{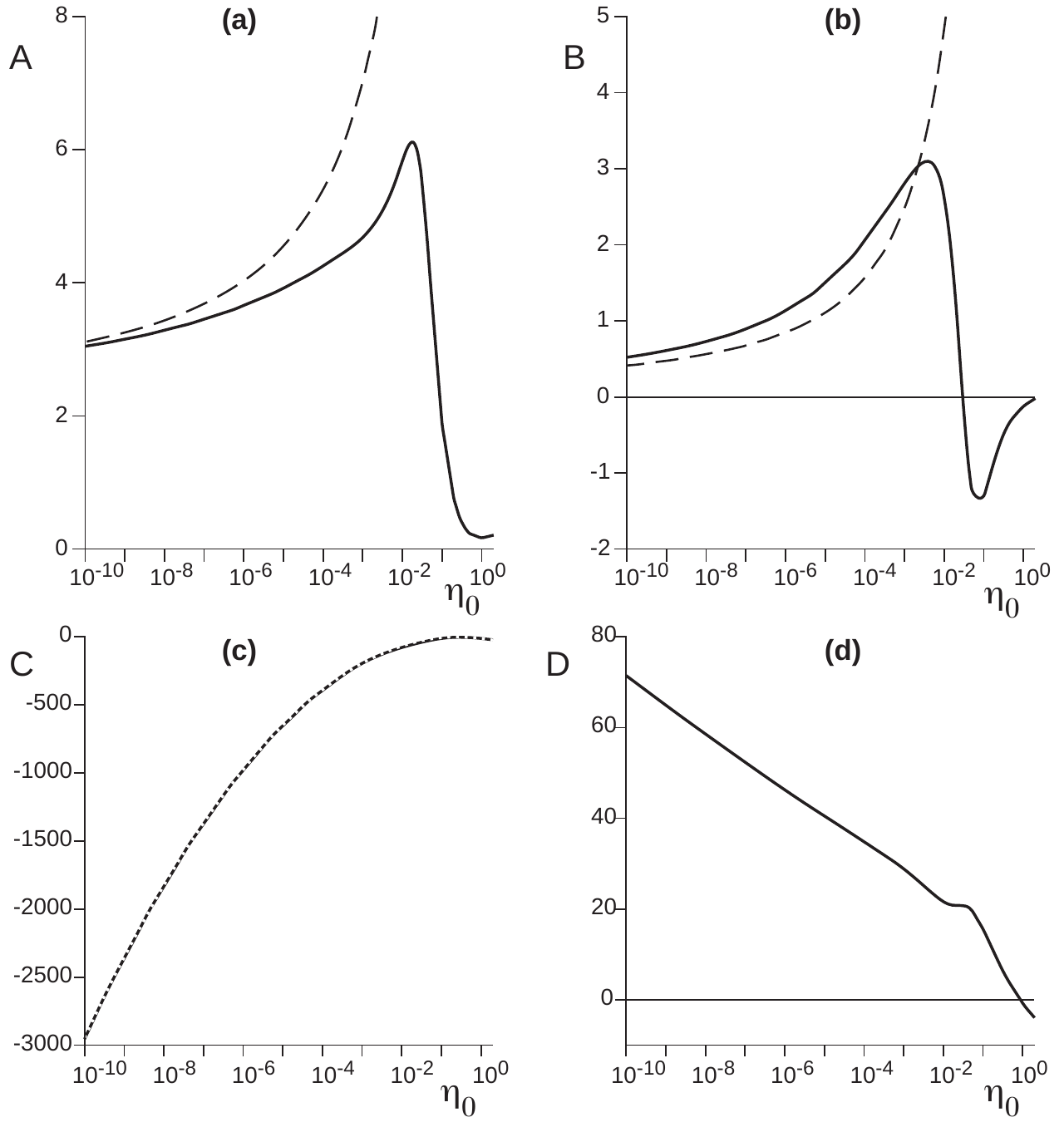}
\caption{Coefficients $A$, $B$, $C$ and $D$ as a function of $\eta_0=kz_0$. These plots show the dependence of the basal shear and normal stresses with the number of decades separating the wavelength $\lambda$ from the soil roughness $z_0$, for a given bump aspect ratio. The solid line corresponds to the results of the model, using the asymptotic matching with the surface layer. The dashed lines represent the analytical formula deduced from \cite{JH75} by \cite{KSH02}. They agree well at very small $\eta_0$.}
\label{ABCDLin}
\end{figure}

In order to visualise the effect of the bottom corrugation on the flow, the flow streamlines are displayed in figure~\ref{Streamlines} (see appendix \ref{streamlines} for explanations about their computation). It can be observed that the velocity gradient is larger on the crest than in the troughs as the streamlines are closer to each other. The flow is disturbed over a vertical distance comparable to the wavelength. A subtler piece of information concerns the position along each streamline at which the velocity is maximum. These points are displayed in the right corner of figure~\ref{Streamlines}. Away from the bottom, they are aligned above the crest of the bump. Very close to it, however, they are shifted upstream. In other words, the fluid velocity is in phase with the topography in the upper part of the flow, but is phase advanced in the inner boundary layer where the shear stress tends to its basal value. In this inner layer, the profile is well approximated by its asymptotic expression (\ref{Uasymp}).

An inspection of the velocity profile evidences two distinct regions (see figures \ref{ModesVerticaux} and \ref{PerfectFluidInner}), in which we recognise those at the basic partitioning of the flow in Jackson \& Hunt work (1975), and subsequent papers. There is an outer region ($\eta \gg k\ell$), where $U$ decreases exponentially with $\eta$ (figure~\ref{PerfectFluidInner}(b)). Seeking for asymptotic solutions decreasing as $e^{-\sigma \eta}$, one has to solve the eigenvalue problem ${\mathcal{P}} \vec{X}=-\sigma \vec{X}$ for asymptotically large values of $\eta$. At the two leading orders, the decrease rate $\sigma$ is given by:
\begin{equation}
2 i \left(\sigma^4+1\right)\kappa^2 \eta + \left(\sigma^2-1\right) \ln\frac{\eta}{\eta_0}=0.
\end{equation}
The asymptotic behaviour is an oscillatory relaxation corresponding to $\sigma=(1\pm i)/\sqrt{2}$. However, the observed decrease corresponds to the intermediate asymptotic regime $\eta < \ln \frac{\eta}{\eta_0}$ for which the solution is $\sigma=1$. This behaviour is reminiscent from that of an inviscid potential flow. In other words, the effect of the turbulent shear stress on the flow disturbance can be neglected.

The intermediate region between the inner and the outer layers is responsible for the asymmetry of the flow as well as the upstream shift of the maximum velocity discussed above. Let us emphasise again that this is the physical key point for the formation of bedforms. One can understand the reason of the phase shift with the following argument. The external layer can be described as a perfect irrotational flow. Since the elevation profile is symmetric, the streamlines are symmetric too, as the flow is solely controlled by the balance between inertia and the pressure gradient induced by the presence of the bump. As a consequence, the velocity is maximum at the vertical of the crest. Now, inside the inner layer, this flow is slowed down by turbulent diffusion of momentum. Focusing on the region of matching between these outer and inner regions, the velocity needs some time to re-adapt to a change of shear stress, due to inertia. Thus, the shear stress is always phase-advanced with respect to the velocity. One concludes that the basal shear stress is phase-advanced with respect to the bump.

As mentioned in the introduction, we are especially interested in the shear stress and pressure distributions on the bottom. We note $\tilde{S}_t(0)=A+iB$ and $\tilde{S}_n(0)=C+iD$. The ratio $B/A$ is the tangent of the phase shift between the shear stress and the topography. It is positive as the shear stress is phase advanced. The four coefficients $A$, $B$, $C$ and $D$ are displayed as a function of $\eta_0$ in figure~\ref{ABCDLin}. Their overall dependence with $\eta_0$ is weak, meaning that the turbulent flow around an obstacle is mostly scale invariant. More precisely, following Jackson \& Hunt's work (\cite{JH75, KSH02}), it has been shown that, for asymptotically small $\eta_0$, one expects logarithmic dependencies:
\begin{equation}
A = \frac{\ln^2\left (\Phi^2/\ln\Phi\right )}{2\ln^3\phi} \left ( 1+\ln\phi+2\ln \frac{\pi}{2} + 4\gamma_E \right )
\quad {\rm and }\quad
B = \pi \, \frac{\ln^2\left (\Phi^2/\ln\Phi\right )}{2\ln^3\phi} \, ,
\label{ABJH}
\end{equation}
where Euler's constant is $\gamma_E \simeq 0.577$, $\phi$ is defined by the equation $\phi\ln\phi = 2\kappa^2\Phi$ and with $\Phi = \pi/(2\eta_0)$. Note that $A$ tends to $2$ and $B$ to $0$ as $\eta_0 \to 0$, as expected when the inner layer thickness $\ell$ vanishes. In this limit, the basal shear stress is directly proportional to the square of the velocity inherited from the outer layer, which is solution of the potential flow problem.

These expressions agree well with our numerical results for very small $\eta_0$. However, for realistic values of $\eta_0$, e.g. $10^{-4}<\eta_0<10^{-2}$, this approximation cannot be accurately used as it leads to errors of order one --~note that Jackson \& Hunt's expressions tend to diverge at larger $\eta_0$. In comparison to $A$ and $B$, we observe that the normal stress coefficients $C$ and $D$ are more robust with respect to the details of the model. In the limit of a perfect flow, the pressure varies as the square of the velocity. Here, one needs to consider the velocity at the scale $\lambda$ of the perturbation, say $u_* \mu$, where the logarithmic factor $\mu$ should be evaluated for $\eta$ of order unity. From this argument, we predict that the pressure coefficient $C$ should scale as the square of $\ln \eta_0$ (a parabola in figure~\ref{ABCDLin}), which is very accurately verified. More precisely, $C=[\mu(1/4)]^2$ is an almost perfect approximation. Finally, it can be observed that the normal stress is also in phase advance with respect to the bottom profile. The coefficient $D$ is positive and shows a linear variation with $\ln \eta_0$.

\section{Effect of the mechanisms controlling the hydrodynamical roughness}
\label{controlzzero}

So far, the computation of the velocity and stress fields has been obtained without any specification of the physics at the scale of $z_0$, as the integration of equation (\ref{systlinplaque}) was started in the inner layer rather than on the bottom. This is of course possible only if this layer is sufficiently thick, i.e. if $\ell$ (or $\lambda$) is much larger than the thickness of the surface layer $h_0$ introduced in section~\ref{unif}. We now discuss several ways to describe the flow inside the surface layer, and investigate the subsequent effect on the shape of the stress coefficients as functions of $\eta_0$. These coefficients should be independent of the physics at work in this surface layer when $\eta_0$ is small enough, but we expect larger differences for larger values of $\eta_0$.

We first present a convenient phenomenological model of geometrically induced roughness, which does not involve additional parameters. Because of its simplicity, it will be used in the next sections, as well as in the second part of the paper. The results will be compared to a rigourous treatment resulting from the weakly non-linear hydrodynamical calculation in section~\ref{NL}. We then consider the case of a viscous surface layer. Inspired from the aeolian transport properties, we finally discuss the focus point assumption as a possible way to describe the situation in which the surface layer is governed by the presence of sediment transport. 
\begin{figure}
\includegraphics{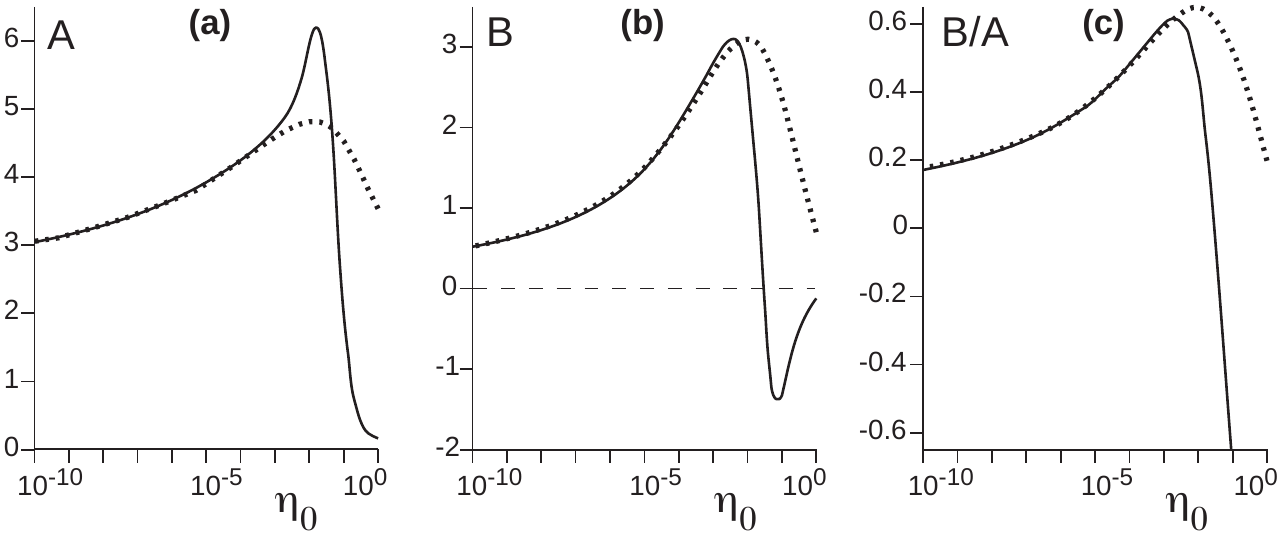}
\caption{Shear stress coefficients $A$ and $B$ (dashed lines) computed with the phenomenological model of geometrically induced roughness (\ref{uzerogeom}). For comparison, the solid lines display the reference case shown in figure~\ref{ABCDLin}.}
\label{ABCDgeom}
\end{figure}
%

\subsection{Geometrically induced roughness}
\label{geom}
For an hydrodynamically rough bottom, the "small scale" roughness elements are larger than the viscous sub-layer. They are submitted to a turbulent drag from the fluid and reciprocally, their presence slows down the flow. The exchanges of momentum in the surface layer are thus dominated by the turbulent fluctuations. Following \cite{R80} and others, a convenient phenomenological model is to define the mixing length involved in the turbulent closure (\ref{tauijPrandtl}) as $L=z_0+z-Z$. In this way, $L$ is still essentially the geometrical distance to the bottom, except that it cannot be smaller than the roughness length. This choice reflects in a intuitive manner the physical picture one can infer from experiments or simulations where square-shaped roughness elements are glued on a flat wall (see e.g. \cite{PSJ69}). We will show in section~(\ref{NL}) that this picture must be refined when dealing with blunt roughness elements of moderate aspect ratio --~for instance the surface of a sand bed.

With this expression for the mixing length, the integration of starting equations in the uniform and steady case gives
\begin{equation}
u_x=\frac{u_*}{\kappa} \ln \left( 1+\frac{z}{z_0} \right),
\label{uzerogeom}
\end{equation}
where the lower boundary condition $u_x=0$ can now be taken in $z=0$. This expression is well approximated by the pure logarithmic profile (eq.~\ref{uzero}) as soon as $z$ is larger than, say, few $z_0$. In other words, for this model, $h_0\sim z_0$.

The above description of the linear analysis, and in particular the expression of the matrix $\mathcal{P}$ and the vector $\mathcal{S}$ involved in (\ref{systlinplaque}), in the case of a wavy bottom is still valid, but now with the following expression for the function
\begin{equation}
\mu(\eta)=\frac{1}{\kappa}  \, \ln \left( 1+\frac{\eta}{\eta_0} \right).
\label{Mu2Etageom}
\end{equation}
The solution of (\ref{systlinplaque}) can again be written as a linear superposition $\vec{X} = \vec{X}_s + S_t(0) \vec{X}_t + S_n(0) \vec{X}_n$, where these three vectors are solutions of 
\begin{eqnarray}
\frac{d}{d \eta}\vec{X}_s= {\mathcal P} \vec{X}_s+\vec{S}
& \qquad \mbox{with} \qquad &
\vec{X}_s (0) =\left (\begin{tabular}{c}
$-\frac{1}{\kappa \eta_0}$ \\ $0$ \\ $0$ \\ $0$
\end{tabular}\right ), \label{equaXsgeom}\\
\frac{d}{d \eta}\vec{X}_t= {\mathcal P} \vec{X}_t
& \qquad \mbox{with} \qquad &
\vec{X}_t (0) =\left (\begin{tabular}{c}
$0$ \\ $0$ \\ $1$ \\ $0$
\end{tabular}\right ), \label{equaXtgeom}\\
\frac{d}{d \eta}\vec{X}_n= {\mathcal P} \vec{X}_n
& \qquad \mbox{with} \qquad &
\vec{X}_n (0) =\left (\begin{tabular}{c}
$0$ \\ $0$ \\ $0$ \\ $1$
\end{tabular}\right ). \label{equaXngeom}
\end{eqnarray}
This decomposition ensures the requirement that both components of the velocity vanish on the bottom, leading to $W(0) = 0$ and $U(0)=-\mu'(0)=-1/(\kappa \eta_0)$. As in the previous section, the coefficients $S_t(0)$ and $S_n(0)$ are found by the upper boundary conditions.

The coefficients $A$ and $B$ resulting from this integration are displayed in figure~\ref{ABCDgeom}. One can see that, for $\eta_0<10^{-3}$, they are not very much different from those obtained in the previous section. However, one can notice significant differences for $\eta_0>10^{-2}$. As the mixing length in the surface layer is larger in this case ($L \sim z_0$) than in the asymptotic case ($L\sim z-Z$), the turbulent `diffusion' is more efficient. This results into a larger phase advance for the shear stress (Fig.~\ref{ABCDgeom}~c). For practical purposes and for later use in the second part of this paper, a very good empirical fit of the coefficients $A$ and $B$ is obtained with
\begin{equation}
A = 2 + \frac{a_1+a_2 R+a_3 R^2+a_4 R^3}{1+a_5 R^2+a_6 R^4} \quad{\rm and }\quad B =  \frac{b_1+b_2 R+b_3 R^2+b_4 R^3}{1+b_5 R^2+b_6 R^4}
\end{equation}
with $\{a_1,a_2,a_3,a_4,a_5,a_6\}=\{1.0702, 0.093069, 0.10838, 0.024835, 0.041603, 0.0010625\}$, $\{b_1,b_2,b_3,b_4,b_5,b_6\}=\{0.036989, 0.15765, 0.11518, 0.0020249, 0.0028725, 0.00053483\}$ and $R=\ln \frac{2\pi}{\eta_0}$.
\begin{figure}
\includegraphics{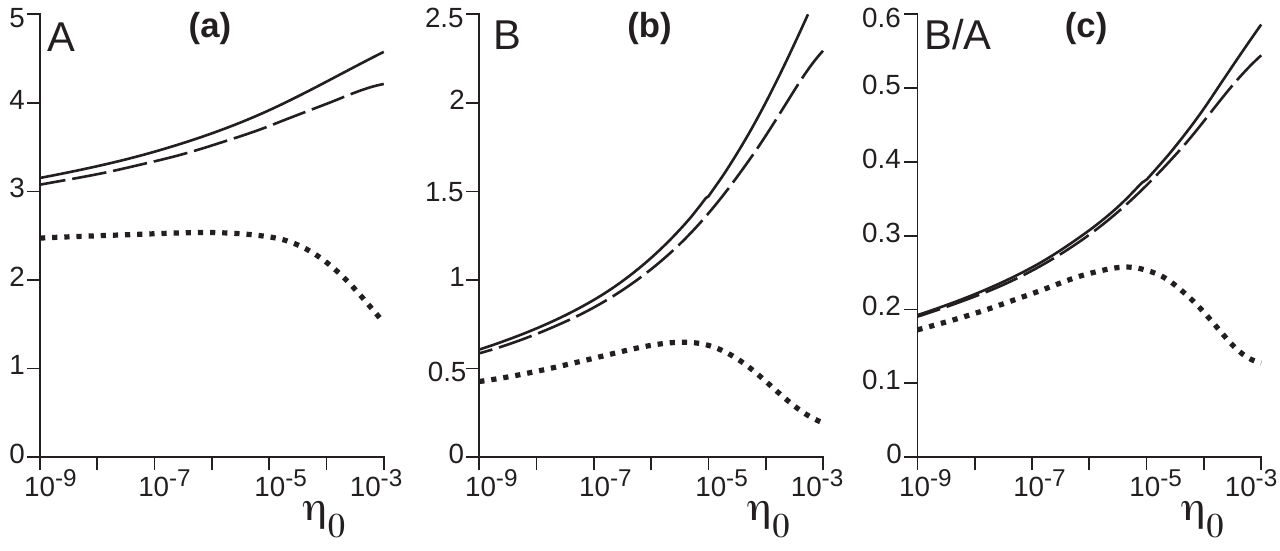}
\caption{Shear stress coefficients $A$ and $B$ computed with a viscous surface layer. Dotted line: $\Rey_t=125$; Dashed line: $\Rey_t=1$. As in figure~\ref{ABCDgeom}, the solid lines display the reference case of figure~\ref{ABCDLin}.}
\label{ABCDHfUfViscous}
\end{figure}
%

\subsection{A viscous surface layer}
In hydraulically smooth situations, it is natural to expect that, very close to the bottom, the flow must be laminar and thus described by the equation
\begin{equation}
\tau = \nu \frac{du_x}{dz} = u_*^2,
\label{viscoustau}
\end{equation}
whose solution is
\begin{equation}
u_x(z) = \frac{u_*^2}{\nu}z.
\label{viscousux}
\end{equation}
We thus neglect here the possibility of a phase shift across the viscous surface layer. The transition from viscous to turbulent regime is governed by the Reynolds number and occurs at a typical value $\Rey_t \simeq 125$. The surface layer thickness can then be easily computed as $h_0 = \frac{\nu}{u_*} \sqrt{\Rey_t}$. At $z=h_0$, both viscous and turbulent expressions for the velocity must coincide:
\begin{equation}
u_h \equiv u_* \sqrt{\Rey_t} = \frac{u_*}{\kappa} \ln \frac{h_0}{z_0} \, .
\label{viscousturbulentmatching}
\end{equation}
From this equality, we can deduce the hydrodynamical roughness seen from the inner layer, due to this viscous surface layer:
\begin{equation}
z_0 = \frac{\nu}{u_*} \sqrt{\Rey_t} \, e^{-\kappa \sqrt{\Rey_t}}.
\label{z0viscous}
\end{equation}
In the case of a sand bed, the transition between the hydrodynamically smooth and rough regimes  occurs when the viscosity induced roughness (eq.~\ref{z0viscous}) is of the order of the geometrically induced roughness ($z_0 \sim d/10$).

With the corresponding value for $\eta_0=kz_0$, we solve equation (\ref{systlinplaque}) in the usual manner, writing the solution in the form of the linear superposition as described above, except that the integration is started at the initial value $\eta=k h_0$, in which we impose that the velocity is parallel to the bed and equal to $u_h$. At linear order, this leads to
\begin{eqnarray}
U(kh_0) & = & -\mu'(kh_0),
\label{Unu} \\
W(kh_0) & = & iu_h/u_* = i \mu(kh_0).
\label{Wnu}
\end{eqnarray}
The resulting shear stress coefficients $A$ and $B$ are displayed in figure~\ref{ABCDHfUfViscous}. As one can expect, in comparison to the reference case, they are smaller for larger values of $\Rey_t$, and all different curves collapse as $\eta_0 \to 0$. The viscous diffusion of momentum is less efficient than that induced by turbulent fluctuations. Moreover, in the Stokes regime, for Reynolds numbers much smaller than $1$, the kinematic reversibility leads to a shear stress in phase with the topography. Consistently, it can be observed in  figure~\ref{ABCDHfUfViscous}(c) that the phase advance is reduced in the hydrodynamically smooth regime. We will show in the second part of this article that this explains the fact that the wavelength at which ripples appear is larger as the Reynolds number decreases.

Experiments in the hydraulically smooth regime have been performed by \cite{ZCH77,ZH79,AH85}, who measured the ionic mobility between two nearby electrodes. This current is assumed to be related, without any spatial or temporal lag, to the basal shear stress. The measured phase shift between the signal and the bottom topography could reach values as high as $80^\circ$. this would correspond to $B/A = \tan(80\pi/180) \simeq 5.67$. Within the present model, the phase shift remains much lower than the measured $80^\circ$.  This unexpected value has been interpreted as the signature of a lag of the laminar-turbulent transition with respect to the Reynolds number criterion $\Rey=\Rey_t$. Further experiments based on a different measure principle are needed to understand this discrepancy.

This viscous surface layer model is an effective way to take bedload transport into account. As a matter of fact, anticipating on the part 2 of this paper where the dynamical mechanisms governing the sediment transport are discussed, transported particles are not passive and exert a stress on the fluid. Close to the transport threshold, their influence on the flow is negligible. However, as their density increases, transport induces a negative feedback on the flow, which should be taken into consideration in the hydrodynamics description. The simplest model of multi-layer sheet flow would be a Newtonian fluid whose viscosity increases with the concentration of moving sediments. In this large shear velocity regime, one thus expects a decrease of the phase-lag responsible for the ripples instability and possibly, a restabilisation of the bed.
\begin{figure}
\includegraphics{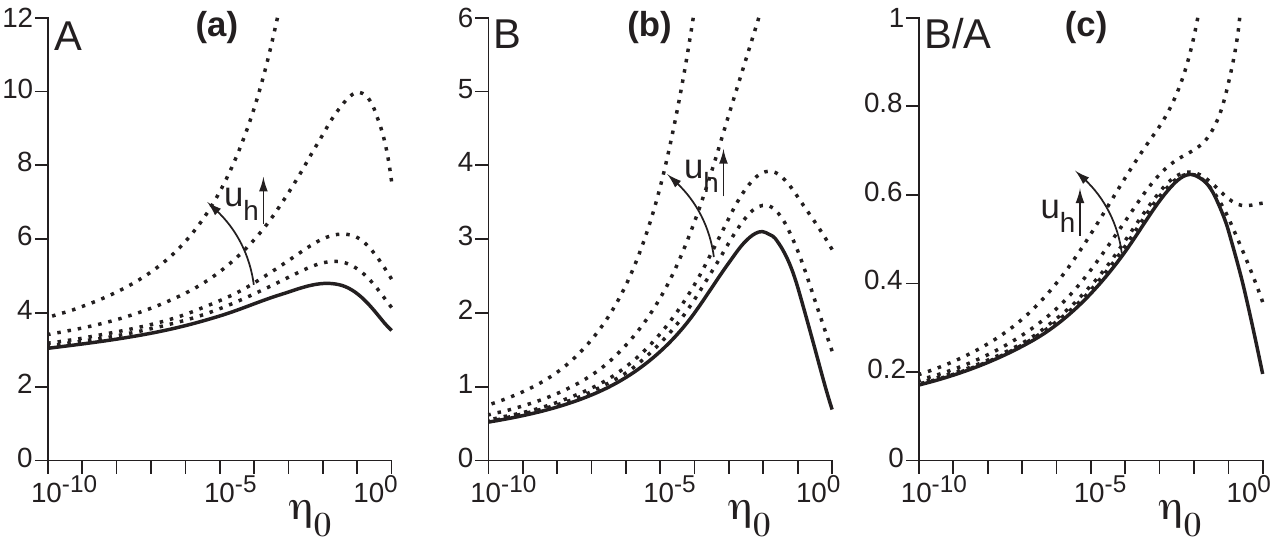}
\caption{(a) Shear stress coefficients $A$ and $B$ computed in the presence of a `focus point' at height $h_0$, where the velocity is $u_h$, as a function of $\eta_0$. $A$ and $B$ are larger for larger values of $u_h/u_*$ ($1$, $2$, $5$ and $10$). However, the ratio $B/A$ is less sensitive to this parameter, up to $\eta_0 \simeq 10^{-3}$. Again, the solid lines display the reference case of figure~\ref{ABCDLin}.}
\label{ABCDHfUf}
\end{figure}
%

\subsection{The focus point assumption}
An alternative manner to take the feedback of the transport on the flow into account can be achieved in analogy with the aeolian case, which provides the archetype of such a situation. In this case, it has been shown that the moving grains slow down the flow in the transport layer, whose thickness $h_0$ is independent of the shear velocity $u_*$. Note that in the subaqueous case, the transport layer thickness is observed to gently increase with the shear stress (\cite{AF77,LvB76}) close to the threshold, in the erosion limited regime (see part 2). Above $h_0$, the effect of the particles on the flow is negligible and one recovers the undisturbed logarithmic velocity profile, but with a roughness larger than that without transport. Below $h_0$, the flow velocity is reduced and is independent of $u_*$ (\cite{UH87,A04}). As shown experimentally by \cite{B41}, the velocity vertical profiles measured for different shear velocities thus cross at the `focus point' $z=h_0$ and $u_x=u_h$. At this point we have
\begin{equation}
\frac{u_h}{u_*}=\frac{1}{\kappa} \ln \frac{h_0}{z_0} \, ,
\label{ufocus}
\end{equation}
which means that the effective roughness in the logarithmic region, due to this transport layer, is
\begin{equation}
z_0 = h_0 \, e^{-\kappa u_h/u_*}
\label{zofocus}
\end{equation}

To determine the flow field in such a situation, the crucial point is to compare $h_0$ with the thickness of the inner layer $\ell$, i.e. the size of the constant stress plateau (see figure~\ref{ModesVerticaux}). If $h_0$ is larger than $\ell$, it means that one cannot reduce the transport issue to a relationship between the sediment flux and the basal shear stress only. In that case, the whole vertical velocity profile, which depends on the entire bottom elevation, is involved. Conversely, for $h_0<\ell$, one can account for transport by modifying the bottom boundary conditions as follows. Following what we have done in the previous sub-section, we can impose that the fluid velocity at $z=Z+h_0$ is parallel to the bed and equal to $u_h$. At the linear order, we then get:
\begin{eqnarray}
U(kh_0) & = & -\mu'(kh_0)
\label{Uf} \\
W(kh_0) & = & iu_h/u_*=i \mu(kh_0).
\label{Wf}
\end{eqnarray}
The result of this choice is shown for the stress coefficients in figure~\ref{ABCDHfUf} for various values of $u_h/u_*$. $A$ and $B$ are larger for increasing focus velocities, or equivalently larger focus altitude. As in the viscous surface layer case, all curves collapse for $\eta_0 \to 0$ because $\ell$ gets larger in this limit (see equation (\ref{ellscaling})). Interestingly, as far as bedforms are concerned (see Part 2), the ratio $B/A$ is much less sensitive to variations of $u_h/u_*$, at least in the region $\eta_0 < 10^{-3}$.

Finally, it should be noted that the focus point model only applies to the momentum limited transport regime (see part 2). Close to the transport threshold, in the erosion limited regime, the feedback of the particle transport on the flow is negligible and the transport should not to be taken into account in the hydrodynamical calculation. This approach should be distinguished from that used by \cite{C04,CS05,CS08}. In those articles, the flow boundary conditions are taken on the bottom, below the transport layer (vanishing velocities), meaning that the feedback of the transport on hydrodynamics is neglected. However, the stresses are evaluated in $h_0$, above the transport layer. This does not constitute a self-consistent choice. Moreover, this introduces a free parameter in the model which can be tuned to choose the phase shift at will. We will present in the second part experimental evidences that this choice is not correct.
\begin{figure}
\includegraphics{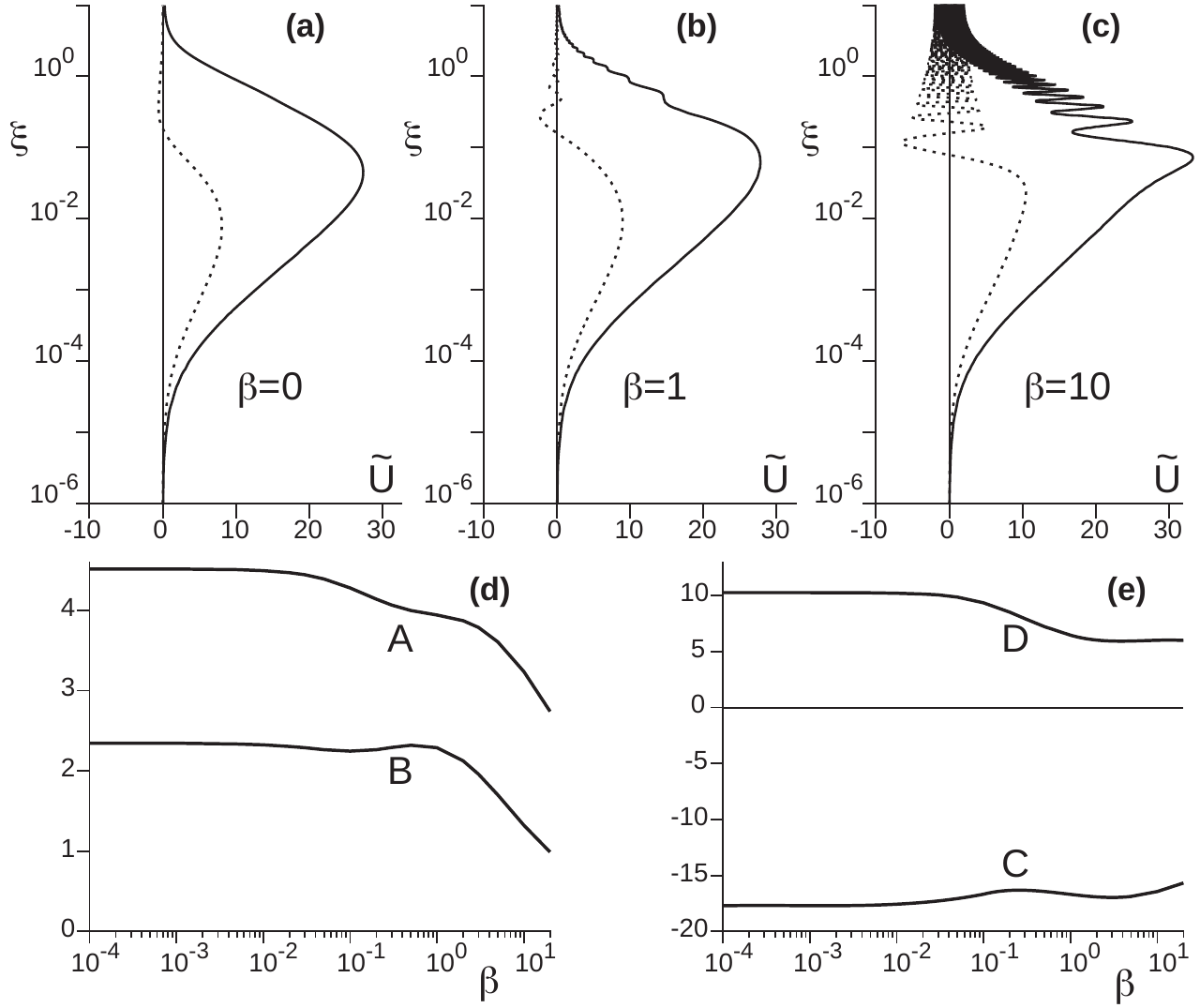}
\caption{Effect of the parameter $\beta$,which is a non dimensional parameter encoding the time lag between a change in the strain rate and that of the Reynolds stress.  (a-c) Vertical profiles of the horizontal component of the velocity $\tilde U=U + \mu'$ for $\eta_0=10^{-4}$ for (a) $\beta=0$, (b) $\beta=1$ and (c) $\beta=10$ respectively. One can see that the profiles develop oscillations as $\beta$ increases, but that the behaviour close to the bottom (in log scale) remains the same. In panels (d) and (e), we plot the coefficients $A$, $B$, $C$ and $D$ \emph{vs} $\beta$ (still for $\eta_0=10^{-4}$). They are weakly affected, meaning again that the behaviour close to the bottom is almost unchanged.}
\label{BetaEffect}
\end{figure}

\subsection{Concluding remarks}
For these three dynamical mechanisms controlling the hydrodynamical roughness $z_0$, we have seen that the asymptotic regime is recovered when the surface layer thickness $h_0$ is smaller than the inner layer thickness $\ell$. As $\ell$ is much smaller than the wavelength $\lambda$ (for standard bedforms, $\ell/\lambda=\go(10^{-3})$), this constitutes a rather restrictive condition. Whenever $h_0$ is larger than $\ell$, specific hydrodynamic models should be derived to determine the relations between stresses and topography.

The phase shift between the basal shear stress and the topography originates from the interplay between inertia and shear stress. The different models of surface layer correspond to different ways of mixing momentum in the direction normal to the wall. Although the argument is general, one sees that the precise value of the phase shift is rather sensitive to the physical origin of the momentum fluxes. In particular, viscous diffusion leads to a much smaller phase advance than turbulent mixing.

\section{Robustness of the results}
\label{robust}

In the same spirit as the previous section, where we have investigated the effect of different ways to treat the surface layer on the stress coefficients $A$ and $B$, we would like now to show the robustness of the results when considering a second order turbulent closure, Reynolds stress anisotropy or a moving bottom.

\subsection{A second order turbulent closure}
As already stated, to solve quantitatively the `dune problem', we need to take correctly into account the effects inducing a phase shift between the stresses and the relief. As a matter of fact, a first order closure assumes that the turbulent energy adapts instantaneously to the mean strain tensor. To take into account the lag between the stress and the strain tensors, one needs to formulate a second order turbulent closure. This can be achieved by deriving dynamical equations for $\tau_{ik}$, which relax the stresses towards their steady state expression prescribed by equation~(\ref{tauijPrandtl}) (see Appendix~\ref{2ndorder}).
\begin{equation}
D_t \tau_{ik}=\partial_t \tau_{ik}+u_j \partial_j \tau_{ik} =\frac{|\dot \gamma|}{\beta} \left[\kappa^2 L^2 \left(\delta_{ij} \frac{1}{3} \chi^2 |\dot \gamma|^2-|\dot \gamma| \dot \gamma_{ij}\right) -\tau_{ij} \right].
\label{2ndME}
\end{equation}
In this relation, the parameter $\beta$ encodes the time lag between an increase of the mean shear strain rate and the corresponding re-adjusment of the fluctuations of the shear stress. We expect $\beta$ to be on the order of unity.

In figure~\ref{BetaEffect}, we show the effect of this new parameter. As expected for inertial effects in a relaxation process, finite values of $\beta$ generate oscillations in the vertical profiles of the velocities and stresses. The example of the horizontal velocity is displayed in the panels (a), (b) and (c). The amplitude and the frequency of these oscillations increase with $\beta$. Interestingly, these oscillations do not affect much the behaviour of the modes close to the bottom. As a consequence, the coefficients $A$, $B$, $C$ and $D$ are weakly affected by $\beta$, see panels (d) and (e). Interestingly, both $A$ and $B$ decrease as $\beta$ increases and their ratio remains roughly constant. $\beta$ has thus a negligible effect on the emergence of bedform, and we shall keep it to zero in the rest of the paper, as well as in part 2.

\begin{figure}
\includegraphics{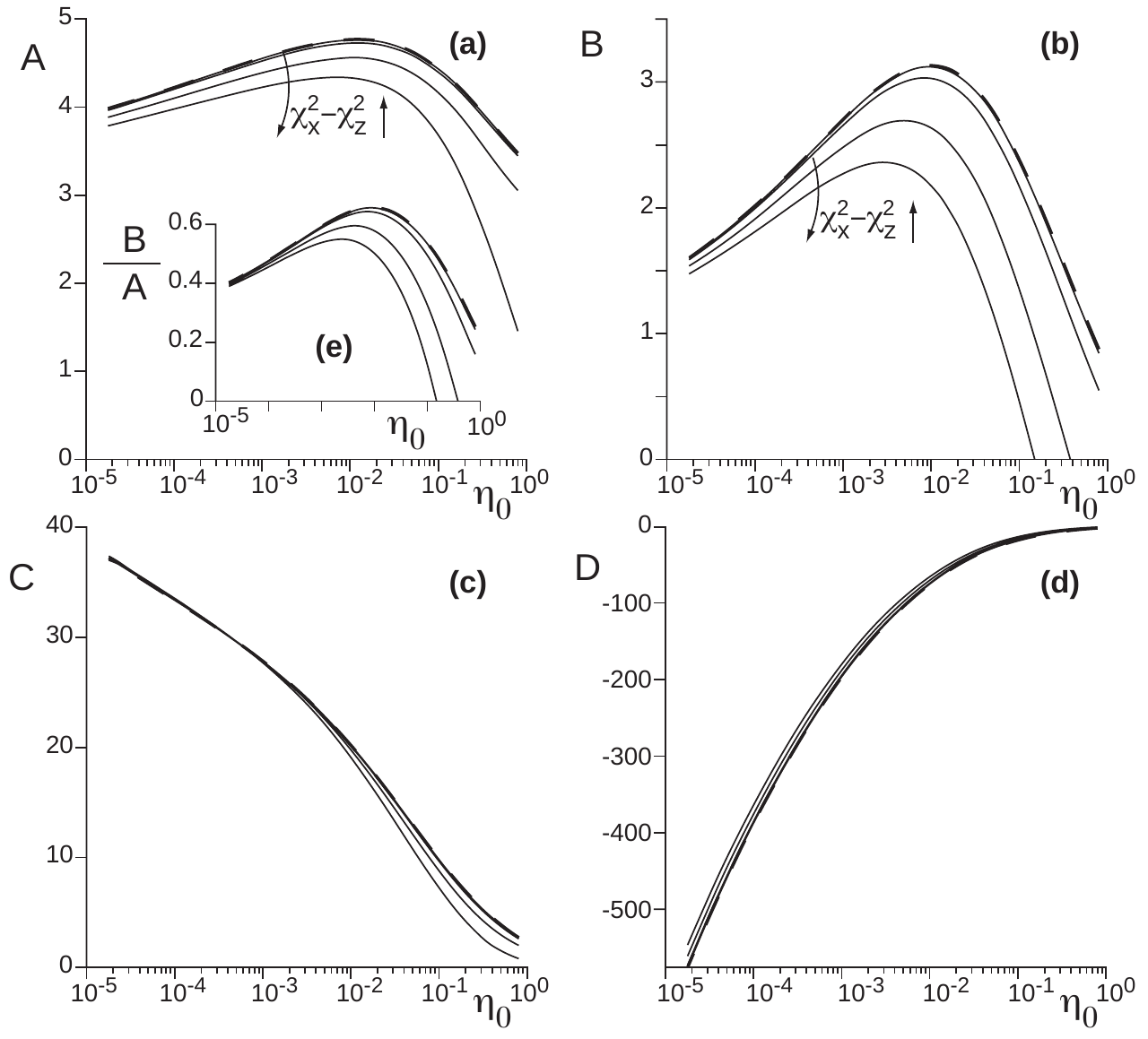}
\caption{Effect of the normal stress anisotropy. (a-d) Coefficients $A$, $B$, $C$ and $D$ as a function of $\eta_0$ for different values of $\chi_x^2-\chi_z^2$ ($0.1$, $1$, $5$ and $10$). The dashed lines correspond to the isotropic case. Inset (e): ratio $B/A$. Arrows indicate increasing normal stress anisotropy.}
\label{ABCDAnis}
\end{figure}
%

\subsection{Reynolds stress anisotropy}
It is an experimental fact that, in a turbulent boundary layer close to a rough wall, the Reynolds stress tensor is \emph{not} isotropic: $\tau_{xx}$ is significantly larger than the other components (\cite{RAR91,SA95}). Besides, anisotropy seems less pronounced for a larger bottom roughness (\cite{KA94,KLMT02}), an issue which is however still matter of debate (\cite{KABA05}).

To account for this Reynolds stress anisotropy, it is easy to generalise the Prandtl-like first order turbulent closure (\ref{tauijPrandtl}) with the following expression:
\begin{equation}
\tau_{ij} =\kappa^2 L^2 |\dot \gamma| \left(\frac{1}{3}\chi_i^2 |\dot \gamma| \, \delta_{ij} - \dot \gamma_{ij} \right),
\label{tauijPrandtl_anisotropic}
\end{equation}
where the value of $\chi_x$ now differs from that of $\chi_z$. Following the above-cited literature, we expect $\chi_x^2 / \chi_z^2$ to be around $1.3$--$1.5$. The modification of the matrix $\mathcal{P}$ due to this new closure is detailed in Appendix~\ref{AvI}. It is shown that the relevant anisotropic parameter entering the equations is $\chi_x^2-\chi_z^2$, for which a realistic value is on the order of unity. As evidenced in figure~\ref{ABCDAnis}, the corresponding values of the functions $A$, $B$, $C$ and $D$ are not much affected by this anisotropy in the relevant range of $\eta_0$. This is particularly true for the coefficients $C$ and $D$, as well as for the ratio $B/A$ as soon as $\eta_0 < 10^{-2}$. The normal stress anisotropy has thus a negligible influence on ripple and dune formation, and we will not take it into account in the rest of the paper, as well as in part 2.

\begin{figure}
\includegraphics{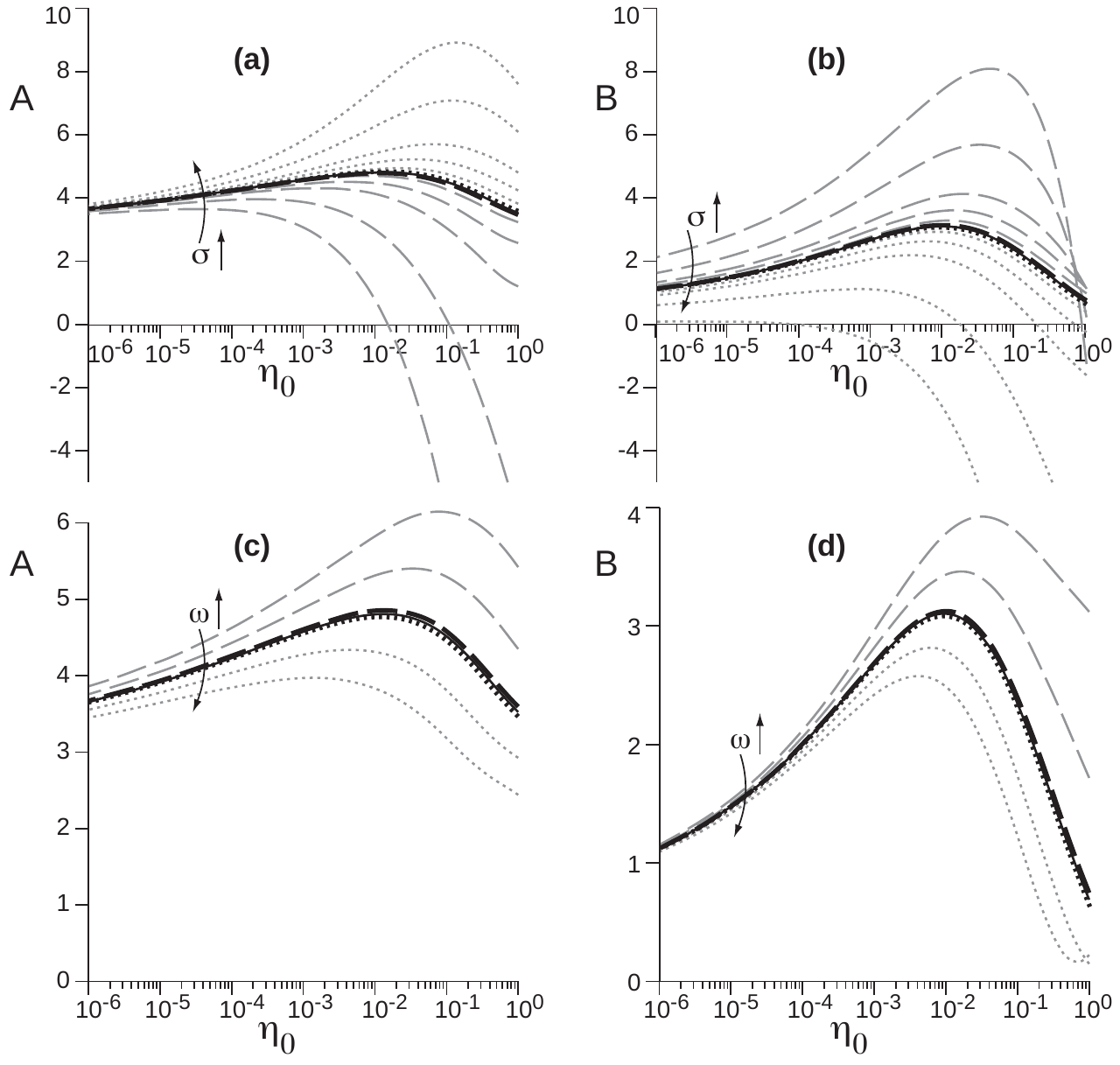}
\caption{Effect of the motion of the bottom. Stress coefficients $A$ and $B$ as a function of $\eta_0$ for different values of the bottom growth rate $\sigma$ (panels (a) and (b)), and different values of the bottom pulsation $\omega$ (panels (c) and (d)). Arrows indicate increasing values of $\sigma$ and $\omega$. In panels (a) and (b), grey dotted lines correspond to $\frac{\sigma}{ku_*}=10$, $5$, $2$, $1$ and $0.5$, the black dotted line being for $\frac{\sigma}{ku_*}=0.1$. The black dashed line corresponds to $\frac{\sigma}{ku_*}=-0.1$, the grey dashed lines being for $\frac{\sigma}{ku_*}=-0.5$, $-1$, $-2$, $-5$ and $-10$. In panels (c) and (d) grey dotted lines correspond to $\frac{\omega}{ku_*}=2$ and $1$, the black dotted line being for $\frac{\omega}{ku_*}=0.1$. The black dashed line correspond to $\frac{\omega}{ku_*}=-0.1$, the grey dashed line being for $\frac{\omega}{ku_*}=-1$ and $-2$. For comparison, in all panels the solid lines correspond to the static case $\sigma=0$, $\omega=0$.}
\label{ABCDSigmaOmega}
\end{figure}
%

\subsection{A moving bottom}
\label{movingboundary}
In order to investigate the effect of a moving bottom on the stress coefficients, we consider a bottom profile of wavevector $k$ like in (\ref{bottomprofile}), but which is now function of both position $x$ and time $t$:
\begin{equation}
Z(x,t) = \zeta\,e^{\sigma t}\,e^{i(kx-\omega t)}\, .
\label{movingbottom}
\end{equation}
In this expression, $\sigma$ represents the growth rate of the profile, and $\omega/k$ its phase velocity. As discussed in \cite{CS05}, this investigation is important as we wish to use the present hydrodynamical study in the context of the formation and development of bedforms, which do have a (small) growth rate and a (small) velocity. Following expression (\ref{movingbottom}), we modify  those for the functions $U$, $W$, $S_t$ and $S_n$ by inserting the extra-factor $e^{(\sigma-i\omega)t}$, along the lines of (\ref{defU}-\ref{defSn}).

In this new case, the linearised equations (\ref{NSxLin}) and (\ref{NSzLin}) of section~\ref{UnboundedCase} must then be modified in the following manner:
\begin{eqnarray}
S_t' &=&  \left ( \frac{\sigma}{ku_*} - i \frac{\omega}{ku_*} + i \mu \right ) U +  \mu' W + i S_n + i S_{xx} - i S_{zz}\, ,
\label{movingNSxLin}
\\
S_n' &=& - \left ( \frac{\sigma}{ku_*} - i \frac{\omega}{ku_*} + i\mu \right ) W + i S_t\, .
\label{movingeNSzLin}
\end{eqnarray}
The linear equation (\ref{systlinplaque}) is then the same, but now with the modified matrix
\begin{equation}
{\mathcal{P}}_{\tilde{t}} =  \left ( \!\!\!\!
\begin{array}{cccc}
0 & -i & \frac{1}{2}\mu' & 0 \\
-i & 0 & 0 & 0 \\
\frac{4}{\mu'}+\frac{\sigma}{ku_*} + i\left( \mu-\frac{\omega}{ku_*} \right) & \mu' & 0 & i \\
0 & -\frac{\sigma}{ku_*} - i\left( \mu-\frac{\omega}{ku_*} \right)  & i & 0
\end{array}
\! \right ) \! .
\label{matrixP_time}
\end{equation}
With the surface layer model described in section~\ref{geom}, the non-slip boundary conditions on the bottom can be written as
\begin{equation}
U(0) = -\mu'(0)
\qquad {\rm et} \qquad
W(0) = \frac{\sigma}{ku_*} - i \frac{\omega}{ku_*}\, .
\label{CLfond_time}
\end{equation}
The result of the integration of this new system is shown in figure~\ref{ABCDSigmaOmega} for the shear coefficients $A$ and $B$. One can see that departure from the static case $\sigma=0$ and $\omega=0$ is noticeable only for values of $\frac{\sigma}{ku_*}$ and $\frac{\omega}{ku_*}$ of order one. The effect of the wave propagation of the bedform can be understood by a simple argument. When the bedforms propagate upstream ($\omega<0$) the relative flow velocity seen by the structure is larger so that it induces a larger shear stress modulation. As $A+iB$ is by definition the basal shear stress rescaled by $u_*^2$, both $A$ and $B$ get larger. Reciprocally, when the bedforms propagate downstream  ($\omega>0$), these coefficients are reduced. Consistently with this argument, the ratio $B/A$ is only weakly affected by $\omega$ (not shown). The growth rate $\sigma$ affects $A$ and $B$ in opposite ways and thus changes the phase shift between the shear stress and the topography. For $\sigma>0$, $A$ is increased while $B$ is reduced. We have not been able to interpret this behaviour in a simple way.

As discussed in part 2, for ripples in water flows these dimensionless numbers are respectively on the order of $10^{-3}$ and $10^{-2}$. They would be even smaller for bedforms of larger wavelength. As a consequence, the assumption that the bottom can be treated as fixed for the study of bedforms is definitively valid (see also the discussion of figure~\ref{DeltaSigmaOmega} below).

\section{Weakly non-linear expansion}
\label{NL}

In this section, we investigate the non-linear effects at finite values of the rescaled bottom corrugation amplitude $k\zeta$. In particular, we wish to address two issues: we wish to relate the hydrodynamic roughness to geometrical quantities; we aim also to describe the first non-linear corrections to the basal stress coefficients $A$, $B$, $C$ and $D$. These results will be used in the second part of this work, to determine the equilibrium height of ripples. In this context, few authors have studied the influence the non-linear terms from hydrodynamics on the shape (\cite{F74}) or the stability (\cite{JM97}) of the bedforms. In contrast to the present paper, however, both of these works describe the turbulent closure with a constant eddy viscosity. In a very empirical manner, \cite{McLS86} computed the flow over two-dimensional bedforms of arbitrary amplitude by the use of a wake function, as described in \cite{SG00}, coupled with a potential flow modified to take into account the velocity logarithmic law. Finally, the linear results of \cite{C04} have also recently been extended to the weakly non-linear situation (\cite{CS08}).

\subsection{Expansion in amplitude}
For this purpose, we perform an expansion with respect to the bottom corrugation amplitude, and introduce non-dimensional the following functions for the different orders:
\begin{eqnarray}
u_x & = & u_* \left[\mu+(k\zeta) e^{ikx} U_1+(k\zeta)^2 U_{0}+(k\zeta)^2  e^{2ikx} U_{2}+(k\zeta)^3 e^{ikx} U_{3}\right], \\
u_z & = & u_* \left[(k\zeta) e^{ikx} W_1+(k\zeta)^2 W_{0}+(k\zeta)^2  e^{2ikx} W_{2}+(k\zeta)^3 e^{ikx} W_{3}\right], \\
\tau_{xz} & = &  - u_*^2 \left[1+(k\zeta) e^{ikx} S_{t1}+(k\zeta)^2 S_{t0}+(k\zeta)^2  e^{2ikx} S_{t2}+(k\zeta)^3 e^{ikx} S_{t3}\right], \\
p+\tau_{zz} & = & p_0+u_*^2 \left[(k\zeta) e^{ikx} S_{n1}+(k\zeta)^2 S_{n0}+(k\zeta)^2  e^{2ikx} S_{n2}+(k\zeta)^3 e^{ikx} S_{n3}\right],\\
\tau_{zz}- \tau_{xx}& = & u_*^2  \left[(k\zeta) e^{ikx} S_{d1}+(k\zeta)^2 S_{d0}+(k\zeta)^2  e^{2ikx} S_{d2}+(k\zeta)^3 e^{ikx} S_{d3}\right].
\end{eqnarray}
Terms in $(k\zeta)^3 e^{3ikx}$, which do not contribute to the harmonic response (i.e. to the stress coefficients), as well as terms of higher order than $(k\zeta)^3$ are neglected. Although the principle of the expansion in amplitude is simple, the actual calculations are painful, and the technical details have been gathered in appendix~\ref{WNLC}. In summary, the non-linear effects result from the expansion of the mixing length (terms in $(k\zeta)^2$) and from the self-interaction of the linear perturbations: in particular, the combination of terms $(k\zeta) e^{ikx}$ generates second order terms in $(k\zeta)^2$.  All involved functions are complex, except $\mu$ and those related to the second order homogeneous corrections (index $0$). To avoid the determination of the asymptotic behaviours in this case, we have chosen to treat the surface layer by the simple phenomenological model of section~\ref{controlzzero}. Plugging the above expressions into the Navier-Stokes and turbulent closure equations, one eventually obtains a linear hierarchy of linear differential equations:
\begin{equation}
\frac{d}{d\eta} \vec{X}_\alpha = {\mathcal{P}_\alpha} \vec{X}_\alpha + \vec{S}_\alpha ,
\end{equation}
where $\vec{X}_\alpha = (U_\alpha, W_\alpha, S_{t\alpha}, S_{n\alpha})$. Of course, $\mathcal{P}_1$ and $\vec{S}_1$ are the matrix and vector of expression (\ref{systlinplaque}). We find that $\mathcal{P}_3=\mathcal{P}_1$, whereas $\mathcal{P}_2$ is slightly different and $\mathcal{P}_0$ is very simple:
\begin{equation}
{\mathcal{P}_1} =  {\mathcal{P}_3} =\left (
\begin{tabular}{cccc}
$0$ & $-i $ & $\frac{1}{2}\mu'$ & $0$ \\
$-i$ & $0$ & $0$ & $0$ \\
$\left(i\mu +\frac{4}{\mu'}\right) $ & $\mu'$ & $0$ & $i$ \\
$0$ & $-\mu i$ & $i$ & $0$
\end{tabular}
\right ),
\label{P1and3}
\end{equation}
\begin{equation}
{\mathcal{P}_2} =  \left (
\begin{tabular}{cccc}
$0$ & $-2i $ & $\frac{1}{2}\mu'$ & $0$ \\
$-2i$ & $0$ & $0$ & $0$ \\
$2\left(i\mu +\frac{8}{\mu'}\right) $ & $\mu'$ & $0$ & $2i$ \\
$0$ & $-2\mu i$ & $2i$ & $0$
\end{tabular}
\right ),
\quad {\rm and} \quad
{\mathcal{P}_0} =  \left (
\begin{tabular}{cccc}
$0$ & $0$ & $\frac{1}{2}\mu'$ & $0$ \\
$0$ & $0$ & $0$ & $0$ \\
$0$ & $0$ & $0$ & $0$ \\
$0$ & $0$ & $0$ & $0$
\end{tabular}
\right ).
\label{P2and0}
\end{equation}
All the heaviness of the method is encoded in the expressions of the right hand terms $\vec{S}_\alpha$: the components of such vectors at a given order depend on the lower order functions $\vec{X}_{\alpha}$ and their derivatives. The integration has thus to follow the hierarchy of the equations, one order after the other.

\begin{figure}
\includegraphics{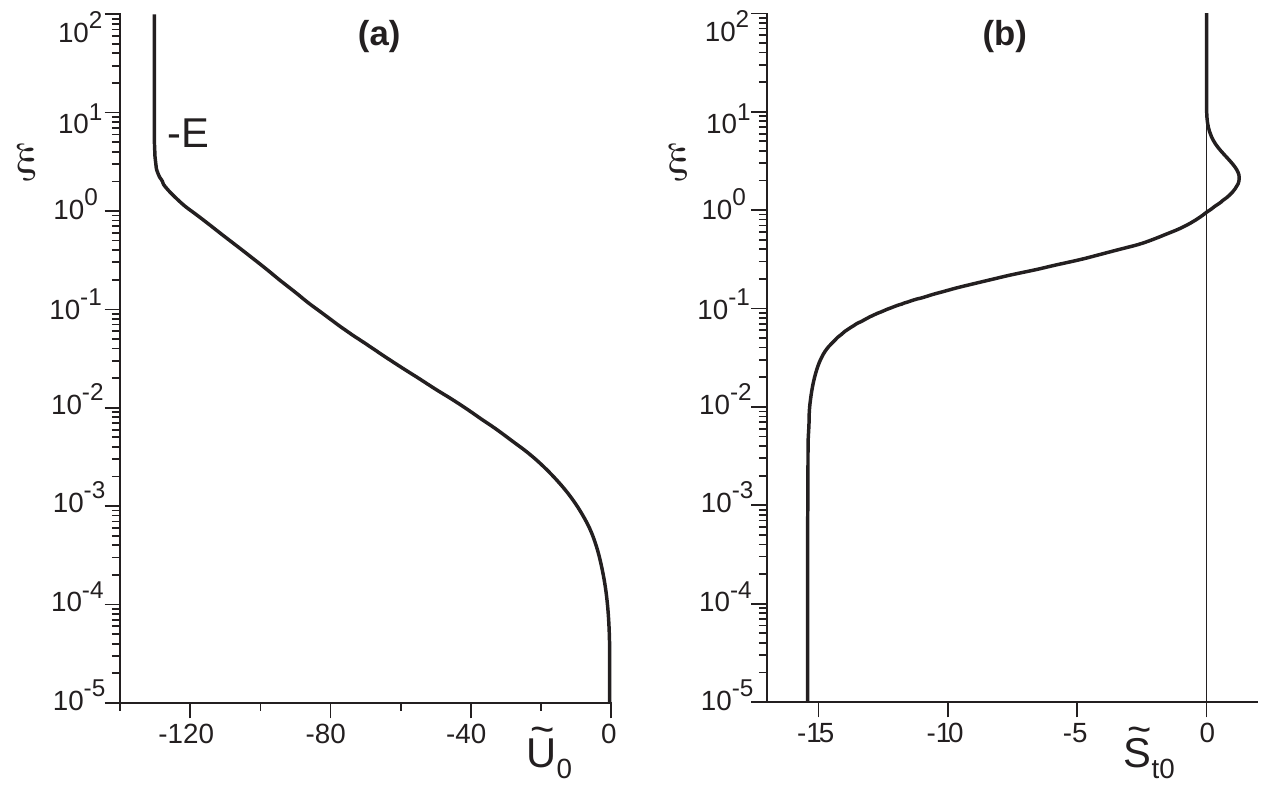}
\caption{Vertical profiles of the homogeneous second order corrections $\tilde{U}_0$ (a) and $\tilde{S}_{t0}$ (b). These curves have been computed with $\eta_0=2. \,10^{-3}$. $\tilde{U}_0$ tends towards a negative constant value $-E$ at large $\xi$, which corresponds to an increased roughness at large distance from the wall. Notice also that, close to the bottom, $\tilde{S}_{t0}$ has a constant value, reminiscent of the inner layer.}
\label{ModeNLZero}
\end{figure}
%

\subsection{Boundary conditions}
The boundary conditions must be specified in order to perform the integration. Following the geometrically induced roughness model (section~\ref{controlzzero}), we require that both components of the velocity should vanish on the bottom. These conditions express easily in the shifted representation, i.e. written in terms of the curvilinear coordinates (see appendix~\ref{rep}): they simply read $\tilde{U}_\alpha(0)=0$ and $\tilde{W}_\alpha(0)=0$. In terms of the Cartesian functions, we get:
\begin{eqnarray}
U_1(0) & = & -\mu'(0), \\
U_0(0) & = & -\frac{1}{4} \mu''(0) - \frac{1}{2} \kappa \mu'^2(0) - \frac{1}{8} \mu'(0) \left [ S_{t1}(0) + S_{t1}^*(0) \right ], \\
U_2(0) & = & -\frac{1}{4} \mu''(0) - \frac{1}{2} \kappa \mu'^2(0) - \frac{1}{4} \mu'(0) S_{t1}(0), \\
U_3(0) & = & -\frac{1}{8} \mu'''(0) + \frac{9}{8} \mu'(0) - \frac{3}{4} \kappa^2 \mu'^3(0) -\frac{1}{16} \left [ \mu''(0) + 2 \kappa \mu'^2(0) \right ] \left [ 2S_{t1}(0)+S_{t1}^*(0) \right ] \nonumber \\
& & + \frac{1}{32} \mu'(0) S_{t1}(0) \left [ S_{t1}(0)+2S_{t1}^*(0) \right ] - \frac{i}{16} \mu'(0) \left [ 2S_{n1}(0)+S_{n1}^*(0) \right ] \nonumber \\
& & - \frac{1}{4} \mu'(0) \left [ 2S_{t0}(0)+S_{t2}(0) \right ],
\end{eqnarray}
and
\begin{eqnarray}
W_1(0) & = & 0, \\
W_0(0) & = & 0, \\
W_2(0) & = & - \frac{i}{2} \mu'(0), \\
W_3(0) & = & -\frac{i}{4} \mu''(0) - \frac{3i}{8} \kappa \mu'^2(0) - \frac{i}{16} \mu'(0) \left [ 2S_{t1}(0)+S_{t1}^*(0) \right ].
\end{eqnarray}
\begin{figure}
\includegraphics{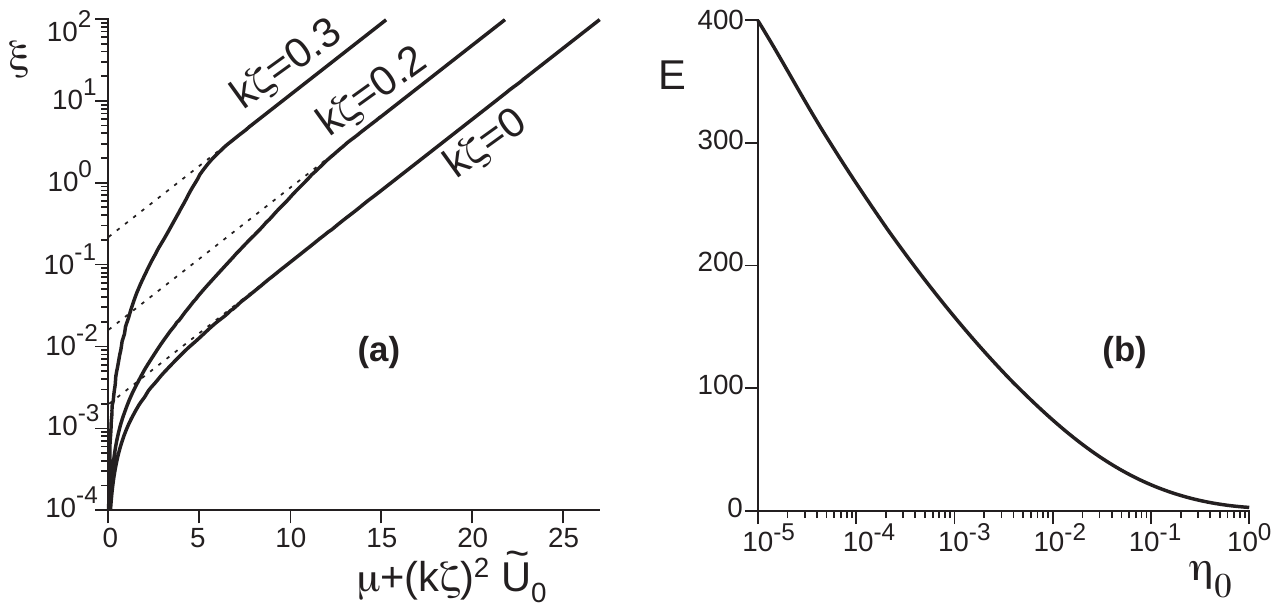}
\caption{(a) Vertical profiles of the homogeneous component of the velocity $\mu+ (k\zeta)^2 \tilde{U}_0$ for $k\zeta=0$,  $k\zeta=0.2$ and $k\zeta=0.3$. The extrapolation to $0$ of the upper part of the curves gives the value of the hydrodynamical roughness seen at a distance from the bottom. (b) Coefficient $E$ as a function of $\eta_0$. A good fit of this function is given by $E=2.75\,(\ln\eta_0-0.62)^2$.}
\label{Rugose}
\end{figure}

As in previous sections, for each order, the solution is expressed as a linear superposition of the form: $\vec{X}_\alpha = \vec{X}_{s\alpha} + a_{t\alpha} \vec{X}_{t\alpha} + a_{n\alpha} \vec{X}_{n\alpha}$, where the different vectors are solutions of the following equations:
\begin{eqnarray}
\frac{d}{d \eta}\vec{X}_{s\alpha}= {\mathcal P}_\alpha \vec{X}_{s\alpha}+\vec{S}_\alpha & \qquad \mbox{with} \qquad &
\vec{X}_{s\alpha} (0) =\left (\begin{tabular}{c}
$U_\alpha(0)$ \\ $W_\alpha(0)$ \\ $0$ \\ $0$
\end{tabular}\right ), \label{equaXsalpha}\\
\frac{d}{d \eta}\vec{X}_{t\alpha}= {\mathcal P}_\alpha \vec{X}_{t\alpha} & \qquad \mbox{with} \qquad &
\vec{X}_{t\alpha} (0) =\left (\begin{tabular}{c}
$0$ \\ $0$ \\ $1$ \\ $0$
\end{tabular}\right ), \label{equaXtalpha}\\
\frac{d}{d \eta}\vec{X}_{n\alpha}= {\mathcal P}_\alpha \vec{X}_{n\alpha} & \qquad \mbox{with} \qquad &
\vec{X}_{n\alpha} (0) =\left (\begin{tabular}{c}
$0$ \\ $0$ \\ $0$ \\ $1$
\end{tabular}\right ). \label{equaXnalpha}
\end{eqnarray}
Again, for the top boundary conditions, we introduce a lid at finite height $H$, impose $S_{t\alpha}(\eta_H)=0$ and $W_\alpha(\eta_H)=0$, and look at the limit $H \to +\infty$, i.e. when the results become independent of $H$. These conditions lead to two equations on $a_{t\alpha}$ and $a_{n\alpha}$, whose solutions give $S_{t\alpha}(0)$ and $S_{n\alpha}(0)$ respectively.

\begin{figure}
\includegraphics{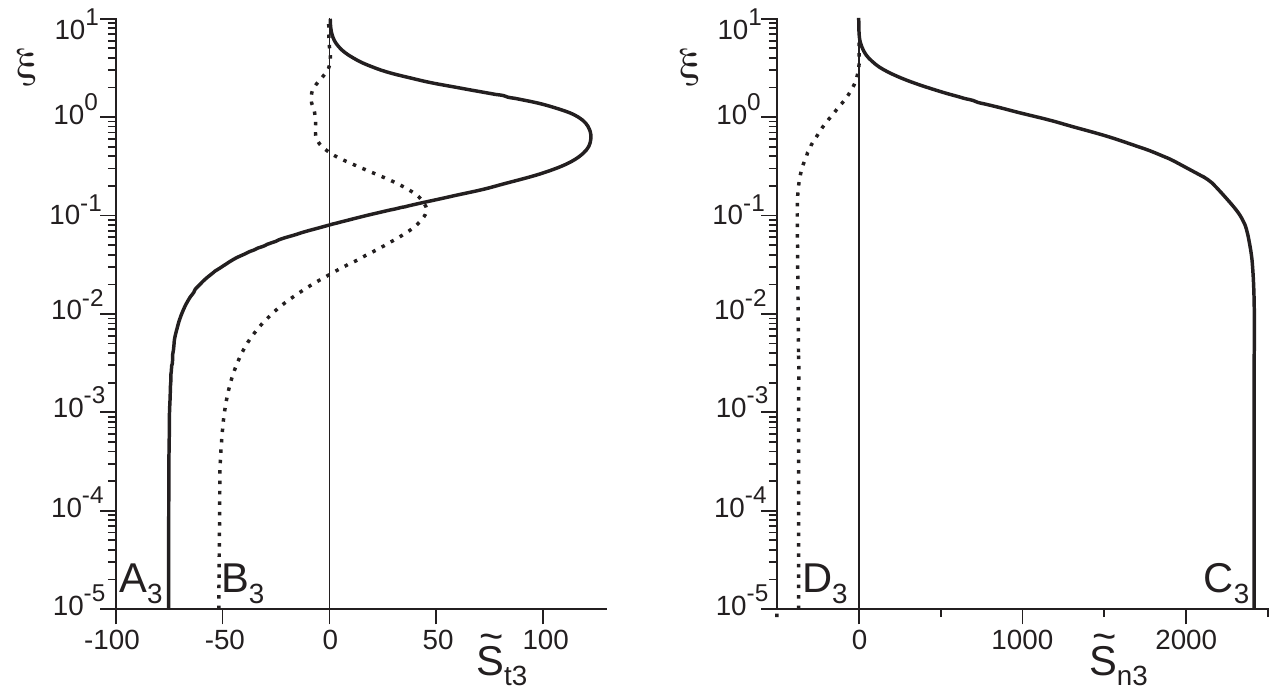}
\caption{Vertical profiles of the third order corrections to the stresses: $\tilde{S}_{t3}$ (a) and $\tilde{S}_{n3}$ (b). The solid lines represent the real parts of the functions, whereas the dashed lines represent the imaginary ones. These curves have been computed with $\eta_0=2. \,10^{-3}$. Notice again the inner layer where the stresses are constant. We write $\tilde{S}_{t3}(0)=A_3+iB_3$ and $\tilde{S}_{n3}(0)=C_3+iD_3$. Note that both $A_3$ and $B_3$ are negative.}
\label{ModeNLThree}
\end{figure}
%

\subsection{Results}
We first consider the corrections to the homogeneous base solution (index $0$). The corresponding velocity profile $\tilde U_0$ and shear stress profile $\tilde{S}_{t0}$ are displayed in figure~\ref{ModeNLZero}. The term in $(k\zeta)^2$ of the velocity decreases continuously from $z \sim z_0$ to $z \sim \lambda$ and tends towards a negative constant $-E$ far from the bottom. Correspondingly, the shear stress decreases and tends to $0$ far from the ground, as requested. The calculation thus predicts an increase of the turbulent drag (i.e. of the basal shear stress) with the corrugation amplitude, due to the non-linearities. In terms of the velocity profile, this translates into a hydrodynamic roughness $z_g$ of geometrical origin. Identifying the expression of the velocity profile far from the bottom with $\frac{u_*}{\kappa} \ln \frac{z}{z_g}$, we simply get:
\begin{equation}
\ln z_g = \ln z_0 + \kappa (k \zeta)^2 E,
\label{defzg}
\end{equation}
As a consequence, $z_g$ increases with $E$ and with the aspect ratio $k \zeta$. For the seek of illustration, several vertical profiles of the homogeneous part of the velocity $\mu+ (k\zeta)^2 \tilde{U}_0$ are plotted in figure~\ref{Rugose}(a) for different values of $k \zeta$. By definition, $z_g$ is the extrapolation of the asymptotic part of the curves to vanishing velocities.

Interestingly, the large scale roughness $z_g$ cannot be related to a single geometrical length, namely to the corrugation amplitude $\zeta$ (\cite{SG00,vR83,RAR91,WN92}). In particular, we predict that the macroscopic roughness $z_g$ associated to a wavy surface still depends on the microscopic roughness $z_0$: as shown in figure~\ref{Rugose}(b), for a given aspect ratio, the apparent roughness $z_g$ is larger for smaller $\eta_0$. Furthermore, expression (\ref{defzg}) is consistent with numerical observations of \cite{TSM89} and computations of \cite{J89}, who respectively report linear (in the range $10^{-8} < \eta_0 < 10^{-3}$) and quadratic variations for the coefficient $E$ as a function of $\ln\eta_0$. Here, the best fit of this function gives $E\simeq2.75\,(\ln\eta_0-0.62)^2$.

\begin{figure}
\includegraphics{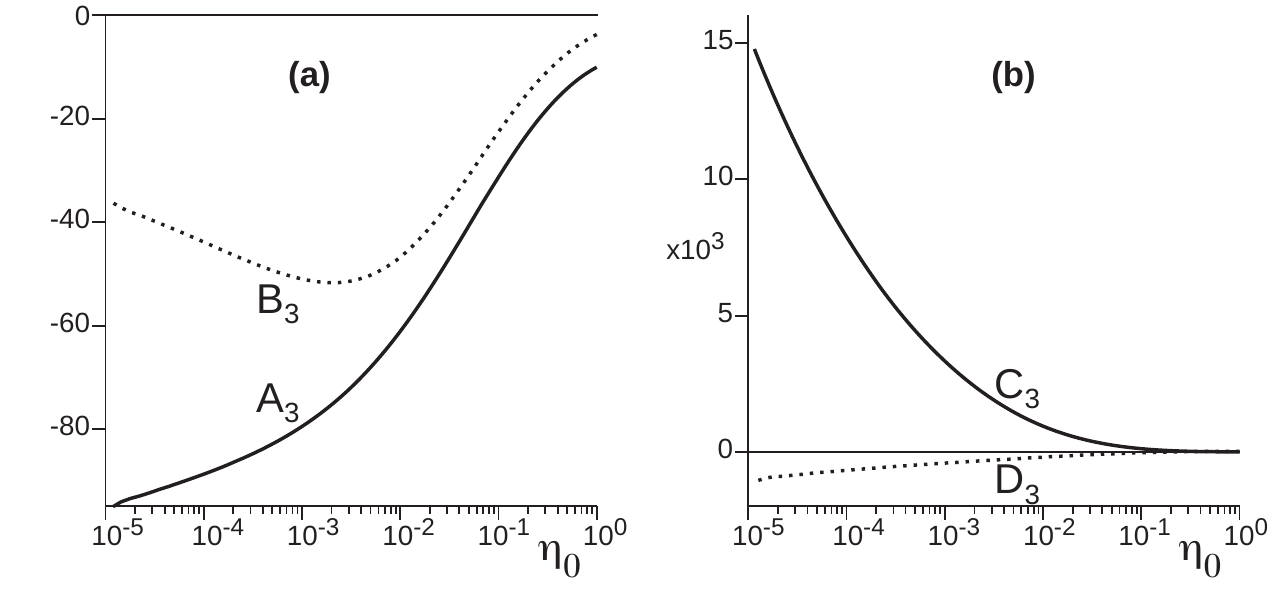}
\caption{Third order stress coefficients $A_3$, $B_3$, $C_3$ and $D_3$ as a function $\eta_0$. Comparing the signs with those of $A=A_1$, $B=B_1$, $C=C_1$ and $D=D_1$, it can be inferred that the non-linearities oppose the linear effects. In particular, as the amplitude increases, the point of maximum shear stress drifts downstream, i.e. $B_1+B_3\,(k\zeta)^2$ decreases.}
\label{ABCDThree}
\end{figure}

The first non-linear corrections to the harmonic terms scale on $(k\zeta)^3$. In figure~\ref{ModeNLThree}, we show the corresponding profiles for the stresses. As in the first order case, there exists a layer close to the bottom where the stresses are almost constant (inner layer). As requested, both components vanish far from the ground. We note $\tilde{S}_{t3}(0)=A_3+iB_3$ and $\tilde{S}_{n3}(0)=C_3+iD_3$ the shear and normal stresses acting on the boundary. These coefficients are plotted as a function of $\eta_0$ in figure~\ref{ABCDThree}. Both $A_3$ and $B_3$ are negative while $A=A_1$ and $B=B_1$ are positive, which means that the first non-linearities oppose the linear effects. We will show in part 2 that this is responsible for the selection of the height of current ripples.

The calculation of the non-linear corrections allows to determine the range of amplitude $\zeta$ for which the linear approximation is valid. The representation of the linear solution with the fixed system of coordinates ($x,z$) is valid only when $\zeta$ is much smaller than $z_0$. However, the representation of the same linear solution in the curvilinear coordinates ($x,z-Z(x)$) is valid up to $\zeta$ of the order of the inner layer thickness $\ell$. It is important to recall that all the descriptions of the flow equivalent at the linear order (including the real solution of the fully non-linear problem)  can have very different domains of validity. 

\begin{figure}
\includegraphics{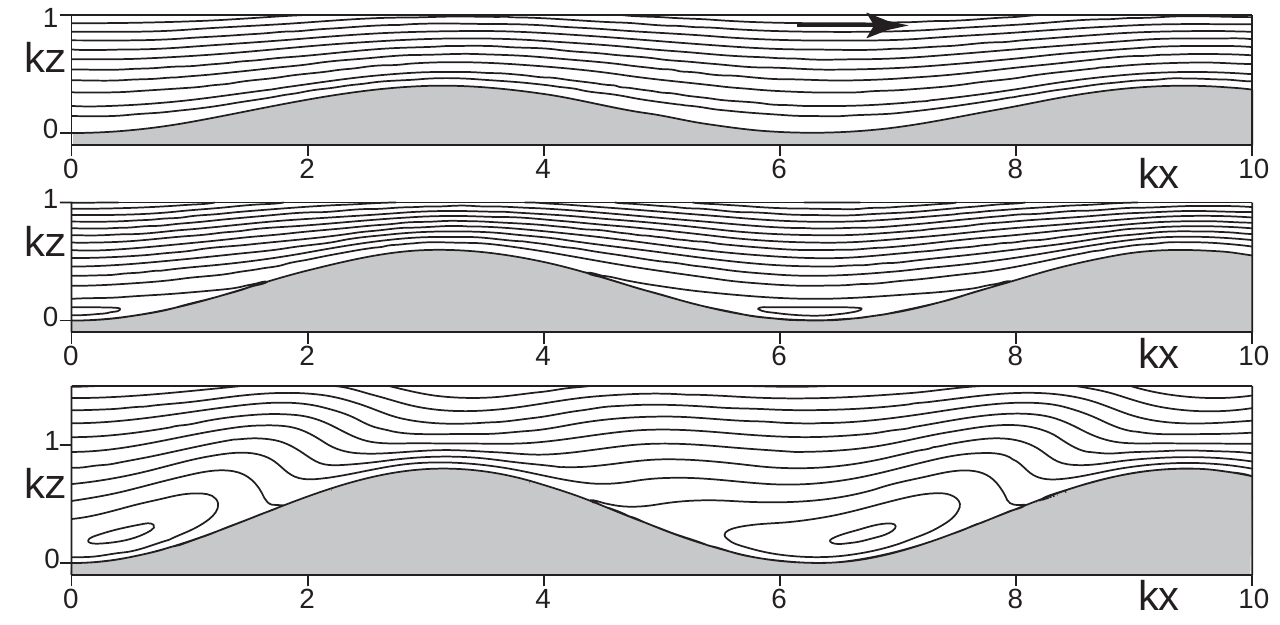}
\caption{Streamlines computed with all non-linear corrections considered here on bumps such that $k\zeta=0.2$, $0.3$ and $0.4$ from top to bottom. The fluid flows from left to right. Note the progressive development of a recirculation bubble for larger acpect ratios.}
\label{StreamlinesNonLinear}
\end{figure}

This weakly non-linear computation is illustrated in figure~\ref{StreamlinesNonLinear}, which shows the streamlines for different aspect ratios. It can be seen that the separation of streamlines and the subsequent formation a  recirculation bubble occurs above an aspect ratio of $\sim 1/13$, in agreement with observations. By comparison, the linear calculation, shown in figure~\ref{Streamlines}, leads to the emergence of a recirculation bubble for an aspect ratio of $1/6$ i.e. twice larger. The non-linear corrections are thus essentials to capture quantitatively the flow features. Panel~\ref{StreamlinesNonLinear}c), computed for an aspect ratio of $1/8$ shows a well-developed recirculation bubble. The distortion of the separation streamline is not realistic, indicating the upper limit of validity of the model. Fortunately, the aspect ratio of ripples and dunes is typically smaller than $1/10$, which falls into the domain of validity of the calculation.

\begin{figure}
\includegraphics{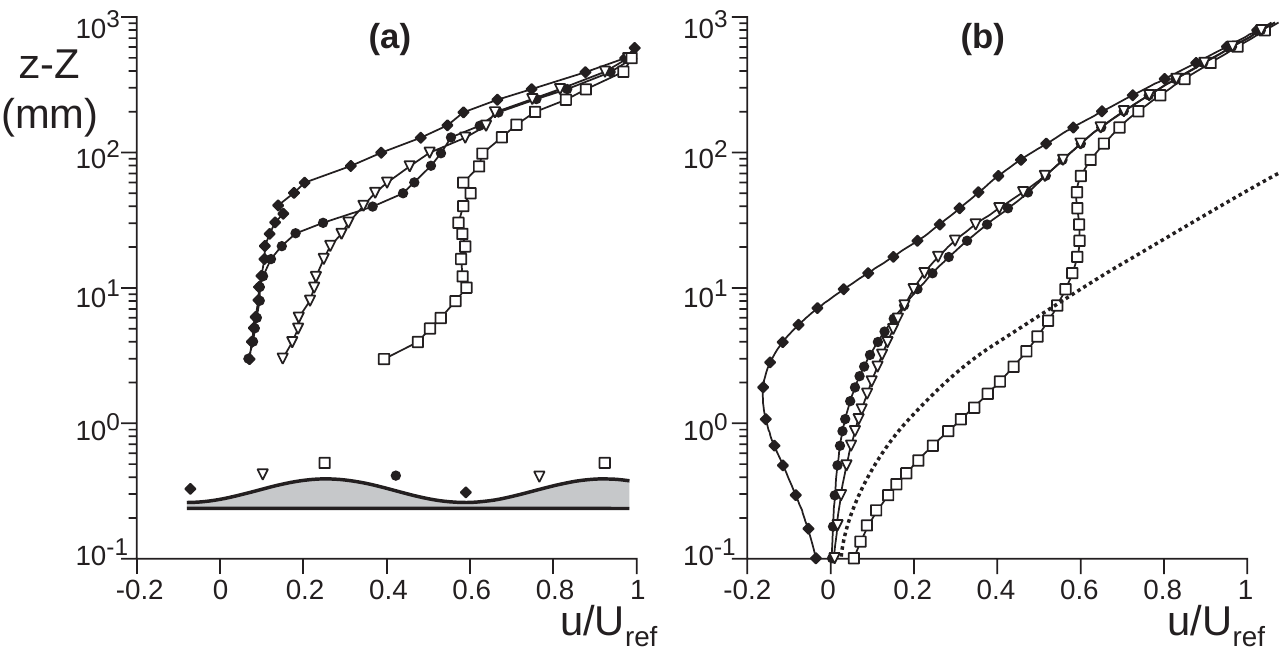}
\caption{(a) Rescaled profiles of the mean velocity modulus, measured by~\cite{GTD96} in the case of a rough wavy bottom of aspect ratio $\sim 1/6$ ($k\zeta=0.5$). $U_{\rm ref}$ is the free stream reference velocity. The symbols correspond different longitudinal locations, as shown in the schematics below the data (crests $\square$, troughs $\blacklozenge$ and half ways up- $\triangledown$ and down-stream $\bullet$). (b) Velocity profile predicted by the present model with a sinusoidal bottom of aspect ratio $1/10$ ($k\zeta=0.31$), plotted with the same symbol code. Dotted line: base velocity profile.}
\label{NonLinearExperiment}
\end{figure}

Several experiments in flumes or wind tunnels have been performed to measure velocity and Reynolds stress profiles over two-dimensional fixed symmetric or asymmetric bedforms (\cite{WN92,NMcLW93,McLNW94,BB95,CNMcLCSM06,V07}). Closer to our calculations, several experiments with a sinusoidal bottom are also reported in the literature (e.g. \cite{ZCH77,ZH79,AH85,BHA84,NH01,PKAR07}). Direct or large eddy numerical simulations of Navier-Stokes equations have also been performed in this geometry (e.g. \cite{dALB97,HS99,SDB01}). For comparison with our model, data from \cite{GTD96} have been chosen. They have been performed in a wind tunnel over sixteen waves with a wavelength $\lambda=609.6$~mm and a trough-to-crest amplitude $2\zeta=96.5$~mm, covered with a carpet to make them aerodynamically rough. This corresponds to an aspect ratio of $1/6$ much too large to be in the domain where the model is quantitative. Unfortunately, we have not found any better data-set for the seek of comparison. The vertical profiles of the velocity measured at different locations are shown in figure~\ref{NonLinearExperiment}(a). More precisely, these authors have measured the average of the instantaneous velocity modulus i.e. a quantity that is always positive and that mixes the average velocity and the fluctuations. Although a recirculation bubble is present, these data cannot show it. We compare these profiles to those computed at the upper limit of validity of the non-linear calculation ($k\zeta=0.3$). Yet, the agreement with our computation is fair; in particular, the way the flow is accelerated over crests and decelerated in troughs in qualitatively well reproduced. The profiles at $\lambda/4$ and $-\lambda/4$ from the crest are close to each other, indicating a re-symmetrisation of the flow by non-linearities. Note that the slight difference between these profiles is qualitatively reproduced by the model. Looking at the upper part of the experimental profiles, one sees that they would extrapolate to $0$ around $10~$mm while the ground roughness is slightly smaller than $1$~mm. These two roughness' are particularly visible on the profile measured on the crest. The model is particularly successful in reproducing this feature.

The non-linear effects on the flow over obstacles are often described in terms of boundary layer separation. It has been proposed by \cite{JZ85} (see also \cite{FRBA90}) that one could still use the linear flow calculation in that case, introducing a fictive surface enveloping the obstacle and the recirculation bubble downstream of it. As such an envelope creates a fictive bump maximum displaced downstream, it artificially moves the point of maximum shear stress on the bump in the same direction (\cite{KSH02,ACD02}). Although this trick is of practical use to simulate dunes, this envelope technique is not based on any firm theoretical ground. The weakly non-linear calculation performed here is thus of extreme interest to incorporate non-linear turbulent effects in dune numerical models in a more controlled way. More generally, it can be used in any problem in which a good approximation of the mean flow is needed at low calculation cost, including separation. For instance, it may find direct applications in the control of turbulence around vehicles. An important limit of such Reynolds averaged calculation is that they do not take vortex shedding into account. 
%

\section{Effect of a free surface}
\label{FS}

In this section, we investigate the effect of the additional presence of a free surface at a finite distance $H$ to the bottom. This situation is relevant to the flow above river dunes (see part 2). We follow the outline of the section~\ref{UnboundedCase}, but staying for easiness with linear calculations in two-dimensional situations.

\subsection{River equilibrium}
In the case of a river inclined at an angle $\theta$ on the horizontal, the shear stress must balance gravity. It thus varies linearly as $\tau_{xz}=g (z-H) \sin \theta$ and vanishes at the free surface. By definition of the shear velocity $u_*$, we also write $\tau_{xz} \equiv u_*^2 (z/H-1)$. In the context of a mixing length approach to describe turbulence, this length should vanish at the free surface. For the sake of simplicity, following the discussion of section~\ref{controlzzero}, we take $L=(z+z_0)\sqrt{1-z/H}$. This choice results in a base flow that is logarithmic as in the unbounded situation:
\begin{equation}
u_x=\frac{u_*}{\kappa} \ln \left(1+\frac{z}{z_0} \right),
\label{uxfreesurface}
\end{equation}
which is consistent with field and experimental observations. The stress balance equation along the $z$-axis allows to get the pressure, which reads:
\begin{equation}
p+\tau_{zz}= p_0 + g  (H-z) \cos \theta=p_0 + \frac{u_*^2}{\tan \theta}\left(1-\frac{z}{H}\right).
\end{equation}
We define the Froude number as the ratio of the surface velocity $u_{\rm surface}$ to the velocity of gravity surface waves in the shallow water approximation:
\begin{equation}
\Fr \equiv \frac{u_{\rm surface}}{\sqrt{gH}} \equiv \frac{1}{\sqrt{gH}} \,  \frac{u_*}{\kappa} \ln \left(1+\frac{H}{z_0} \right)=\frac{1}{\kappa} \ln \left(1+\frac{H}{z_0} \right) \, \sqrt{\sin \theta}.
\end{equation}
In the literature, the Froude number is sometime defined as the ratio of the mean velocity to the velocity of gravity waves. We will justify this choice in the next paragraph. The Froude number of natural sandy rivers lies in general between $0.1$ and $0.3$ as they flow on very small slopes. Larger Froude numbers are reached in flume experiments.
%

\subsection{Disturbances}
In the same manner as in section~\ref{UnboundedCase}, we consider now a wavy bottom $Z=\zeta e^{ikx}$. We note again $\eta=kz$ and $\eta_H=kH$. We write the first order corrections to the base flow as
\begin{eqnarray}
u_x & = & u_* \left[\mu +k\zeta e^{ikx} U \right], \label{defUFS} \\
u_z & = & u_* k\zeta e^{ikx} W, \label{defWFS} \\
\tau_{xz} & = & \tau_{zx}= - u_*^2 \left[1 - \frac{\eta}{\eta_H} + k\zeta e^{ikx} S_t \right], \label{defStFS} \\
p+\tau_{zz} & = & p_0+u_*^2 \left[\frac{1}{\tan\theta} \left(1-\frac{\eta}{\eta_H}\right)+k\zeta e^{ikx} S_n \right], \label{defSnFS}
\end{eqnarray}
where, in accordance with equation (\ref{uxfreesurface}), the function $\mu$ is defined by the relation (\ref{Mu2Etageom}). The free surface is also disturbed by the presence of the non-uniform bottom, and we denote $H+\Delta(x)$ the flow depth at the position $x$. The modified expression for the mixing length then reads
\begin{equation}
L=(z_0+z-Z) \sqrt{\frac{H+\Delta - z}{H+\Delta-Z}} \, .
\end{equation}
Linearising the free surface profile as $\Delta(x)=\delta \zeta e^{ikx}$, one can expand $L$ to the first order as
\begin{equation}
kL=(\eta+\eta_0) \sqrt{1-\frac{\eta}{\eta_H}} \left \{ 1 - k \zeta e^{ikx} \left[\frac{1}{\eta+\eta_0} -\frac{1}{2\eta_H} - \delta \, \frac{\eta}{2\eta_H^2 \left (1-\frac{\eta}{\eta_H} \right )}
\right ] \right \}.
\end{equation}

The shear stress closure as well as the Reynolds averaged Navier-Stokes equations can be linearised in the same way as before, and we finally get at the first order in $k\zeta$ a system of differential equations which can be written under the following form:
\begin{equation}
\frac{d}{d\eta} \vec{X} = {\mathcal{P}} \vec{X}+\vec{S}+\delta \vec{S_\delta},
\label{systlinfreesurface}
\end{equation}
with
\begin{eqnarray}
{\mathcal{P}} & = & \left (
\begin{tabular}{cccc}
$0$ & $-i$ & $\frac{\mu'}{2\left (1-\frac{\eta}{\eta_H} \right )} $ & $0$ \\
$-i$ & $0$ & $0$ & $0$ \\
$\frac{4}{\mu'}\left (1-\frac{\eta}{\eta_H} \right )+i \mu$ & $\mu'$ & $0$ & $i$ \\
$0$ & $-\mu i$ & $i$ & $0$
\end{tabular} \right ),
\\
\vec{S} & = & \left (\begin{tabular}{c}
$\kappa \mu'^2-\frac{\mu'}{2\eta_H}$ \\ $0$ \\ $0$ \\ $0$
\end{tabular}\right ), \quad{\rm and} \quad \vec{S_\delta}  = \left (\begin{tabular}{c}
$-\frac{\eta \mu'}{2\eta_H^2 \left (1-\frac{\eta}{\eta_H} \right )}$ \\ $0$ \\ $0$ \\ $0$
\end{tabular}\right ).
\end{eqnarray}
%

\subsection{Resolution of the linearised equations}
Again, making use of the linearity of the equations, we seek the solution under the form $\vec{X} = \vec{X}_0 + a_t \vec{X}_t + a_n \vec{X}_n + \delta \vec{X}_\delta$, where the vector $\vec{X}_\delta$ is solution of equation:
\begin{eqnarray}
\frac{d}{d \eta}\vec{X}_\delta= {\mathcal P} \vec{X}_\delta + \vec{S}_\delta
\qquad & \mbox{with} & \qquad
\vec{X}_\delta (0) =\left (\begin{tabular}{c}
$0$ \\ $0$ \\ $0$ \\ $0$
\end{tabular}\right ),
\end{eqnarray}
while $\vec{X}_0$, $\vec{X}_t$ and $\vec{X}_n$ are still solutions of equations~
(\ref{equaXsgeom})-(\ref{equaXngeom}). The bottom boundary conditions $U(0)=-1/(\kappa\eta_0)$ and $W(0)=0$ are then automatically satisfied. At the free surface, we impose the material nature of the surface, $W(\eta_H)=i \mu(\eta_H) \delta$, and vanishing stresses: $S_t(\eta_H)=\delta/\eta_H$ and $S_n(\eta_H)=\delta/( \eta_H\,\tan \theta)$. These last three conditions select the coefficients $a_t$ and $a_n$ as well as the value of $\delta$. Finally note that the analytical approximation of the solution close to the bottom in the limit $\eta_0 \to 0$ is the same as in the unbounded case -- it does not depend on the position of the upper boundary -- and expressions (\ref{Uasymp})-(\ref{Snasymp}) are thus still correct in the limit $H \gg z_0$.

\begin{figure}
\includegraphics{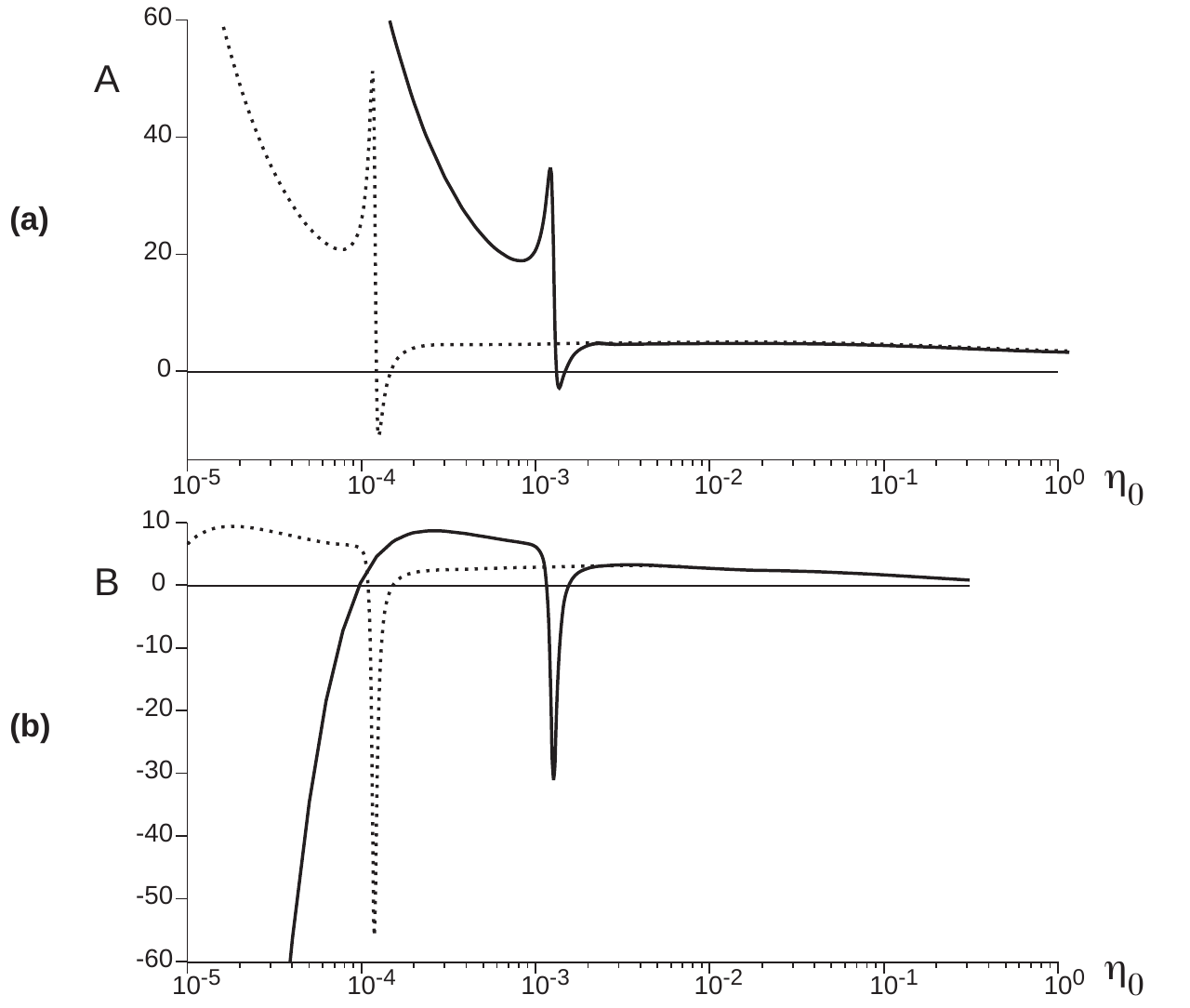}
\caption{$A$ and $B$ as functions of $\eta_0$ for $\Fr=0.9$ and $H/z_0=10^3$ (solid line) or $H/z_0=10^4$ (dotted line). In the right part of the plots, the curves collapse on the shape displayed in panels (a) and (b) of figure~\ref{ABCDgeom}. They differ at smaller $\eta_0$, showing a resonance peak and a divergence at $\eta_0 \to 0$.}
\label{StretchLimo}
\end{figure}
\begin{figure}
\includegraphics{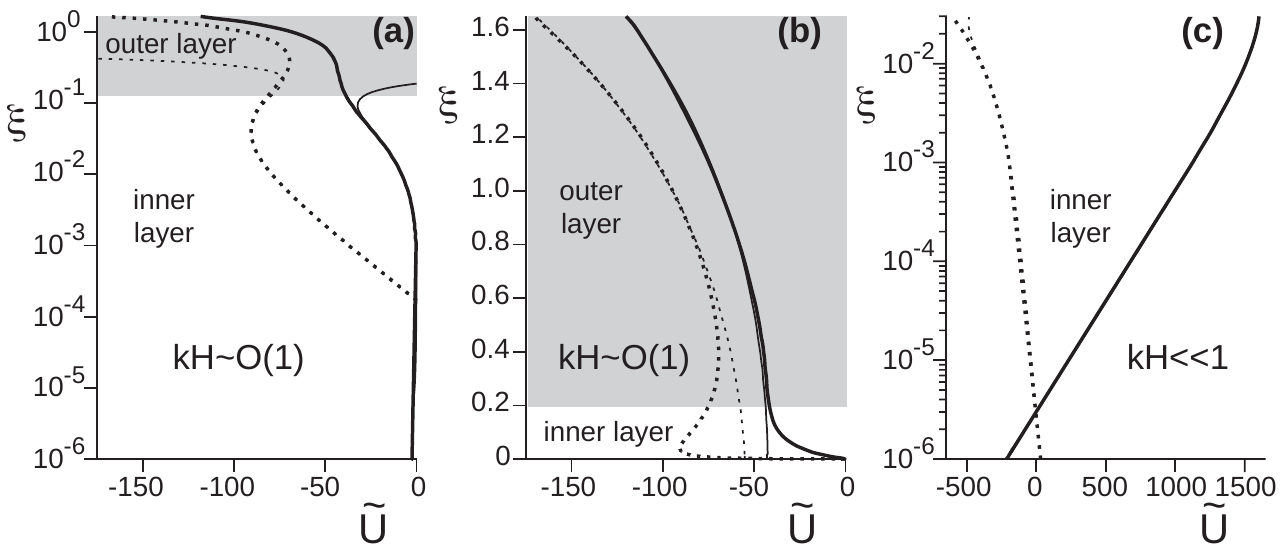}
\caption{Relative importance of the inner (white) and outer (grey) layers for  $kH=1.65$ (a)-(b), and $kH=0.03$ (c), with $H/z_0=10^{4}$. The velocity profiles (bold lines) are compared to their asymptotic behaviour in the inner and outer layers (thin lines). The solid lines represent the real part of the functions, and the dotted lines the imaginary ones. (a) and (b) show the very same profile, but with a logarithmic scale in (a) to emphasize the inner region. The thin lines in (a) and (c) represent the asymptotic behaviour in the inner layer. Those in (b) correspond to a sum of an increasing and a decreasing exponential of the form $\exp(\pm \eta)$, as for an inviscid potential flow.}
\label{ModesSurfaceLibre}
\end{figure}
%

\subsection{Results}
In order to evidence the role of the free surface, we have plotted the stress coefficients $A$ and $B$ as functions of $\eta_0$ in figure~\ref{StretchLimo}, for different values of $H/z_0$. For a large enough wave-number $k$ (a small enough wavelength $\lambda$), one recovers the plots of the panels (a) and (b) of figure~\ref{ABCDgeom}, independently of $H/z_0$. This means that for a bottom wavelength much smaller than the flow depth $H$ (i.e. for subaqueous ripples), the free surface has a marginal effect and the results of section~\ref{UnboundedCase} apply. For smaller $\eta_0$, however, the curves exhibit a peak, whose position depends on the value of $H/z_0$, followed by a diverging behaviour when $\eta_0 \to 0$. As discussed below, this peak can be ascribed to a resonance of standing waves at the free surface, excited by the bottom topography, meaning that the proper scale is now $H$ and not $z_0$. As the ratio $\lambda/H$ is the key parameter separating ripples from dunes, we shall turn extensively in part 2 to this point.

The analysis of velocity profiles for different values of $kH$ gives the following physical picture. For $kH > 1$, as for the unbounded case the flow can be thought of as being divided into two regions: an inner layer close to the bottom where it can be described by the equilibrium approximation and an outer layer behaving like an inviscid potential flow, where the profiles can be decomposed into the sum of decreasing and increasing exponentials $e^{\pm \eta}$. For smaller values of $kH$, this outer region progressively vanishes and the whole flow is controlled by the inner layer.
\begin{figure}
\includegraphics{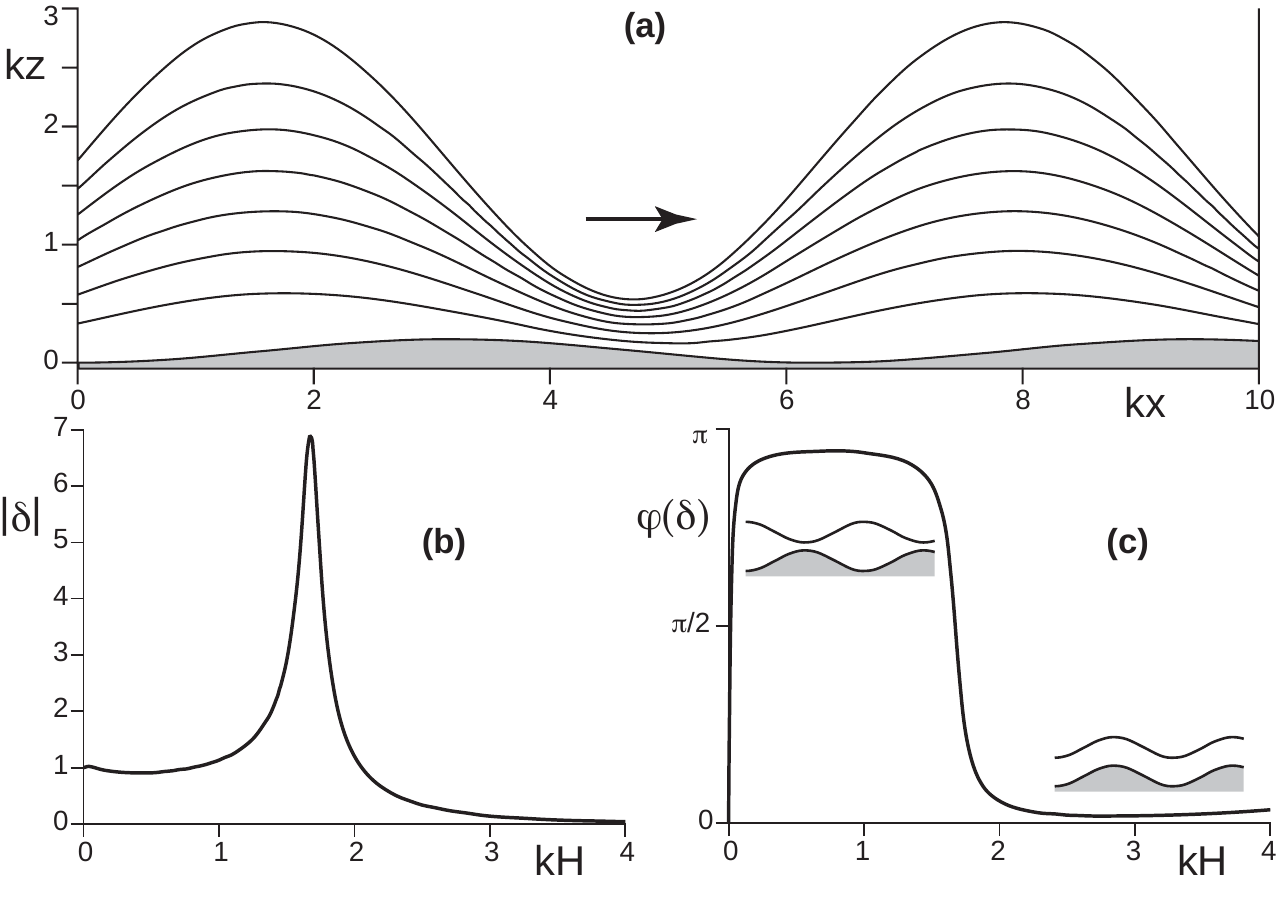}
\caption{(a) Streamlines of a flow over a sinusoidal bottom close to the free surface resonance conditions ($\varphi=\pi/2$). The flow is from left to right. Note the squeezing of the lines \emph{downstream} the crest of the bump. Amplitude $|\delta|$ (b) and phase $\varphi(\delta)$ (c) of the free surface as a function of $kH$ for $\Fr=0.8$. The peak in amplitude and the phase shift from $0$ to $\pi$ correspond to the resonance. The two schematics illustrate the situations in phase or in antiphase.}
\label{ResonanceFS}
\end{figure}
\begin{figure}
\includegraphics{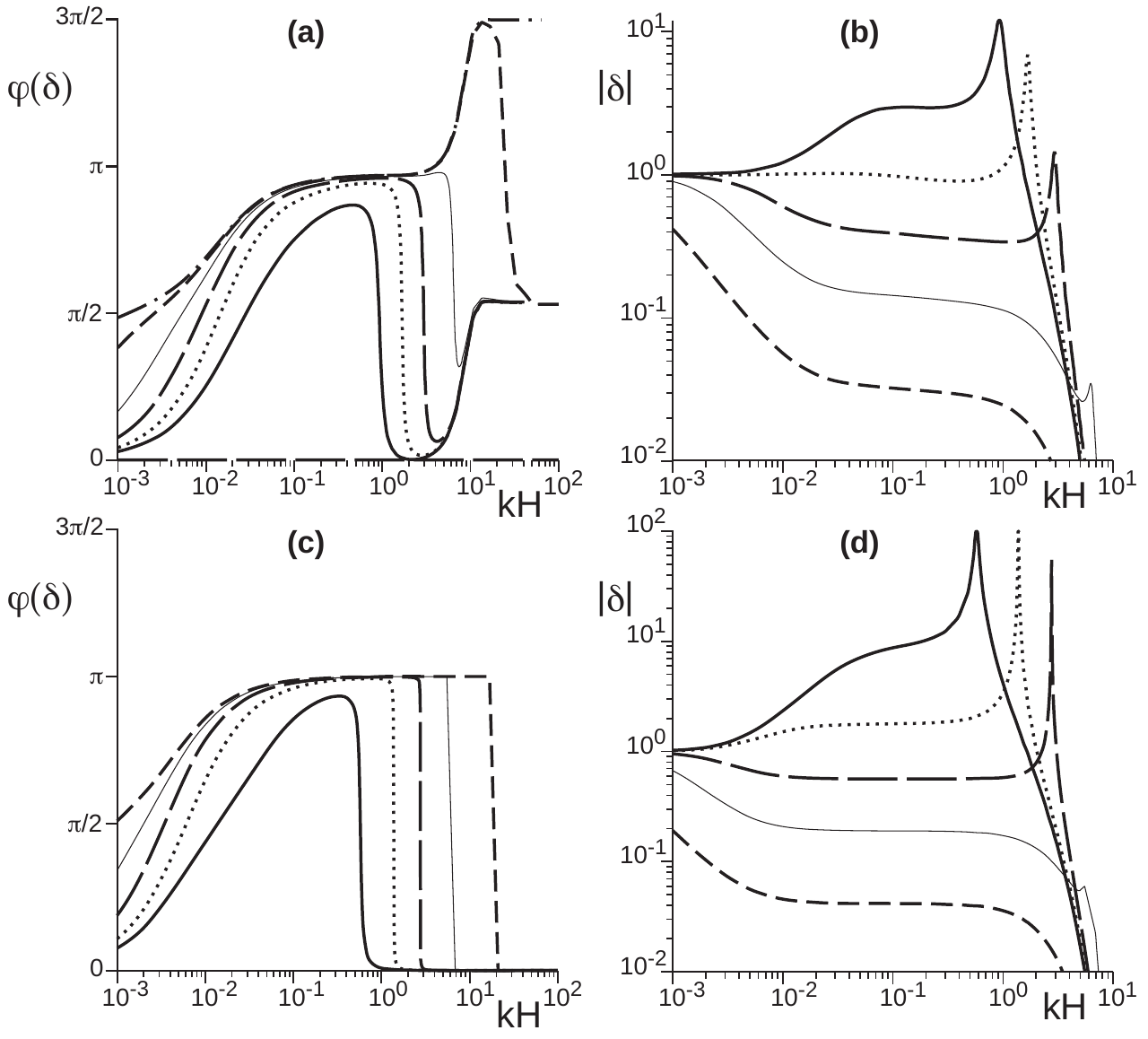}
\caption{The phase (a) and amplitude (b) of the rescaled free surface deformation $\delta=\Delta/\zeta$ as a function of $kH$ for $\Fr \to 0$ (dotted dashed line), $\Fr = 0.2$ (dashed line), $\Fr = 0.4$ (thin solid line), $\Fr = 0.6$ (long dashed line), $\Fr=0.8$ (dotted line) and $\Fr=1$ (solid line), and $H/z_0 = 10^{3}$. Crossing the resonance, the phase shifts from $0$ to $\pi$. (c) and (d), same for the friction force analytical model.}
\label{DeltaSurfaceLibre}
\end{figure}

We display the phase and amplitude of the free surface as a function of $kH$ in figure~\ref{ResonanceFS}(b-c). The peak in amplitude accompanied by the phase shift of $\pi$ are the signature of a surface wave resonance. The source of disturbances is of course the corrugation of the bed. For $kH$ larger than its resonant value, the bottom and the free surface are in phase; conversely, for $kH$ below the resonance, they are in antiphase. In between, at the resonance, the phase shift is $\varphi=\pi/2$ (figure~\ref{ResonanceFS}(a)) so that the streamlines are squeezed downstream to the crest. This resonance is model-independent as it comes from a very robust physical mechanism. As the fluid flows over the periodic bottom, gravity surface waves are excited at the wavelength $\lambda$. The latter propagate at the velocity:
\begin{equation}
c\simeq u_{\rm surface} \pm \sqrt{\frac{g}{k}\,\tanh(kH)}=\Fr\sqrt{gH} \pm \sqrt{\frac{g}{k}\,\tanh(kH)}
\end{equation}
with respect to the bottom (see the friction force model derived in Appendix~\ref{FFC}). As in the sound barrier phenomenon, the wave energy induced by the bottom disturbances accumulate when this velocity vanishes i.e. for:
\begin{equation}
\Fr=\sqrt{\frac{\tanh(kH)}{kH}}
\end{equation}
In the shallow water approximation ($kH\ll1$), this resonant condition gives $\Fr=1$ as standardly obtained in hydraulics. In the deep water approximation, it gives $\Fr=1/\sqrt{kH}$ or equivalently $kH=1/\Fr^2$ (\cite{K63}). So, for a bottom of wavelength $\lambda$, the flow is subcritical at low $\Fr$ and low $kH$ and supercritical at large $\Fr$ and $kH$. Ignoring dissipation, the Bernoulli relation states that the sum of the gravitational potential energy $\rho g \Delta$ and the kinetic energy $\frac{1}{2}\rho u_{\rm surface}^2$ is constant along the free surface. The subcritical regime corresponds to deep slow flows dominated by gravity: as the velocity increases over a bump, the corresponding increase of kinetic energy must be balanced by a loss of gravitational potential energy. As a consequence, the free surface is pinched over the bump ($\varphi=\pi$, see figure~\ref{ResonanceFS}c). The supercritical regime corresponds to thin rapid flows dominated by kinetic energy. By conservation of the flow rate, a pinch of the free surface would lead to an increased velocity. As the bump pushes up the free surface, the corresponding gain of potential energy should be balanced by a decrease of kinetic energy which is achieved by a deformation of the free surface in phase with the bump ($\varphi=0$, see figure~\ref{ResonanceFS}c). In summary, the free surface responds in phase with the excitation at small wavelength and becomes delayed as $\lambda/H$ increases. As in a standard second order linear system, the disturbance and the system response are in quadrature at the resonance.

The phase $\varphi$ and the rescaled amplitude $|\delta|$ of the free surface are displayed in figure~\ref{DeltaSurfaceLibre}(a)-(b) for different values of $\Fr$. In panels~\ref{DeltaSurfaceLibre}(c)-(d), they are compared to the analytical predictions obtained using a much simpler closure (Appendix~\ref{FFC}). One can see that the amplitude of the resonance increases with the Froude number. For very small $\Fr$, the phase curve is more complicated to interpret, but note that this corresponds to a vanishing amplitude $\delta$: the resonance essentially disappears. For $kH \to 0$, the free surface amplitude seems to converge to some finite value, but the phase slowly goes back to $0$. This gentle crossover is indeed expected at very large wavelength, a situation for which the free surface must follow the bottom topography.

\begin{figure}
\includegraphics{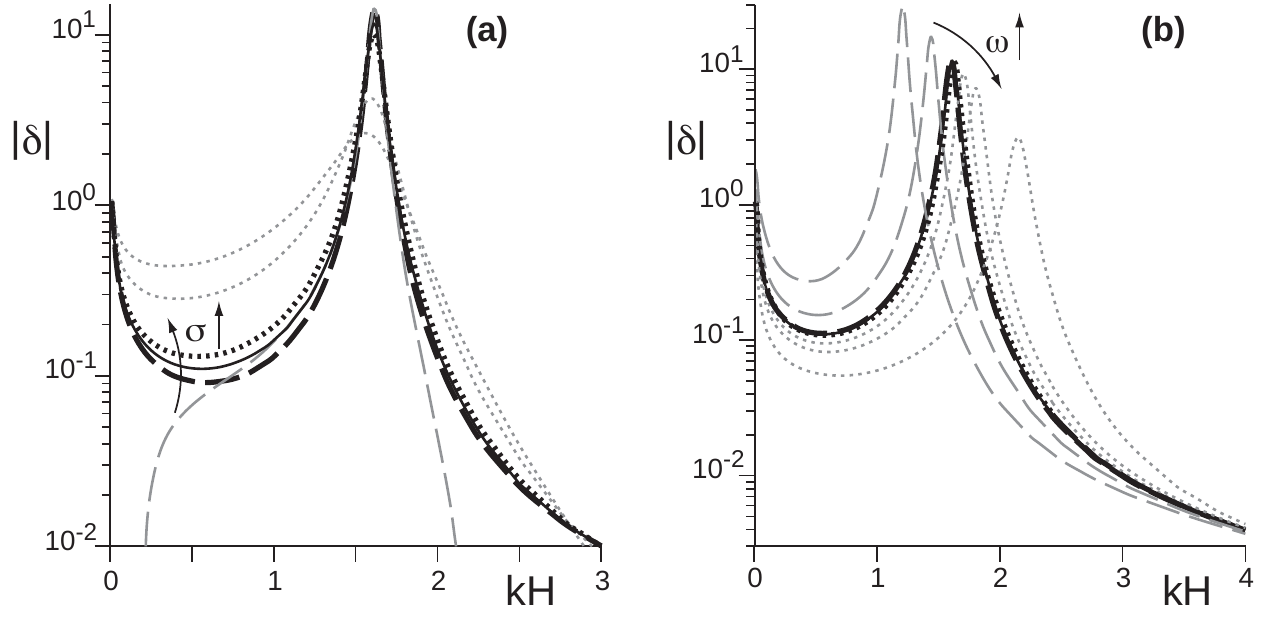}
\caption{Amplitude of the free surface $|\delta|$ as a function of $kH$ for different values of the growth rate $\sigma$ (a) and the pulsation $\omega$ (b) of the bottom boundary. Arrows indicate increasing values of $\sigma$ and $\omega$. These graphs have been computed with $H/z_0=10^4$ and $\Fr=0.8$. In panel (a), the grey dashed line is for $\frac{\sigma}{ku_*}=-1$, the back dashed line for $-0.1$, the black dotted line for $0.1$ and the two grey dotted lines for $1$ and $2$. In panel (b), the two grey dashed lines is for $\frac{\omega}{ku_*}=-5$ and $-2$, the black dashed line for $-0.1$, the black dotted line for $0.1$ and the three grey dotted lines for $1$, $2$ and $5$. For comparison, in all panels the solid lines correspond to the fixed case $\sigma=0$, $\omega=0$.}
\label{DeltaSigmaOmega}
\end{figure}

It is interesting to investigate how the resonance is affected by the fact that the bottom moves i.e. can grow or propagate. Following the notations introduced in~\ref{movingboundary}, we display in figure~\ref{DeltaSigmaOmega} the amplitude of the free surface $|\delta|$ as a function of $kH$ for different values of the growth rate $\sigma$ and the pulsation $\omega$. For positive growth rates $\sigma$, the Q-factor of the resonance gets smaller but the resonant wavenumber is not affected. A bottom propagating at the velocity $\omega/k$ moves the resonant peak along the $kH$-axis. For a positive propagation velocity $\omega/k$ the surface velocity with respect to the bottom and thus the effective Froude number get reduced. As a consequence, the resonant wavelength gets smaller --~and $kH$ larger. Conversely, the peak moves to smaller wavenumber for an upstream moving bottom. As in sub-section~\ref{movingboundary}, these effects are noticeable for values of $\frac{\sigma}{ku_*}$ and $\frac{\omega}{ku_*}$ of order one, while realistic values are respectively on the order of $10^{-2}$ and $10^{-3}$. For the ripples and dunes problem (part 2), the bedform motion can thus be safely ignored in the hydrodynamical calculation and the effect of free surface interpreted in terms of surface standing waves.

\begin{figure}
\includegraphics{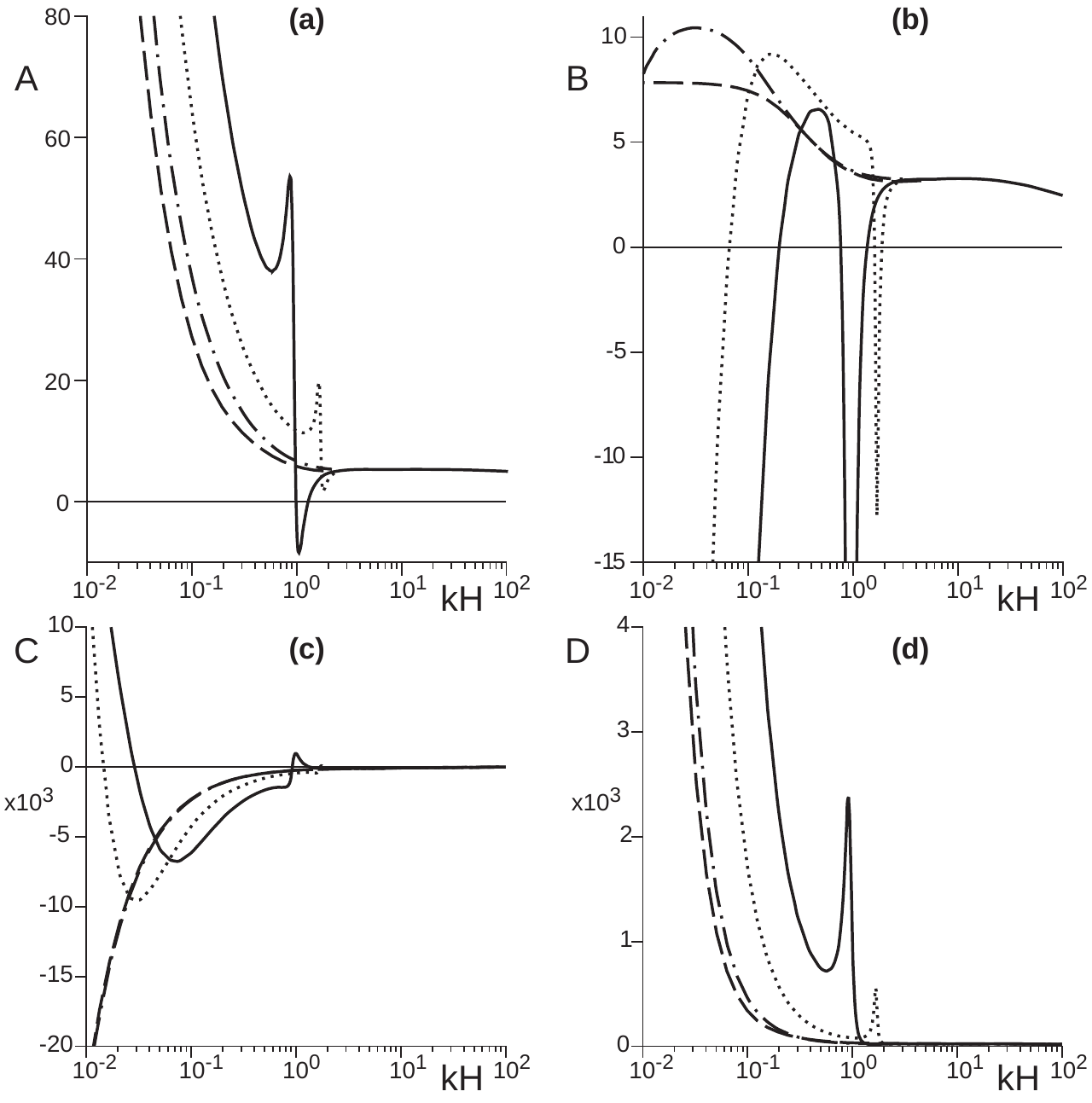}
\caption{$A$, $B$, $C$ and $D$ as functions of $kH$, for $\Fr = 0.1$ (dotted dashed line), $\Fr=0.8$ (dotted line) and $\Fr=1$ (solid line). The dashed lines correspond to a rigid boundary at the same height $H$. The plots have been computed for $H/z_0 = 10^{3}$. Comparing a free surface to a rigid boundary condition (or to the case $H \ll \lambda$), it can be inferred that the hydrodynamics is controlled by the surface waves. In particular, the resonance leads to a drop of the shear stress component $B$ i.e. to a downstream shift of the point of maximum shear stress.)}
\label{ABCDFreeSurface}
\end{figure}

The basal shear stress and pressure and subsequently the coefficients $A$, $B$, $C$ and $D$ are modified by the presence of the free surface when $kH$ is of order one and below. In figure~\ref{ABCDFreeSurface}, the coefficients are plotted as functions of $kH$ for different values of the Froude number. One can see that the resonance peak is more pronounced for larger $\Fr$ -- they are actually not visible when $\Fr$ is too small. In agreement with the streamlines of figure~\ref{ResonanceFS}(a), which shows a squeezing downstream the bump crest, the peak of $B$ is negative, corresponding to a phase delay of the stress with respect to the bottom. Furthermore, the curves corresponding to the presence of a rigid lid at the same height $H$ do not exhibit these peaks. Finally, the diverging behaviour of $B$ as $kH \to 0$ is also a free surface effect as, in the same limit, $B$ reaches a plateau in the case of a rigid top boundary. The behaviours of $A$, $B$, $C$ and $D$ at small $kH$ (below the resonant condition) can be determined analytically using the simple closure proposed in the Appendix~\ref{FFC}. For small $kH$, we get $A \propto 1/(kH)$, $B \propto -1/(kH)^2$, $C \propto 1/(kH)^3$ and $D \propto 1/(kH)^2$. These scalings fit fairly well the solutions of the full equations.

As a conclusion, there are two situations in which the excitation of standing waves by the topography affects significantly the characteristics of the inner layer: (i) around the resonance, since the surface wave amplitude is very large and (ii) for vanishing $kH$, when the distance $H$ between the topography and the free surface becomes so small that the inner layer invades the whole flow. 
%

\section{A qualitative summary of the results}
\label{qualitativesummary}

As this article is based on a rather technical ground, it is useful to sum up, in a qualitative manner, our key results and to put them in perspective with respect to the second part of the paper. In the context of the formation of ripples and dunes from a flat sand bed submitted to a turbulent flow, a central issue is the description of the basal shear stress, which controls bed load transport. Due to the scale separation between the typical evolution time of the bedforms and that of the flow, the bottom can be considered as quasi-static. The hydrodynamics can be investigated independently of the transport issue.

In the traces of the seminal work of \cite{JH75},  three main regions can be evidenced in the turbulent flow over a wavy bottom.
\begin{itemize}
\item An outer layer, away from the bottom, in which the flow is well described by inviscid potential equations, i.e. where the perfect flow approximation is valid.  The streamlines follow the topography so that the velocity is in phase with the bottom.
\item An inner layer, which corresponds to the region where the inertial terms of the Navier-Stokes equation are negligible, and thus where the longitudinal pressure gradient is balanced by the transverse mixing of momentum due to turbulent fluctuations i.e. by the Reynolds shear stress transverse gradient. The thickness $\ell$ of the inner layer is related to wavelength by $\lambda \sim \ell\,\ln^2(\ell/z_0)$. At the transition between the inner and outer layers, the fluid velocity is slowed down by the shear stress. Due to inertia, the velocity is always phase delayed with respect to the shear stress. As the velocity is inherited from the outer layer, the shear stress is phase-advanced with respect to the topography.  
\item A thin surface layer of thickness $h_0$, which is responsible for the hydrodynamical roughness $z_0$ seen from the inner layer. The dominant physical mechanism at work in this surface layer can be of different nature. For instance, $z_0$ can result from the mixing due to roughness elements, the predominance of viscous dissipation, or the presence of bed-load transport. Specific scaling laws for $z_0$ and $h_0$ are obtained for each of theses cases.
\end{itemize}

The linear relationship between the stresses and the bottom profile can be encoded into two complex coefficients $A+iB$ and $C+iD$. The tangent of the phase between the stresses and the bottom is given by the ratios $B/A$ and $D/C$ respectively. These coefficients are key inputs for the sediment transport issue (see part 2), and are needed to compute the dispersion relation of the bedforms, i.e. the growth rate as a function of the bedform wavenumber, which tells whether a given wavelength is stable or unstable. Their dependencies with respect to several important parameters such as the length scale ratios $z_0/\lambda$ or $H/\lambda$ have been determined. Their sensitivity to different ways of modelling turbulence or of imposing the bottom boundary conditions have also been discussed. Our conclusions regarding these functions can be summed up as follows:
\begin{itemize}
\item  $A$ and $B$ are generically positive, whereas $C$ is positive and $D$  negative. This means that the shear stress profile reaches its maximum before the crests of the bumps. This effect can be visualised on the streamlines, which are squeezed at this pint of maximum shear. For the normal stresses, the situation is opposite: the pressure maximum is slightly delayed after the crests. As the pressure is almost constant across the inner boundary layer, this phase shift $D/C$ is less pronounced by an order of magnitude. $A$, $B$, $C$ and $D$ have weak dependencies on the ratio $\lambda/z_0$. The shear stress phase shift $B/A$ vanishes for asymptotically small $kz_0$ and gently increases with $\ln(kz_0)$. As the inner layer thickness $\ell$ becomes of the order of the surface layer thickness $h_0$, the phase shift drops.
\item These features are robust  to (i) the turbulent closure, (ii) the existence of a Reynolds stress anisotropy, (iii) the motion (growth or propagation) of the bottom. 
\item Much more important is the role of the free surface in the case of a water depth $H$ on the order of the bottom wavelength $\lambda$. The undulations of the bottom excite standing gravity waves at the free surface. Resonant conditions are reached when these waves precisely propagate at a velocity equal to that of the flow, i.e. when $\Fr^2 \simeq \tanh(kH)/(kH)$. At the resonance, the response of the free surface is in quadrature with the disturbance. When the Froude number is large enough, the deformation $|\delta|$ of the free surface at the resonance is so large that it has a strong effect on the flow close to the bottom. In particular, the streamlines are squeezed downstream the crests of the bump, so that the shear stress becomes phase-delayed with respect to the topography ($B<0$). 
\item Another effect due to the presence of a free surface is found for $kH \to 0$. In this limit, the water depth becomes very thin in comparison to the wavelength, and the inner layer invades the whole flow (i.e. $\ell \simeq H$). In this situation, the shear stress and the bottom profiles tend to be in phase so that $B/A \to 0$. Moreover, the shear stress becomes phase delayed ($B<0$) below a threshold value of $kH$ that increases with the Froude number. 
\item The shear stress profile is insensitive to the mechanisms at work in the surface layer provided that its thickness $h_0$ is smaller than the inner layer thickness $\ell$: the hydrodynamical roughness $z_0$ is the single quantity inherited from the surface layer. The asymptotic calculation performed by \cite{JH75} is recovered but only for asymptotically large $\ln(\lambda/z_0)$, a limit hardly reached in real problems. When $h_0$ is comparable or larger than $\ell$, the shear stress coefficients $A$ and $B$ are smaller in the hydraulically smooth regime where the roughness is due to viscosity than in the hydraulically rough regime. In the later case, the mixing of momentum in the surface layer is dominated by the turbulent fluctuations induced by the roughness elements. In the ripples and dunes problem, these results directly apply to the case where sediments are transported with negligible feedback on the flow (erosion limited transport, see part 2). When the extra-stress due to bedload transport is significant (momentum limited transport),  we predict that $A$ and $B$ become larger.  but with an almost identical phase shift $B/A$. 
\item The normal stress profile is almost independent of the surface layer model.
\end{itemize}

Beyond the linear case, we have expended the hydrodynamical calculation to the third order in $k\zeta$. In contrast to the linear calculation, the separation of streamlines,  when the shear stress and the pressure gradient are antagonist, and the associated formation of a recirculation bubble are obtained for realistic values of the aspect ratio.  At the quadratic order, we get a correction to the mean velocity profile which corresponds to a roughness $z_g$ at large scale due to bottom corrugation. This roughness of geometrical origin depends on the `microscopic' roughness $z_0$ inherited from the surface layer and is not simply proportional to the amplitude of the bottom profile. We get the scaling law: $z_g=z_0\,\exp(\kappa\,(k\zeta)^2\,E)$, where the factor $E$ weakly depends on $\lambda/z_0$. Looking at the shear stress modulation, we have shown that the phase shift between the shear stress and the bottom profile is reduced by non-linear corrections and change sign at some particular aspect ratio. In the second part of this paper, this non-linear analysis is used to predict the selection of the aspect ratio of mature ripples and dunes.

\vspace*{0.3cm}

\noindent
\rule[0.1cm]{3cm}{1pt}

The understanding of the qualitative reason for the upstream shift of the maximum shear stress on a bump has been meditated with A.B. Murray. This work has benefited from the financial support of the french minister of research. We thank l'\'Eouv\'e for hospitality, where part of this paper was written.

\appendix
\section{An anisotropic turbulent closure}
\label{AvI}

The calculations of section~\ref{UnboundedCase} can be generalised in the case of the following anisotropic stress-strain relationship:
\begin{equation}
\tau_{ij} =\kappa^2 L^2 |\dot \gamma| \left(\frac{1}{3}\chi_i^2 |\dot \gamma| \, \delta_{ij} - \dot \gamma_{ij} \right).
\label{tauijPrandtl_AvI}
\end{equation}
At the linear order, the velocity, pressure and stress fields read:
\begin{eqnarray}
u_x & = & u_* \left[\mu+k\zeta e^{ikx} U \right],
\label{defU_AvI}\\
u_z & = & u_* k\zeta e^{ikx} W,
\label{defW_AvI}\\
\tau_{xz} & = & \tau_{zx}= - u_*^2 \left[1+k\zeta e^{ikx} S_t\right],
\label{defSt_AvI}\\
p+\tau_{zz} & = & p_0+u_*^2 \left[ \frac{1}{3}\chi_z^2 + k\zeta e^{ikx} S_n\right],
\label{defSn_AvI}\\
\tau_{zz} & = & u_*^2  \left[ \frac{1}{3}\chi_z^2 + k\zeta e^{ikx} S_{zz}\right],
\label{defSzz_AvI}\\
\tau_{xx} & = & u_*^2  \left[ \frac{1}{3}\chi_x^2 + k\zeta e^{ikx} S_{xx}\right],
\label{defSxx_AvI}
\end{eqnarray}
and the stress equations can be simplified into
\begin{eqnarray}
\mu' S_t & = & 2  (U'+iW)-2 \kappa^2 (\eta+\eta_0) \mu'^3,
\label{relaxSt_AvI}\\
\mu' S_{xx} & = & -2 i U+ \frac{2}{3}\chi_x^2(U'+iW) - \frac{2}{3}\chi_x^2\kappa \mu'^2,
\label{relaxSxx_AvI}\\
\mu' S_{zz} & = & - 2 W'+\frac{2}{3}\chi_z^2(U'+iW) - \frac{2}{3}\chi_z^2\kappa \mu'^2.
\label{relaxSzz_AvI}
\end{eqnarray}
The normal stress difference is this time:
\begin{equation}
S_{xx}-S_{zz} = \frac{-4 iU}{\mu'} + \frac{2}{3}\frac{\chi_x^2-\chi_z^2}{\mu'}(U'+iW-\kappa \mu'^2) \, ,
\label{diffSxxSzz_AvI}
\end{equation}
so that on gets the following four closed equations:
\begin{eqnarray}
U' & = & - i W +\frac{1}{2}\mu' S_t+\kappa \mu'^2,
\label{equaU'_AvI}\\
W' & = & -i U,
\label{equaW'_AvI}\\
S_t' & = & \left(i\mu +\frac{4}{\mu'}\right) U+  \mu' W + \frac{i}{3}(\chi_x^2-\chi_z^2)S_t + i S_n,
\label{equaSt'_AvI}\\
S_n' & = & -i \mu  W +i S_t.
\label{equaSn'_AvI}
\end{eqnarray}
As before, they can be written in the usual compact matrix form (\ref{systlinplaque}), now with
\begin{equation}
{\mathcal{P}} =  \left ( \!\!\!\!
\begin{tabular}{cccc}
$0$ & $-i $ & $\frac{1}{2}\mu' $ & $0$ \\
$-i$ & $0$ & $0$ & $0$ \\
$\left(i\mu +\frac{4}{\mu'}\right) $ & $\mu'$ & $\frac{i}{3}(\chi_x^2-\chi_z^2)$ & $i$ \\
$0$ & $-\mu i$ & $i$ & $0$
\end{tabular}
\! \right ) \! .
\label{matrixP_AvI}
\end{equation}
%

\section{A second order turbulent closure}
\label{2ndorder}

\subsection{Relaxation equation}
The dynamical equations governing the second-order moments $\tau_{ik}$ can be derived rigorously. Under the assumption of turbulence isotropy at the dissipative scale, it can be written under the form:
\begin{equation}
D_t \tau_{ik}=\partial_t \tau_{ik}+u_j \partial_j \tau_{ik} =-\tau_{kj} \partial_j u_i-\tau_{ij} \partial_j u_k-\partial_j \phi_{ik} - \pi_{ik}  - \frac{2}{3} \delta_{ik} \epsilon.
\end{equation}
$\epsilon$ is the dissipation rate; $\phi_{ik}=\overline{u'_i u'_j u'_k}$ is the spatial flux of turbulent energy induced by fluctuations; the pressure term $\pi_{ik}=\overline{u'_k \partial_i p'}+\overline{u'_i \partial_k p'}$ conserves energy and is usually responsible for the isotropisation of fluctuations.

We wish to get a stress tensor that relaxes towards its steady state expression prescribed by equation~(\ref{tauijPrandtl}). For dimensional reasons, we write the relaxation rate under the form $|\dot \gamma|/\beta$, where $\beta$ is a phenomenological constant, and keep the mixing length $L$ fixed by the geometrical distance to the wall. The second moment equation then takes the form of a first order relaxation equation:
\begin{equation}
D_t \tau_{ik}=\partial_t \tau_{ik}+u_j \partial_j \tau_{ik} =\frac{|\dot \gamma|}{\beta} \left[\kappa^2 L^2 \left(\delta_{ij} \frac{1}{3} \chi^2 |\dot \gamma|^2-|\dot \gamma| \dot \gamma_{ij}\right) -\tau_{ij} \right].
\label{2ndME_2ndorder}
\end{equation}
Setting $\beta=0$, one recovers the stationary solutions (\ref{tauijPrandtl}). A finite value of $\beta$ introduces a lag between a change of the flow velocity field and the point/time at which the Reynolds stress readapts to this change.

\subsection{Equations for 2D steady flows}
For 2D steady situations, the stress relaxation equations are the following:
\begin{eqnarray}
u_x \partial_x \tau_{xz}+u_z \partial_z \tau_{xz} &=&\frac{|\dot \gamma|}{\beta}  \left[-\kappa^2 L^2 |\dot \gamma| \dot \gamma_{xz}-\tau_{xz} \right],
\label{relaxxz_2ndorder}\\
u_x \partial_x \tau_{xx}+u_z \partial_z \tau_{xx} &=&\frac{|\dot \gamma|}{\beta} \left[-\kappa^2 L^2 |\dot \gamma| \dot \gamma_{xx}+ \frac{1}{3} \kappa^2 \chi^2 L^2 |\dot \gamma|^2-\tau_{xx} \right],
\label{relaxxx_2ndorder}\\
u_x \partial_x \tau_{zz}+u_z \partial_z \tau_{zz} &=&\frac{|\dot \gamma|}{\beta} \left[-\kappa^2 L^2 |\dot \gamma| \dot \gamma_{zz}+ \frac{1}{3} \kappa^2 \chi^2 L^2 |\dot \gamma|^2-\tau_{zz} \right].
\label{relaxzz_2ndorder}
\end{eqnarray}
At linear order, they simplify into:
\begin{eqnarray}
( \mu'+i\beta \mu) S_t & = & 2  (U'+iW)-2 \kappa^2 (\eta+\eta_0) \mu'^3, \label{relaxSt_2ndorder}\\
(\mu'+i \beta \mu) S_{xx} & = & -2 i U+ \frac{2}{3}\chi^2(U'+iW) - \frac{2}{3}\chi^2\kappa \mu'^2,
\label{relaxSxx_2ndorder}\\
(\mu'+i \beta \mu) S_{zz} & = & -2 W'+\frac{2}{3}\chi^2(U'+iW) - \frac{2}{3}\chi^2\kappa \mu'^2.
\label{relaxSzz_2ndorder}
\end{eqnarray}
Taking the difference of equations (\ref{relaxSxx_2ndorder}) and (\ref{relaxSzz_2ndorder}), one can compute
\begin{equation}
S_{xx}-S_{zz}= \frac{-4 iU}{\mu'+i\beta \mu}
\end{equation}
to obtain four closed equations:
\begin{eqnarray}
U'&=&- i W +\frac{ \mu'+i\beta \mu}{2} S_t+\kappa \mu'^2,
\label{equaU'_2ndorder}\\
W'&=&-i U,
\label{equaW'_2ndorder}\\
S_t' &=& \left(i\mu +\frac{4}{\mu'+i\beta \mu}\right) U+  \mu' W + i S_n,
\label{equaSt'_2ndorder}\\
S_n' &=&-  i \mu  W +i S_t.
\label{equaSn'_2ndorder}
\end{eqnarray}
As before, they can be written in the usual compact matrix form (\ref{systlinplaque}), now with
\begin{equation}
{\mathcal{P}} =  \left ( \!\!\!\!
\begin{tabular}{cccc}
$0$ & $-i $ & $\frac{\mu'+i\beta \mu}{2} $ & $0$ \\
$-i$ & $0$ & $0$ & $0$ \\
$\left(i\mu +\frac{4}{\mu'+i\beta \mu}\right) $ & $\mu'$ & $0$ & $i$ \\
$0$ & $-\mu i$ & $i$ & $0$
\end{tabular}
\! \right ) \! .
\label{matrixP_2ndorder}
\end{equation}
%

\section{Representation of the disturbances}
\label{rep}

\subsection{Linear order}
Recall that $Z(x)=\zeta e^{ikx}$ is the bottom profile whose wavelength is $\lambda=2\pi/k$, and $\eta=kz$ the dimensionless vertical coordinate. At the linear order, we write all the relevant quantities under the form:
\begin{equation}
f=\bar{f}(\eta)+k\zeta e^{ikx} f_1(\eta).
\end{equation}
An alternative is to use the curvilinear coordinates $\xi = \eta-kZ$ and write the field $f$ as:
\begin{equation}
f=\bar{f}(\xi)+k\zeta e^{ikx} \tilde{f}_1(\xi).
\end{equation}
We call these expressions respectively the `non-shifted' and `shifted' representations of $f$. They lead to the same linearised equations as they are related to each other at the linear order by 
\begin{equation}
\tilde{f}_1=f_1+\bar{f}'.
\end{equation}
In practice, this is especially relevant for $U$, for which $\bar{f}=\mu$ is not constant: $\tilde{U}=U+\mu'$ is the shifted representation of the first order correction to the horizontal velocity. However, $\tilde{W}=W$, $\tilde{S}_t=S_t$ and $\tilde{S}_n=S_n$. Importantly, note that the range in $\eta$ for which these two representations are valid is not the same \emph{a priori}.

\subsection{Representations for the non-linear expansion}
All fields are expanded up to the third order in $k\zeta$, neglecting also non-harmonic terms in $(k\zeta)^3 e^{\pm i3kx}$. Non-shifted representation of the streamwise velocity:
\begin{equation}
\frac{u_x}{u_*} = \mu(\eta)+(k\zeta) e^{ikx} U_1(\eta)+(k\zeta)^2 U_{0}(\eta)+(k\zeta)^2  e^{2ikx} U_{2}(\eta)+(k\zeta)^3 e^{ikx} U_{3}(\eta).
\end{equation}
Shifted representation of the same quantity:
\begin{equation}
\frac{u_x}{u_*} = \mu(\xi)+(k\zeta) e^{ikx} \tilde U_1(\xi )+(k\zeta)^2 \tilde U_0(\xi )+(k\zeta)^2  e^{2ikx} \tilde U_2(\xi )+(k\zeta)^3 e^{ikx} \tilde U_3(\xi).
\end{equation}
Expanding the functions $\tilde{f}_\alpha(\eta-k\zeta e^{ikx})$ with respect to $k\zeta$, one can relate a representation to the other as:
\begin{eqnarray}
U_1 &= &\tilde U_1-\mu', \\
U_0 &= &\tilde U_0+\frac{1}{4}\mu''-\frac{1}{4}(\tilde U_1'+\tilde U_1'^*), \\
U_2 &= &\tilde U_2+\frac{1}{4}\mu''-\frac{1}{2}\tilde U_1', \\
U_3 &= &\tilde U_3-\frac{1}{8}\mu'''-\tilde U_0'-\frac{1}{2}\tilde U_2'+\frac{1}{4}\tilde U_1''+\frac{1}{8}\tilde U_1''^*.
\end{eqnarray}
Conversely:
\begin{eqnarray}
\tilde U_1&=&U_1+\mu', \\
\tilde U_0&=&U_0+\frac{1}{4}\mu''+ \frac{1}{4}\left(U_1' +  U_1'^*\right), \\
\tilde U_2&=&U_2+\frac{1}{4}\mu'' +\frac{1}{2} U_1', \\
\tilde U_3&=&U_3+\frac{1}{8}\mu'''+\frac{1}{4} U_1''+\frac{1}{8} U_1''^*+ U_0'+\frac{1}{2} U_2'.
\end{eqnarray}
The passage from a representation to the other for the other fields works the same, except that there is no zeroth order (function $\mu$) in the expressions.

\section{Stream function}
\label{streamlines}

To compute the streamlines, we introduce the so-called stream function $\Psi(x,z)$, defined by $\dr \Psi / \dr x = - u_z$ and $\dr \Psi / \dr z = u_x$. This function is such that $\vec{u} \cdot \vec{\nabla}\Psi = 0$, so that the iso-contours $\Psi={\rm Cst}$ precisely show the streamlines. Using the continuity equation (\ref{NScont}), it is easy to show that a solution is $\Psi=\int \! d\breve{z} \, u_x$. This integral is computed between $\breve{z}=Z$ (the bottom) and $\breve{z}=z$. We note $\xi=\eta-kZ$ the rescaled distance to the bottom. Restricting to the linear order, with the relation $U_1=iW_1'$ (equation (\ref{equaW'})), we end up with
\begin{equation}
\Psi = \frac{u_*}{k} \left \{ (\xi+\eta_0) \mu(\xi) - \frac{1}{\kappa}\xi + k\zeta e^{ikx} \left [ iW_1(\xi)+\mu(\xi) \right ] \right \},
\end{equation}
where the function $\mu$ is given by expression (\ref{Mu2Etageom}).

In the situation with a free surface, one can use the following representation for the field $f$:
\begin{equation}
f=\bar{f} (\xi) + k\zeta e^{ikx} \tilde{f}_1 (\xi), \quad {\rm with} \quad \xi = \eta_H \frac{z-Z}{H+\Delta-Z} \, .
\end{equation}
This curvilinear variable $\xi$ vanishes on the bottom $z=Z$, and $\xi=\eta_H$ at the surface $z=H+\Delta$. The new function $\tilde{f}_1$ is related to those of the non-shifted representation $\bar{f}$ and $f_1$ as:
\begin{equation}
\tilde{f}_1 (\xi) = f_1 (\xi) + \left ( 1 + (\delta-1)\frac{\xi}{\eta_H} \right ) \bar{f}' (\xi).
\end{equation}
For $f=u_x$, we have $\bar{f}=\mu$ and $f_1=U_1=iW_1'$. Consequently, the new stream function is
\begin{equation}
\Psi_{FS} = \frac{u_*}{k} \left \{ (\xi+\eta_0) \mu(\xi) - \frac{1}{\kappa}\xi + k\zeta e^{ikx} \left [ iW(\xi)+\mu(\xi)+(\delta-1)\frac{\xi \mu(\xi)}{\eta_H} \right ] \right \}.
\end{equation}
One can check that the free surface is indeed a streamline itself, as one of the top boundary conditions is $W_1(\eta_H)=i\mu(\eta_H)\delta$.

In the non-linear (and unbounded) case, we get:
\begin{eqnarray}
\Psi_{NL}
&=& \frac{u_*}{k} \left \{ (\xi+\eta_0) \mu - \frac{1}{\kappa}\xi + (k\zeta) e^{ikx} \left( iW_1+\mu \right) \right .
\nonumber\\
&+& (k\zeta)^2 \left(\int_0^\xi U_0(\breve{\xi}
)d\breve{\xi}+\frac{1}{4}\mu'+ \frac{1}{4}\left(U_1 +  U_1^*\right) \right)
\nonumber\\
&+& (k\zeta)^2  e^{2ikx} \left(\frac{i}{2}~W_2+\frac{1}{4}\mu'
+\frac{1}{2} U_1\right)\nonumber\\
&+& \left .(k\zeta)^3 e^{ikx} \left(iW_3+\frac{1}{8}\mu''+\frac{1}{4}
U_1'+\frac{1}{8} U_1'^*+ U_0+\frac{1}{2} U_2\right) \right \}
\end{eqnarray}
%

\section{Weakly non-linear calculations}
\label{WNLC}

Definition of the different functions involved in the expansion:
\begin{eqnarray}
u_x & = & u_* \left[\mu+(k\zeta) e^{ikx} U_1+(k\zeta)^2 U_{0}+(k\zeta)^2  e^{2ikx} U_{2}+(k\zeta)^3 e^{ikx} U_{3}\right], \\
u_z & = & u_* \left[(k\zeta) e^{ikx} W_1+(k\zeta)^2  e^{2ikx} W_{2}+(k\zeta)^3 e^{ikx} W_{3}\right], \\
\tau_{xz} & = &  - u_*^2 \left[1+(k\zeta) e^{ikx} S_{t1}+(k\zeta)^2 S_{t0}+(k\zeta)^2  e^{2ikx} S_{t2}+(k\zeta)^3 e^{ikx} S_{t3}\right], \\
p+\tau_{zz} & = & p_0+u_*^2 \left[(k\zeta) e^{ikx} S_{n1}+(k\zeta)^2 S_{n0}+(k\zeta)^2  e^{2ikx} S_{n2}+(k\zeta)^3 e^{ikx} S_{n3}\right],\\
\tau_{zz}- \tau_{xx}& = & u_*^2  \left[(k\zeta) e^{ikx} S_{d1}+(k\zeta)^2 S_{d0}+(k\zeta)^2  e^{2ikx} S_{d2}+(k\zeta)^3 e^{ikx} S_{d3}\right].
\end{eqnarray}

Expansion of the mixing length:
\begin{equation}
\kappa^2(kL)^2 =\frac{1}{\mu'^2} -\frac{2\kappa}{\mu'}(k\zeta) e^{ikx}+\frac{\kappa^2}{2} (k\zeta)^2+\frac{\kappa^2}{2} (k\zeta)^2 e^{2ikx}.
\end{equation}

Expansion of the strain tensor components:
\begin{eqnarray}
\dot \gamma_{xx}&=&2\left((k\zeta) e^{ikx} iU_1+(k\zeta)^2  e^{2ikx} 2iU_{2}+(k\zeta)^3 e^{ikx} iU_{3}\right), \\
\dot \gamma_{zz}&=&-\dot \gamma_{xx}, \\
\dot \gamma_{xz}&=&\mu'+(k\zeta) e^{ikx} (U_1'+iW_1)+(k\zeta)^2 U_0'+(k\zeta)^2  e^{2ikx} (U_2'+2iW_{2}) \nonumber \\
&+&(k\zeta)^3 e^{ikx} (U_{3}'+iW_{3}),
\end{eqnarray}
which gives for the strain modulus:
\begin{eqnarray}
|\dot \gamma|&=&\mu'+(k\zeta) e^{ikx} (U_1'+iW_1)+(k\zeta)^2 \left[U_0'+\frac{U_1U_1^*}{\mu'}\right]+(k\zeta)^2  e^{2ikx}\left[U_2'+2iW_{2}-\frac{U_1^2}{\mu'}\right] \nonumber \\
&+&(k\zeta)^3  e^{ikx}\left[U_3'+iW_{3}-\frac{U_1 U_1^*}{\mu'^2}(U_1'+iW_1)+\frac{U_1^2}{2\mu'^2}(U_1'^*-iW_1^*)+\frac{4U_1^*U_2}{\mu'}\right],
\end{eqnarray}
and then
\begin{eqnarray}
& & \kappa^2 (kL)^2 |\dot \gamma|=\frac{1}{\mu'}+(k\zeta) e^{ikx} \left[\frac{1}{\mu'^2}(U_1'+iW_1)-2\kappa\right]\nonumber\\
&+&(k\zeta)^2 \left[\frac{\kappa^2\mu'}{2}-\frac{\kappa}{2\mu'}(U_1'+U_1'^*+iW_1-iW_1^*)+\frac{1}{\mu'^2}\left(U_0'+\frac{U_1U_1^*}{\mu'}\right)\right]\nonumber\\
&+&(k\zeta)^2 e^{2ikx}\left[\frac{\kappa^2\mu'}{2}-\frac{\kappa}{\mu'}(U_1'+iW_1)+\frac{1}{\mu'^2}\left(U_2'+2iW_{2}-\frac{U_1^2}{\mu'}\right)\right]\nonumber\\
&+&(k\zeta)^3 e^{ikx}\left[\frac{\kappa^2}{4}\left(2U_1'+U_1'^*+2iW_1-iW_1^*\right)-\frac{\kappa}{\mu'} \left(2U_0'+\frac{U_1(2U_1^*-U_1)}{\mu'}+U_2'+2iW_{2}\right)\right.\nonumber\\
&+&\left.\frac{1}{\mu'^2}\left(U_3'+iW_{3}-\frac{U_1 U_1^*}{\mu'^2}(U_1'+iW_1)+\frac{U_1^2}{2\mu'^2}(U_1'^*-iW_1^*)+\frac{4U_1^*U_2}{\mu'}\right)\right].
\end{eqnarray}

Expressions of the different functions corresponding to the normal stresses $\tau_{zz}-\tau_{xx}$:
\begin{eqnarray}
S_{d1}&=& \frac{4i}{\mu'}  U_1, \\
S_{d0}&=& \frac{i}{\mu'^2}\left(U_1(U_1'^*-iW_1^*)-U_1^*(U_1'+iW_1)\right)+2i\kappa(U_1^*-U_1), \\
S_{d2}&=&\frac{8i}{\mu'} U_2+\frac{2i}{\mu'^2}U_1(U_1'+iW_1)-4i\kappa U_1, \\
S_{d3}&=&\frac{4i}{\mu'} U_3 +\frac{4i}{\mu'^2}U_2(U_1'^*-iW_1^*)-8i\kappa U_2+i\kappa^2\mu' (2U_1-U_1^*)+\frac{6i}{\mu'^3}U_1^2U_1^* \\
&+&\frac{2i\kappa}{\mu'}(U_1'+iW_1)(U_1^*-U_1)-\frac{2i\kappa}{\mu'}U_1(U_1'^*-iW_1^*)+\frac{4i}{\mu'^2}U_1U_0'-\frac{2i}{\mu'^2}U_1^*(U_2'+2iW_{2}). \nonumber
\end{eqnarray}

Expressions of the different $\vec{S}_\alpha$:
\begin{eqnarray}
\vec{S}_1 & = & \left (
\begin{tabular}{l}
$\kappa \mu'^2$ \\ $0$ \\ $0$ \\ $0$
\end{tabular}
\right ), \\
\vec{S}_0 & = & \left (
\begin{tabular}{l}
\begin{tabular}{ll}
$-\frac{\kappa^2\mu'^3}{4}$ &
$ - \,\,\, \frac{1}{4\mu'}(U_1'+iW_1)(U_1'^*-iW_1^*)$ \\
\quad &
$+ \,\,\, \frac{\kappa \mu'}{2}(U_1'+U_1'^*+iW_1-iW_1^*) - \frac{1}{2\mu'}U_1U_1^*$
\end{tabular}
\\
$0$
\\
$\frac{1}{4}(W_1U_1'^*+W_1^*U_1')$
\\
$\frac{i}{2} (U_1W_1^*-U_1^*W_1)$
\end{tabular}
\right ), \\
\vec{S}_2 & = & \left (
\begin{tabular}{l}
$-\frac{\kappa^2\mu'^3}{4} - \frac{1}{4\mu'}(U_1'+iW_1)^2 + \kappa \mu' (U_1'+iW_1) + \frac{1}{2\mu'}U_1^2$
\\
$0$
\\
$\frac{1}{2} W_1 U_1' + \frac{i}{2} U_1^2 + \frac{4}{\mu'2} U_1(U_1'+iW_1) - 8\kappa U_1$
\\
$0$
\end{tabular}
\right ), \\
\vec{S}_3 & = & \left (
\begin{tabular}{l}
\begin{tabular}{ll}
$2\kappa \mu' U_0$ &
$+ \,\,\, \frac{\kappa}{2} U_1(2U_1^*-U_1) + \kappa \mu'(U_2'+2iW_{2}) - \frac{2}{\mu'} U_1^*U_2$ \\
\quad &
$- \,\,\, \frac{\mu'}{2}(U_1'+iW_1)\left[\kappa^2\mu'-\frac{\kappa}{2\mu'}(U_1'+2U_1'^*+iW_1-2iW_1^*)+\frac{2}{\mu'^2}U_0'\right]$ \\
\quad &
$- \,\,\, \frac{\mu'}{4}(U_1'^*-iW_1^*)\left[\kappa^2\mu'+\frac{2}{\mu'^2}(U_2'+2iW_{2})\right]$
\end{tabular}
\\
$0$
\\
\begin{tabular}{ll}
$i U_0 U_1$ &
$+ \,\,\, \frac{i}{2} U_2 U_1^* + U_0' W_1 + \frac{1}{2} W_1^* U_2' + \frac{1}{2} W_2 U_1'^*$ \\
\quad &
$+ \,\,\, \frac{4}{\mu'^2}U_2(U_1'^*-iW_1^*)-8\kappa U_2+\kappa^2\mu' (2U_1-U_1^*)+\frac{6}{\mu'^3}U_1^2U_1^*$ \\
\quad &
$+ \,\,\, \frac{2\kappa}{\mu'}(U_1'+iW_1)(U_1^*-U_1)-\frac{2\kappa}{\mu'}U_1(U_1'^*-iW_1^*)+\frac{4}{\mu'^2}U_1U_0'$ \\
\quad &
$- \,\,\, \frac{2}{\mu'^2}U_1^*(U_2'+2iW_{2})$
\end{tabular}
\\
$-i U_0 W_1 - \frac{3i}{2} W_2 U_1^* + \frac{3i}{2} U_2 W_1^*$
\end{tabular}
\right ).
\end{eqnarray}
%

\section{A friction force closure}
\label{FFC}

Several of the free surface effects can be recovered within a simple friction force model, for which analytical expressions of the linear solution of the flow can be derived. In particular, the resonance condition as well as the behaviour of the basal stress coefficients $A$, $B$, $C$ and $D$ for $kH \to 0$ can be found and interpreted.

\subsection{Reference state}
We start from the Navier-Stokes equations for a perfect flow, with a crude additional turbulent friction term as an approximation of the stress derivatives:
\begin{eqnarray}
\partial_x u_x+ \partial_z u_z & = & 0, \\
u_x \partial_x u_x +u_z \partial_z u_x & = & -\partial_x p+g \sin \theta - \Omega \frac{u_x}{H} u_x , \\
u_x \partial_x u_z+u_z \partial_z u_z & = & -\partial_z p- g \cos \theta - \Omega \frac{2 u_x}{H} u_z,
\end{eqnarray}
Physically, the force applied to a fluid particle is directly related to the relative velocity with respect to the ground. At an angle $\theta$, the following plug flow is an homogeneous solution of the above equations:
\begin{eqnarray}
u_{x}&=&\overline{u}=\sqrt{\frac{g H \sin \theta}{\Omega}} \, ,
\label{ux0_FFC}\\
u_{z} & = & 0 \, ,
\label{uz0_FFC} \\
p&=&g \cos \theta (H-z).
\label{p0_FFC}
\end{eqnarray}
In order to estimate the value of the friction coefficient, one can make use of the fact that typical turbulent velocity vertical profiles are logarithmic. However, as the logarithm varies slowly when $z$ is much larger than $z_0$, we write $\overline{u} \sim \frac{1}{H} \int_0^H \! dz \, u_x(z) \sim \frac{u_*}{\kappa} \left ( \ln \frac{H}{z_0} -1 \right )$. Identifying the shear stress on the bottom as $u_*^2=gH\sin\theta$, we finally get with the relation (\ref{ux0_FFC})
\begin{equation}
\Omega \sim \left ( \frac{\kappa}{\ln \frac{H}{z_0}-1} \right )^2 \, .
\label{lambda_FFC}
\end{equation}
For $H/z_0$ in the range $10^3$-$10^4$, we get a typical value for $\Omega$ on the order of few $10^{-3}$. We now normalize quantities by $\overline{u}$ and $H$ and get a single non-dimensional (Froude) number:
\begin{equation}
\Fr=\frac{\overline{u}}{\sqrt{g H \cos \theta}} \, .
\label{Froude_FFC}
\end{equation}
%

\subsection{Disturbance}
The starting equations can be linearised around the above reference state. Looking a the flow over a corrugated bottom $Z(x)=\zeta e^{ikx}$, it is easy to show that the solution is of the following form
\begin{eqnarray}
u_x & = & \overline{u} + \overline{u}k\zeta e^{ikx} \left [ -a_+ e^{kz} + a_- e^{-kz} \right ], \\
u_z & = & \overline{u}ik\zeta e^{ikx} \left [ a_+ e^{kz} + a_- e^{-kz} \right ], \\
p & = & g \cos \theta (H-z) + \overline{u}^2 (kH - i 2\Omega) \frac{\zeta e^{ikx}}{H} \left [ a_+ e^{kz} - a_- e^{-kz} \right ],
\end{eqnarray}
where $a_+$ and $a_-$ must be determined by the boundary conditions. This exponential form is characteristic of potential flows.

\subsection{Boundary conditions}
We require that the velocity normal to the bottom vanish. Following the notations of the main part of the paper, we define $\Delta$ such that the free surface is at the altitude $H+\Delta$. It is a material line where the pressure vanishes. The three boundary conditions are then:
\begin{eqnarray}
u_z(z=0)&=&i \overline{u} k\zeta e^{ikx}, \\
u_z(z=H)&=&i \overline{u} \delta k\zeta e^{ikx}, \\
p(z=H)&=& \frac{\overline{u}^2}{H \Fr^2} \, \delta \zeta e^{ikx},
\end{eqnarray}
where, as before, $\delta$ is defined as $\Delta(x)=\delta \zeta e^{ikx}$. The constants $a_+$ and $a_-$, as well as $\delta$ are thus solutions of
\begin{eqnarray}
a_+ + a_- & = & 1 \, , \label{FSFF1}\\
a_+ e^{kH} + a_- e^{-kH} & = & \delta \, , \label{FSFF2}\\
a_+ e^{kH} - a_- e^{-kH} & = & \frac{\delta}{(kH-i 2 \Omega) \fr^2} \, , \label{FSFF3}
\end{eqnarray}
from which we get:
\begin{eqnarray}
a_+ & = & \frac{1}{2} \left [ 1 -
\frac{(kH-i2\Omega)\tanh kH - \frac{1}{\fr^2}}
     {(kH-i2\Omega) - \frac{1}{\fr^2} \tanh kH}
\right ], \\
a_- & = & \frac{1}{2} \left [ 1 +
\frac{(kH-i2\Omega)\tanh kH - \frac{1}{\fr^2}}
     {(kH-i2\Omega) - \frac{1}{\fr^2} \tanh kH}
\right ].
\end{eqnarray}
%

\subsection{Basal shear stress and pressure}
The shear stress is not part of the variables of this model, but we can consistently define it as $\tau= -\Omega u_x^2$. Looking at the shear stress $\tau_b$ and normal stress $p_b$ on the bottom, in accordance with the notations of the previous sections of the paper, we introduce the coefficients $A$, $B$, $C$ and $D$ as
\begin{eqnarray}
\tau_b & = & -\Omega \overline{u}^2 \left [ 1 + (A+iB) k\zeta e^{ikx} \right ], \\
p_b & = & gH\cos\theta + \Omega \overline{u}^2 (C+iD) k\zeta e^{ikx},
\end{eqnarray}
which gives
\begin{eqnarray}
A & = & 2 \, \frac{\left [ (kH)^2 + 4\Omega^2 + \frac{1}{\fr^4} \right ] \tanh kH - \frac{1}{\fr^2} \, kH [\tanh^2 kH+1]}
{ \left ( kH-\frac{1}{\fr^2}\tanh kH \right )^2 + 4\Omega^2} \, , \\
B & = & \frac{2\Omega}{\fr^2} \, \frac{[\tanh^2 kH-1]}
{\left ( kH-\frac{1}{\fr^2}\tanh kH \right )^2 + 4\Omega^2} \, , \\
C & = & \frac{1}{2\Omega} \left ( -A-\frac{2\Omega B}{kH} \right ), \\
D & = & \frac{1}{2\Omega} \left ( -B+\frac{2\Omega A}{kH} \right ).
\end{eqnarray}
It is worth noting that the friction force model predicts negative values of $B$ for any $kH$. This means that there is always a phase delay of the shear stress with respect to the bottom, which is a clear disagreement with the full solution. In order to fix this flaw, one would need to empirically introduce an imaginary part to $\Omega$. Finally, this discrepancy shows that a precise description of the phase between the basal friction and the relief is a subtle and difficult issue that fully justify the use of a rigorous but heavy formalism.


\end{document}